\documentclass[12pt]{article}
\usepackage{amsmath}
\usepackage{amssymb}
\usepackage{subfigure}
\usepackage{psfrag}
\usepackage{epsfig}
\usepackage{graphicx}
\usepackage{tikz}
\usepackage{cite}
\usepackage{hyperref}
\usepackage[font=footnotesize,labelfont=bf]{caption}

\usepackage[numbers,sort&compress]{natbib}

\newcount\xrefpos \xrefpos=0
\newcount\yrefpos \yrefpos=0
\newcount\xput \xput=0
\newcount\yput \yput=0

\def\refpos#1 #2 #3{\global\xrefpos=#1 \global\yrefpos=#2
                         \rlap{$\smash{#3}$}}
\def\put #1 #2 #3{\xput=#1 \yput=#2
                  \advance\xput by -\xrefpos
                  \advance\yput by -\yrefpos
                  \rlap{\kern\the\xput truebp
                        \vbox to 0pt{\vss\hbox{$\displaystyle #3$}
                        \kern\the\yput truebp}}}
\def\beginlabels\refpos#1\endlabels{\hbox{$\refpos#1$}}

\newcommand{\ba}{\begin{eqnarray}}
\newcommand{\ea}{\end{eqnarray}}
\newcommand{\beq}{\begin{equation}}
\newcommand{\eeq}{\end{equation}}

\usepackage{amsfonts,amssymb}
\begin{document}

 \begin{center}
 {\Large \bf \    Fermi Liquids from D-Branes}

\bigskip
\bigskip
\bigskip
\bigskip

\vspace{3mm}

Moshe Rozali\footnote{email: rozali@phas.ubc.ca}, Darren Smyth\footnote{email: dsmyth@phas.ubc.ca}

\bigskip\medskip
\centerline{\it Department of Physics and Astronomy}
\smallskip\centerline{\it University of British Columbia}
\smallskip\centerline{\it Vancouver, BC V6T 1Z1, Canada}

 \bigskip\bigskip\bigskip

 \end{center}
 
 \abstract{We discuss finite density configurations on probe D-branes, in the presence of worldvolume fermions. To this end we consider a phenomenological model whose bosonic sector is governed by the DBI action, and whose charged sector is purely fermionic. In this model, we demonstrate the existence of a compact worldvolume embedding, stabilized by a Fermi surface on the D-brane. The finite density state in the boundary QFT is a Fermi-like liquid. We comment on the possibility of realizing non-Fermi liquids in this setup.}
 \newpage

\section{Introduction and Outline}
        The gauge-gravity duality \cite{Witten:1998qj,Maldacena:1997re} has become a useful tool in the study of strongly coupled quantum systems.  Gravitational duals have been useful in the qualitative study of systems as diverse as QCD and high temperature superconductors. In particular, QFT-gravity duals have been constructed which realize some of the fascinating physics of non-Fermi liquids in 2+1 dimensions. Following the realization that probe fermions in the AdS Reissner-Nordstrom black hole exhibit features characteristic of a non-Fermi liquid \cite{Liu:2009dm,Cubrovic:2009ye}, much effort was devoted to the difficult task of going beyond the probe limit. See in particular \cite{Sachdev:2011ze,Allais:2012ye,Allais:2013lha}  for works most relevant to our current endeavour.

While the above gravity duals utilize the bulk closed-string sector, many holographic models also utilize open string sectors, i.e. probe D-branes \cite{Karch:2002sh} which may be embedded in an ambient space-time without back-reacting on the geometry. In this work we use such D-brane constructions to  study  a new class of  holographic matter resulting from the inclusion of worldvolume fermions. In the spirit of bottom-up holography we  consider a model which includes the minimal set of ingredients
needed to construct the state we are interested in. Our bosonic fields are then a gauge field and an embedding function, governed by the Dirac-Born-Infeld (DBI) action. These are accompanied by charged world-volume fermionic matter, which for the sake of simplicity we choose to be a massive Dirac fermion on the D-brane. 

The bulk solutions in our model consist of a compact (``Minkowski") brane embeddings whose gauge field
and embedding function are coupled to a finite density of charged
 fermions on the world-volume. Such compact embeddings are known to be unstable if the charged matter is bosonic. This instability can be understood as a result of Bose-Einstein condensation of the charged bosons at the point of the brane cap off. This will manifest itself geometrically in the brane embedding being pulled towards the interior of the geometry \cite{Kobayashi:2006sb}. The new ingredient for us is Fermi statistics, resulting in an additional effective pressure, the Fermi pressure of the worldvolume Fermi surface. By constructing the state numerically, we explicitly show that for sufficiently dense fermions, such pressure  can stabilize the compact brane  embedding.

The state in the dual field theory is shown to be a Fermi liquid, in that it has a sharp Fermi surface (at zero temperature), and the low energy fermionic degrees of freedom have a quasiparticle description.  The resulting Fermi-like liquid is similar to that constructed in \cite{Sachdev:2011ze}. We demonstrate the existence of a Fermi surface and discuss some qualitative features of the quasiparticle scattering rate. We also identify limits of parameter space where perturbation theory is likely to break down, resulting in a qualitative change in the nature of the fermionic state.
We therefore propose that this state may  provide a useful starting point for the construction of non-Fermi liquids. Such a construction will need to tackle the difficult issues addressed in \cite{Allais:2012ye,Allais:2013lha} in the gravitational context.

The layout of this paper is as follows. In section \ref{section1} we  introduce the bosonic and fermionic components of our action and derive the equations of motion and boundary conditions. We also take this opportunity to discuss the various parameters and couplings in our phenomenological probe brane action. In  section 3 we explain our numerical process, especially the unique features  associated with imposing Fermi statistics, following the discussion of \cite{Sachdev:2011ze}. We then present our solutions for the bulk fields, first  in the probe limit and then including the backreaction of the fermions on the  brane embedding.  We focus especially on identifying limits where the perturbative expansion in $\frac{1}{\mathcal{N}}$ , utilized here, is likely to break down. In section 4 we analyze   the state of the dual QFT. We demonstrate the existence of a Fermi surface via examination of the retarded Green's function and discuss the equation of state. We conclude with some remarks on potential avenues for  future research.

\bigskip                        
\section{Setup: Equations and Boundary Conditions} \label{section1}

We take a phenomenological, bottom-up approach to the problem of constructing finite density fermionic states on probe D-branes. Thus, in constructing our finite density state we do not commit to the matter content and full set of couplings resulting from any specific brane configuration. Rather, we take the minimal set of ingredients necessary to construct the state we are interested in. Here we enumerate those basic ingredients needed for the construction. We comment below on the expected impact of varying the matter content and couplings of our phenomenological model.
 \bigskip
\subsection{Bosonic sector}

The starting point for our phenomenological model is a Dp-Dq system in the holographic decoupling limit: a single (or a few) Dq-probe branes in the near-horizon geometry of a stack of Dp-branes. Since we would like to study a $2+1$ dimensional QFT, we choose $p=2$.  The simplest type of probe brane is $q=p+4$, a system that has been extensively studied starting with \cite{Karch:2002sh}. We choose therefore to study the worldvolume dynamics of a single D6 brane in the near horizon geometry of a stack of D2-branes\footnote{We will be interested in the region of parameter space for which the resulting geometry is well-described by type IIA supergravity in ten dimensions.}. This brane configuration was discussed in the holographic context, e.g. in\cite{Karch:2009eb,Mateos:2007vn,Kobayashi:2006sb}.

This construction provides the basic elements needed to study holographic finite density matter: a world volume gauge field which can be sourced by a chemical potential, and an embedding function which can adjust to the presence of finite density matter. These are analogous to the metric and bulk gauge field in holographic constructions utilizing the bulk closed string sector, the main difference being the DBI action controlling their couplings. Here we explore consequences of these differences.

Finite density holographic matter on D-branes, in the absence of charged fermions, was discussed in \cite{Kobayashi:2006sb}.  In the presence of non-zero density only the ``black hole" embedding exists. This is an embedding for which the probe brane is extended through the horizon of the bulk black hole (or the Poincare horizon if zero temperature is considered). The instability of a compact, ``Minkowski" embedding is intuitive: In the presence of finite density, and therefore finite electric flux, sources for the electric field are needed for a compact embedding. The analysis in \cite{Kobayashi:2006sb} then shows that the available (bosonic) charged sources are strings connecting the brane to the horizon, which inevitably pull the brane embedding towards the horizon. One of the consequences of our construction is that this outcome may be avoided in the presence of charged fermion sources. The reasoning behind this is also intuitive: if the charged sources are fermions,  the Pauli exclusion principle dictates they form a Fermi surface. The resulting Fermi pressure counteracts the pull towards the horizon, potentially resulting in a stable configuration. Here we give an example of such a configuration.
 
 The near horizon geometry of the D2-brane stack, at zero temperature, is:
\begin{align}\label{background_metric}
ds^2 =\frac{ -dt^2 + dx^2+dy^2 }{u^{5/2}}+\frac{du^2}{u^{3/2}} + u^{1/2} \left(d\mathcal{S}_3^2  + \sin^2 (\theta)  \,d \tilde{\mathcal{S}}_3^2\right)
 \end{align} 
where we use the function $0<u<\infty$ as the holographic radial coordinate with the boundary located at $u=0$, and we set the spacetime radius of curvature to unity. The coordinates $t, x,y$ parametrize the boundary field theory directions. In addition to the curved metric, the spacetime has a non-trivial dilaton profile and RR flux. The D6-brane is extended in the field theory and holographic radial directions, and wraps three of the six compact directions. It is convenient then to introduce coordinates such that the sphere wrapped by the D6-brane is  $\tilde{\mathcal{S}_3}$. This sphere is fibred over the sphere $\mathcal{S}_3$ on which the brane is localized. The location of the brane is specified by an angle\footnote{In
the least action solution we expect the brane location in the
other angular directions on $\mathcal{S}_3$ to stay constant.} $\theta$, and the volume of $\tilde{\mathcal{S}_3}$ depends on that location as indicated. 

The embedding of the D6-brane may then be specified by giving its location $\theta$ as function of the holographic radial coordinate $u$. As explained in \cite{Frolov:2006tc}, for branes that do not cross the horizon a more natural parametrization near the cap-off point is provided by $u(\theta)$. Here we choose this parametrization globally, resulting in a somewhat unusual form of the DBI action. The coordinate $\theta$ ranges between its value at the cap-off point $\theta=0$ and its asymptotic value $\theta=\frac{\pi}{2}$ as $u\rightarrow 0$. In this parametrization the  induced metric on the brane is:
\begin{equation}\label{induced_metric}
ds_{induced}^2 =  \frac{-dt^2+dx^2+dy^2}{u^{5/2}}+du^2 u^{5/2} \left(\frac{u^2}{u'^2}+1\right)
 +\sin ^2(\theta ) \sqrt{ u} d \tilde{\mathcal{S}}_3^2 
\end{equation}

 From this one can determine \cite{Frolov:2006tc} that the worldvolume caps off smoothly at finite value of the radial coordinate if $u(\theta=0)=u_0$ and $u'(\theta=0)=0$, and that this point is reached when the radius of the $\mathcal{S}_3$ goes to zero, i.e. when $\theta=0$, as indicated. Near the boundary, $u(\frac{\pi}{2})=0$, the embedding function has the near boundary expansion:
\begin{align}\label{met_exp}
u(\theta) & \simeq m_0  \left(\frac{\pi }{2}-\theta \right)+ \chi \left(\frac{\pi }{2}-\theta \right)^3+...
\end{align}
where $m_0$ and $\chi$ are constants. For a constant worldvolume gauge field, a solution for the embedding equation with the desired properties is $u(\theta)=m_0 \cos \theta$.

In addition to the embedding function, we will need to turn on the gauge potential on the D6-brane, which we denote by  $G(\theta )$. The action for the two bosonic fields is then proportional to the DBI action:
\begin{align}\label{DBI}
\mathcal{L} = \frac{\sin ^3(\theta ) \sqrt{u'(\theta )^2+u(\theta )^2- u(\theta )^4 G'(\theta )^2}}{L^3 u(\theta )^5}
\end{align}
where we have chosen $\alpha'=\frac{1}{2 \pi}$. Near the cap-off point smoothness requires that $G'(0)=0$, while near the boundary:
\beq
G(\theta) \simeq \mu +\rho \left(\frac{\pi }{2}-\theta \right)^2 +...
   \eeq
   
   Our boundary conditions near the asymptotic boundary are therefore $G(\frac{\pi}{2})= \mu$ and $u'(\frac{\pi}{2})= m_0$ where $m_0$ and $\mu$ are parameters of our solution.
 \bigskip
 \subsection{Fermionic sector}
We now turn our attention to the fermionic sector. We work with fermions localized in the field theory and radial directions, described by the Dirac action coupled to the bosonic sector in a  manner described below. In a ``top-down" context, charged fermions can arise when considering multiple probe D-branes. Consider for example an additional D6-brane with the same configuration as above, but localized at $\theta=0$ . When the two D6-branes are separated we have a massless Abelian gauge field (the ``relative" gauge group) and charged fermions with respect to that gauge field. In this setup, however, there are additional fields, including charged bosons. The presence of light charged bosons is likely to result in Bose-Einstein condensate being the dominant phase at low temperatures. However, this conclusion can be avoided by various additional couplings or other complications. As such couplings will unnecessarily complicate our analysis, we take a phenomenological approach (commonly used in constructing gravity duals) and use the minimal matter content and couplings required to construct the state we are interested in. We keep the above ``top-down" context only as a motivation for our construction.

Our fermionic action is then as follows
\begin{align}\label{ferm_action}
\mathcal{S} & = -i\,\beta \sqrt{-\gamma} (\bar{\psi} \Gamma^{M} \mathcal{D}_{M} \psi - m(\theta) \bar{\psi} \psi) \\ \nonumber
\mathcal{D}_M & = \partial_M +\frac{1}{4} \omega_{abM} \Gamma^{ab} - i q A_M \\ \nonumber
\Gamma^{M} &=\Gamma^{a} e_{a}^{M} \\ \nonumber
m(\theta)  &=m_{\psi}+ u(\theta)^{  \frac{1}{4} }  \sin (\theta )
\end{align}
where $M$ refers to bulk space-time indices and $a,b$ to tangent space indices. Here $q$ is the electric charge, $\gamma$ is the determinant of the induced metric, and $\omega$ is the spin connection. The coupling to the induced metric $\gamma_{ab}$ and gauge field follows  from symmetries, which determine the form of the covariant derivative we use, $\mathcal{D}_M$. In addition to the bare Dirac mass $m_{\psi}$, we choose to add a Yukawa coupling giving the fermions an effective mass  proportional to the radius of the $\tilde{\mathcal{S}}_3$. This coupling is motivated by the ``top-down" context described above. The tuneable parameter $\beta$ controls the backreaction of the fermions on the bosonic sector\footnote{The parameter
$\beta$ is defined only with respect to a specific normalization
of the fermionic fields, which we choose as $\int_0^{\pi/2} \mathrm{d}\theta
 \bar{\psi} \psi =1$.}, and is typically take to be fairly small. Note that in the brane setup $\beta \propto g_{str}$ since it arises as the ratio of the coefficient of the Dirac action to the brane tension (which we chose to normalize to unity in the bosonic action (\ref{DBI})).

To solve the fermionic equations we follow the procedure first outlined in \cite{Iqbal:2009fd, Liu:2009dm}, but using the notations of \cite{Faulkner:2009wj, Iizuka:2011hg, Sachdev:2011ze}. First, it is convenient to rescale the fermions  in order to remove the spin connection from the Dirac equation:
\begin{align}\label{ferm_resc}
\psi &=(-\gamma \gamma^{u u})^{-\frac{1}{4}} e^{-i \omega t+i k_i x^i}  \Phi
\end{align}
where $\gamma^{uu}$ refers to the radial component of the induced metric given in equation \eqref{induced_metric} and $w, k_i$ are the frequency and spatial momenta, respectively (with $i=x,y$). We then divide the four components of the Dirac spinor to normalizable and non-normalizable modes. Specifically, if we define\footnote{Underlined indices for the gamma matrices are tangent space indices.} $P_{\pm}=\frac{1}{2} (1 \pm \Gamma^{\underline{u}})$ then
\begin{align}
\phi_{\pm}\equiv P_{\pm} \Phi \equiv\begin{pmatrix}
  y_{\pm} \\
  z_{\pm}  \\
   \end{pmatrix}
\end{align}
are normalizable and non-normalizable modes of the Dirac fermion. Furthermore, we can use rotational invariance in the boundary directions to choose the momentum  $k_x=k, k_y=0$. This choice allows us to organize the four Dirac equations as a decoupled pair of two ordinary differential equations. Choosing real gamma matrices as in \cite{Faulkner:2009wj, Iizuka:2011hg, Sachdev:2011ze}, the decoupled components can be organized into 
\begin{align}
\Phi_{1}=\begin{pmatrix}
  i y_{-} \\
  z_{+}  \\
 \end{pmatrix},
 \quad 
 \Phi_{2}=\begin{pmatrix}
  -i z_{-} \\
  y_{+}  \\
 \end{pmatrix}
 \end{align}

It is therefore sufficient to consider only one of $\Phi_{1}$ and $\Phi_{2}$ when constructing our state and calculating correlation functions. Information regarding the other spinor may be extracted using rotational invariance as in \cite{Faulkner:2009wj,Iizuka:2011hg}. We therefore choose to work with, in a slight change of notation, $\bar{f}=(f_1 ,f_2)\equiv(z_{-},y_{+})$,  reducing the fermionic problem to the solution of two ordinary differential equations.

We now consider the boundary conditions for the fermionic fields $(f_1 ,f_2)$. Near the asymptotic boundary there are two independent modes, normalizable and non-normalizable. For the former, the component $f_1$ is constant near the boundary at $\theta = \frac{\pi}{2}$, and $f_2$ is then determined by the equations of motion. For the latter, non-normalizable mode, the roles of $f_1$ and $f_2$ are interchanged. As we are seeking normalizable solutions we require $f_2(\frac{\pi}{2})=0$.

Turning to the behaviour of the fermions in the IR: Conservation of the symplectic norm requires that either $f_1$ or $f_2$ must be zero at the cap-off. This statement can be verified by integrating the symplectic flux   $\psi^{\dagger} \Gamma^0 \Gamma^{\mu}  \psi n_{\mu} $ over the hypersurface located at $u=u_0$. To obtain a non-trivial solution we then set $f_1(0)=0$ for regularity in the interior.  Since we have two first-order ordinary differential equations, our boundary value problem is then well-posed.
     \bigskip
\subsection{Parameters and Limits}\label{prob_lim_sec}
Our equations and boundary conditions have the following parameters:
\begin{itemize}
\item The chemical potential, $\mu$.
\item The source term for the scalar field dual to the embedding function, $m_0$.
\item The fermion bare mass $m_\psi$.
\item The electric charge $q$. 
\item The relative strength of the contribution of the Dirac and DBI actions to the total action. This is determined by the parameter $\beta$.
\end{itemize}

In our numerical construction, described below, we vary all five parameters independently. We divide these into two groups- the thermodynamic variables, $(\mu, m_0)$, and the mass and coupling terms which define our theory, $(\beta, q, m_{\psi})$. The physical significance of these parameters is as follows:

\begin{itemize}
\item Varying the chemical potential controls the fermionic density on the brane. As we are working at zero temperature the chemical potential sets the bulk Fermi energy. 

\item Varying $m_0$ controls how far the embedding proceeds into the bulk before capping off. Increasing $m_0$  moves the the cap-off point further into the IR. As such embeddings require larger charge densities to support them we will see that increasing $m_0$ also has the effect of increasing the charge density.

\item The parameter $\beta$ controls the backreaction of the fermionic charge density on the bosonic fields,  i.e. $\beta=0$ corresponds to decoupled fermions\footnote{This is similar  to  \cite{Faulkner:2009wj} where the Dirac equation was considered on a Reissner-Nordstrom black hole background.}. Here we work with a finite but small $\beta$.

\item It is convenient to rescale the matter fields and charge as $ \mathcal{\bar{H}} \rightarrow \mathcal{H} \epsilon$, $ q \rightarrow \frac{e}{\epsilon}$, where $\mathcal{H}$ represents the gauge field or fermion fields and $e$ is fixed. We then trade the charge $q$ for the parameter $\epsilon$, small $\epsilon$ corresponds to large charge. The limit $\epsilon \rightarrow 0$ is  the commonly used probe limit, as used for example in constructing holographic superconductors \cite{Hartnoll:2008kx}. Setting $\epsilon=0$  decouples the embedding function from the  Maxwell-Dirac sector. Here we work with small but finite $\epsilon$. 

\item Varying the bare mass terms of the fermions $m_{\psi}$ corresponds to setting their mass at the cap-off point. 
\end{itemize}

We comment further below on the parameter dependence of the state we construct, and the limits and range of values of these parameters in our numerical simulations. 

 \bigskip
\section{Bulk Fermi Surface}\label{sols_and_anal}

\subsection{Iteration Procedure}\label{iteration}

The main novelty in solving the equations of our setup is the implementation of Fermi statistics. The requirement that the fermions form a Fermi surface is a non-local constraint, effectively rendering the problem an integro-differential system of equations, instead of a set of local ordinary differential equations. We therefore solve our system, as in \cite{Sachdev:2011ze}, by an iterative process we now describe. Further details of our implementation are found in appendix B.

For every step in the iteration process, the fermion equations are formulated in a background of the bosonic fields. One solves the Dirac equation in that background to find the complete set of eigenstates of the Dirac operator. The state of the fermions in a fixed chemical potential (and zero temperature) is a filled Fermi surface: all states with energy below $\mu$ are filled. These eigenstates come in bands; we work with parameter ranges such that only a small number of these bands (approximately 1 to 10) are filled. This has the interpretation of multiple (but order one in the large $N$ limit) Fermi surfaces in the dual QFT. The inclusion of several bands is necessary when one tunes parameters towards states of larger charge densities. The state of the fermions is iteratively adjusted  mainly through the change in the eigenstates of the Dirac operator.

Given the state of the fermions, the bosonic equations are sourced by specific fermion bilinears, obtained from varying the action (\ref{ferm_action}) with  respect to the bosonic fields (the specific form of these bilinears and further details regarding their properties can be found in appendix A). Briefly, we identify these sources as the charge density, $Q$, which sources the gauge equation and embedding equation. In addition the embedding equation is sourced via effective stress energy terms $T_r$, and $T_M$ which we label as the ``radial" and ``Minkowski" stresses. As in  \cite{Sachdev:2011ze}, we evaluate these sources at leading order in $\frac{1}{N}$, i.e. in the classical limit in the bulk. This approximation does not include the higher order (and much more complex) renormalization of the coupling constants in the brane action. We comment below on regimes of solutions where such higher order effects may become relevant and their potential implications for the iterative solution. 

Once the adjusted bosonic background is obtained, we turn back to the fermion eigenvalue problem, and iterate to convergence.
Details of our numerical algorithm can be found in appendix B: both the fermionic eigenvalue problem and the bosonic boundary value problem were discretized using pseudo-spectral collocation methods on a Chebyshev grid. The iterative procedure described above was repeated until the residuals for the bosonic system were driven to sufficiently low values and the change in solution between successive steps, for the gauge and embedding functions, was sufficiently small.

 This solution method was found to converge for regions of parameter space where the backreaction was small relative to the radial scale of the probe limit embedding. We will therefore restrict our investigations to regimes of small densities, leaving investigation of higher density regimes for future work.

\subsection{Solutions in the probe limit}
In this section we examine the behaviour of the system in the probe limit, corresponding to taking $\epsilon=0$. This decouples the gauge and embedding equations and freezes the embedding function to be  $u(\theta)=m_0 \cos(\theta)$. Allowing for the probe limit, our model still has four tuneable parameters- namely $\beta$, $m_0$, $m_{\psi}$ and $\mu$. To study the effect of these parameters, we vary individual parameters while leaving the others fixed. In order to visualize the bulk solutions we display the gauge field, the fermionic dispersion relationship, and the charge $Q$, for each of the parameter variations.
\begin{figure}
\center
    \vspace*{-1.5cm}
   \includegraphics[width=4cm]{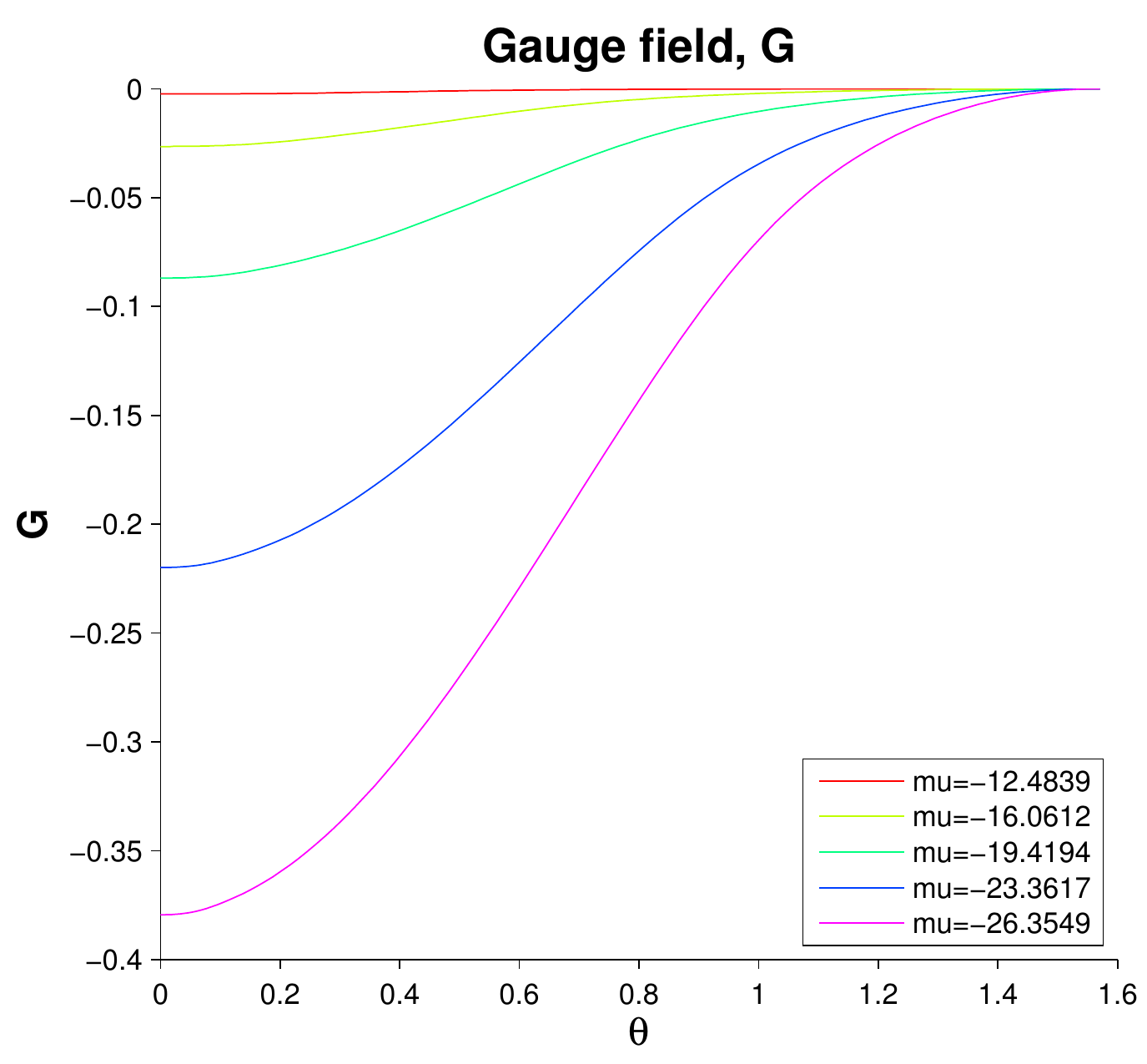}
   \includegraphics[width=4cm]{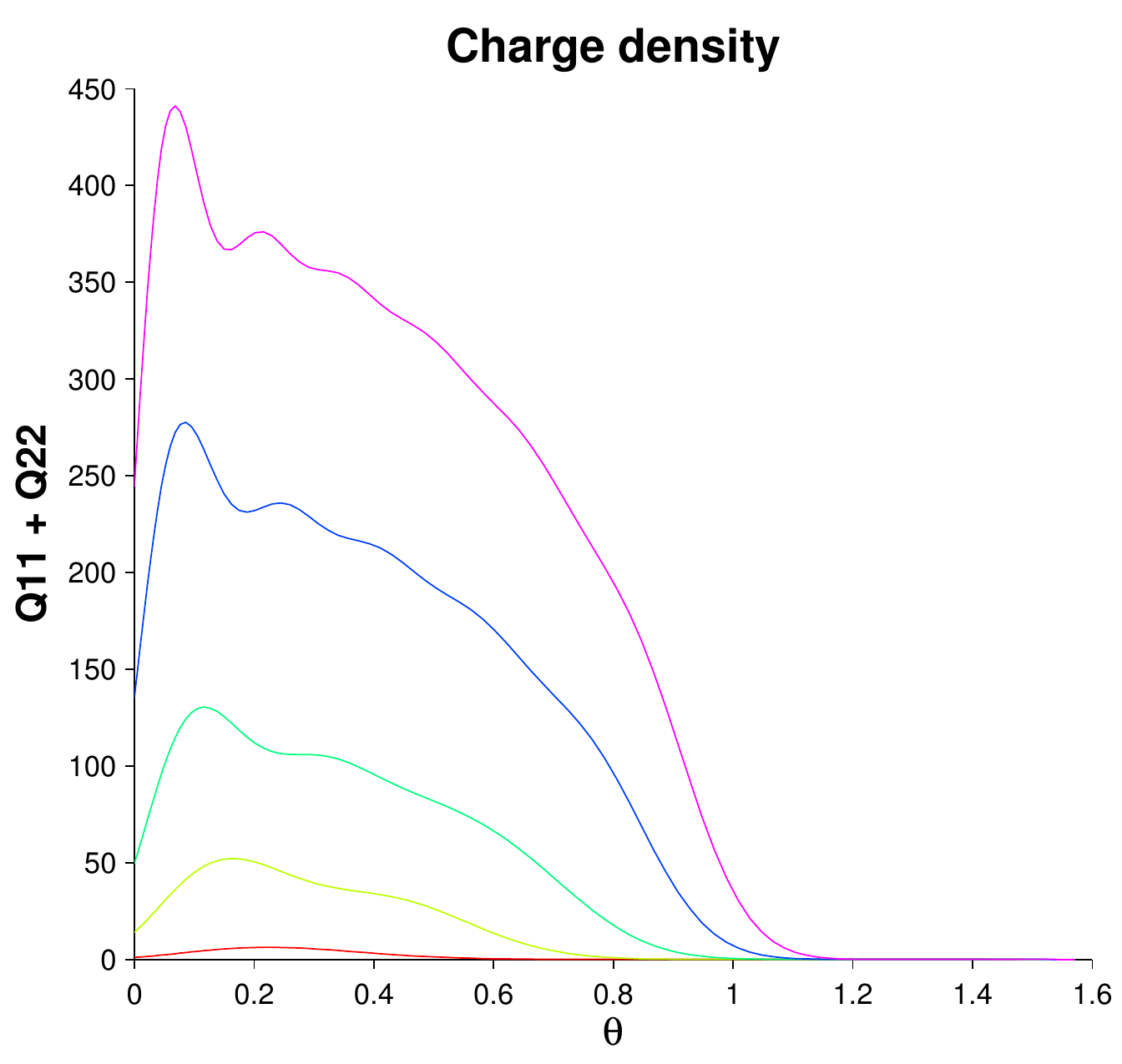}
   \includegraphics[width=4cm]{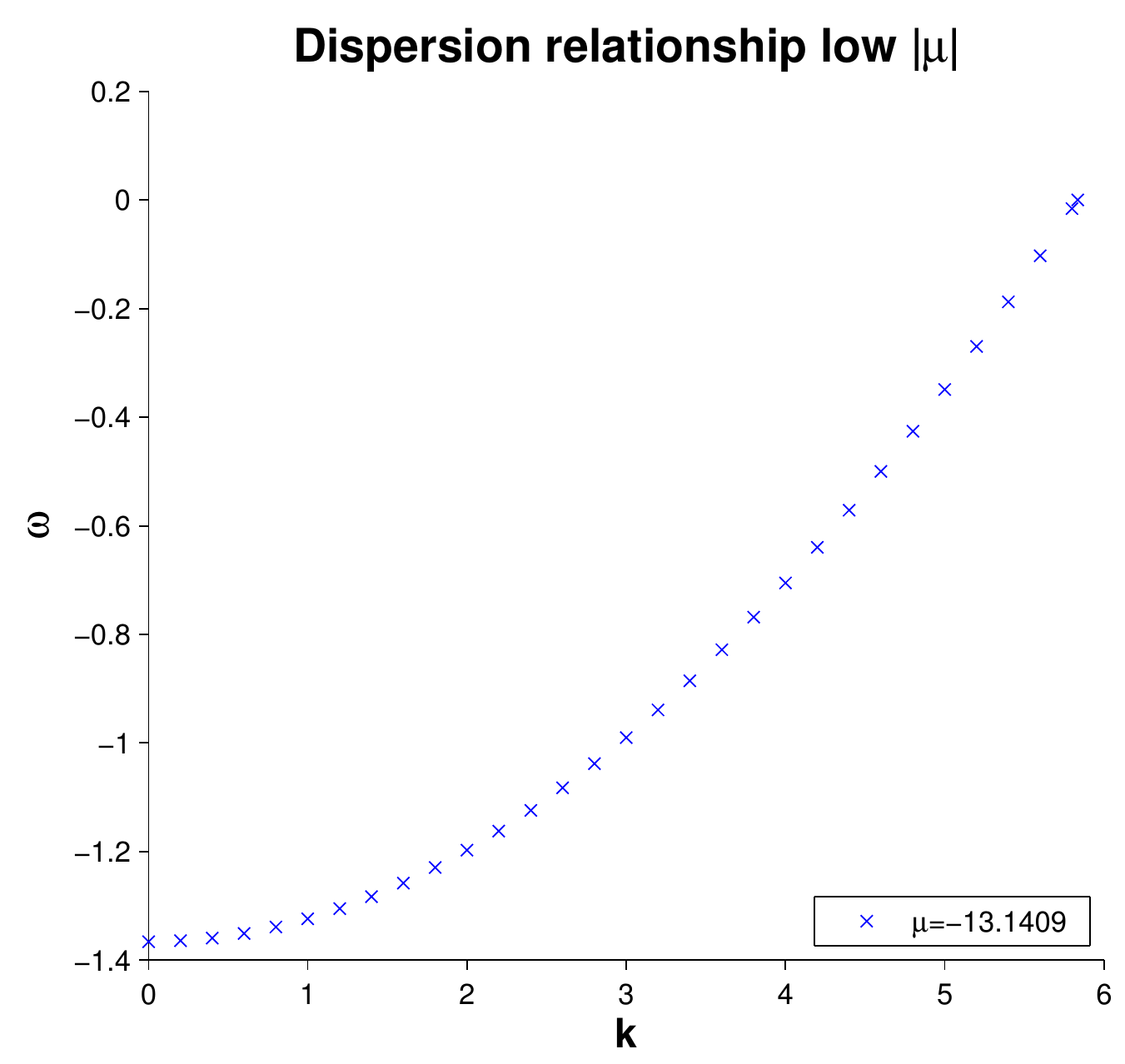}
   \includegraphics[width=4cm]{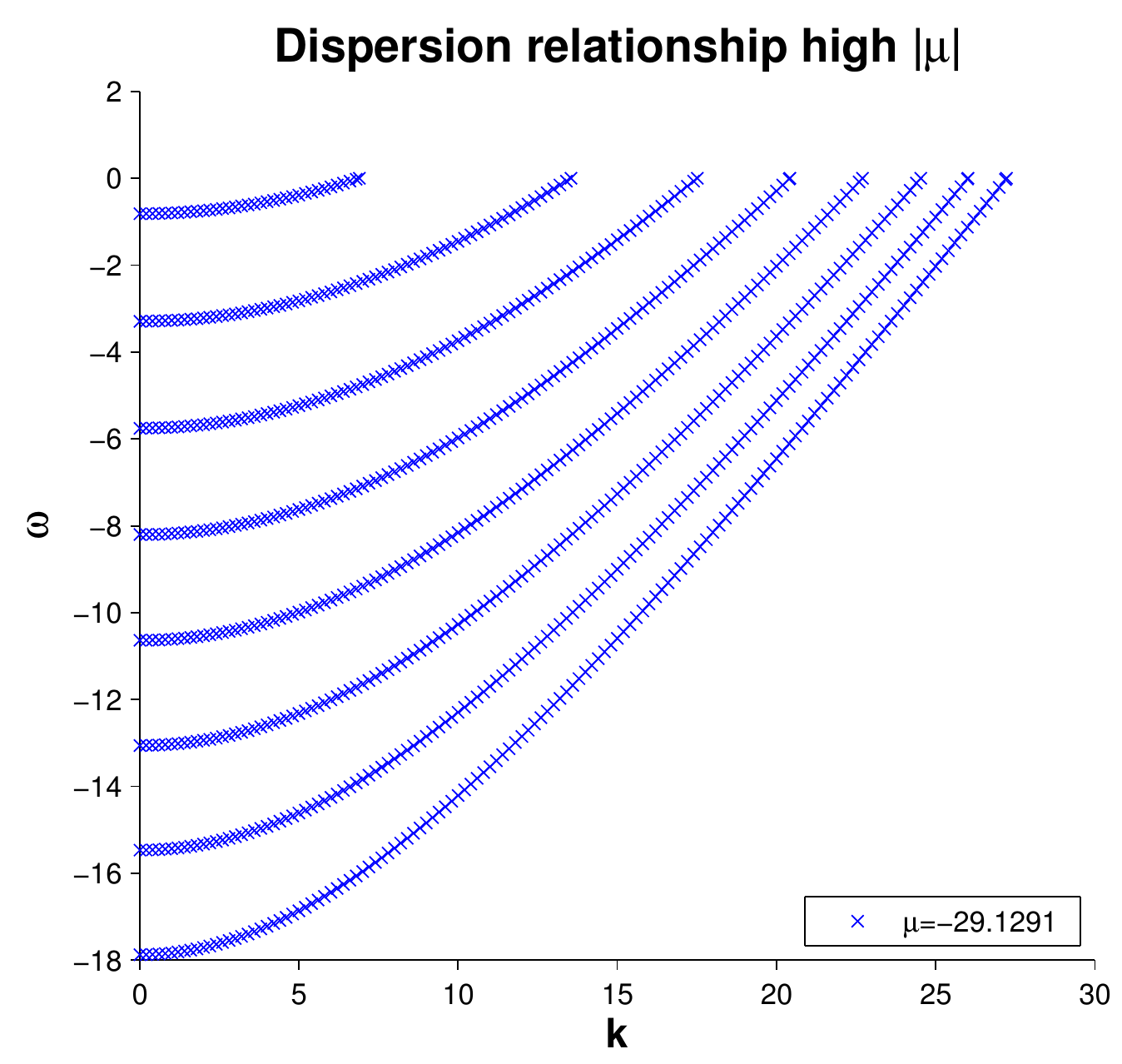}
    \caption[]{Bulk field profiles resulting from varying $\mu$
with $m_0=1$, $\beta=-0.01$ and $m_{\psi}=10$. We note that the shape
of the gauge field profile does not change greatly relative to
the scale set by the chemical potential, as $\mu$ is increased.
This is illustrated by subtracting the value of the chemical
potential in each case. The dispersion relationship curves are
for filled states lying below the Fermi surface.}
\label{fig:mu_probe}
\end{figure}

We start by varying the chemical potential $\mu$.  Plots for bulk quantities for various values of $\mu$ are displayed in figure \eqref{fig:mu_probe}. We note that, as expected, increasing $\mu$ results in larger charge densities. Note that the charge density is finite at the cap-off (unlike the case with bosonic charge carriers, where the cap-off point supports all of the charge). This is an indication that such a charge density  may be able support the compact embedding once we include backreaction.  As expected
the number of bands increases with $\mu$, for large $\mu$ up
to 10 bands may be filled. As these higher bands correspond to higher modes of the bulk fermion eigenfunctions they contribute a modulated component to the charge density. This may be seen in the above figures.
\begin{figure}
  \vspace*{0cm}
\center
   \includegraphics[width=4cm]{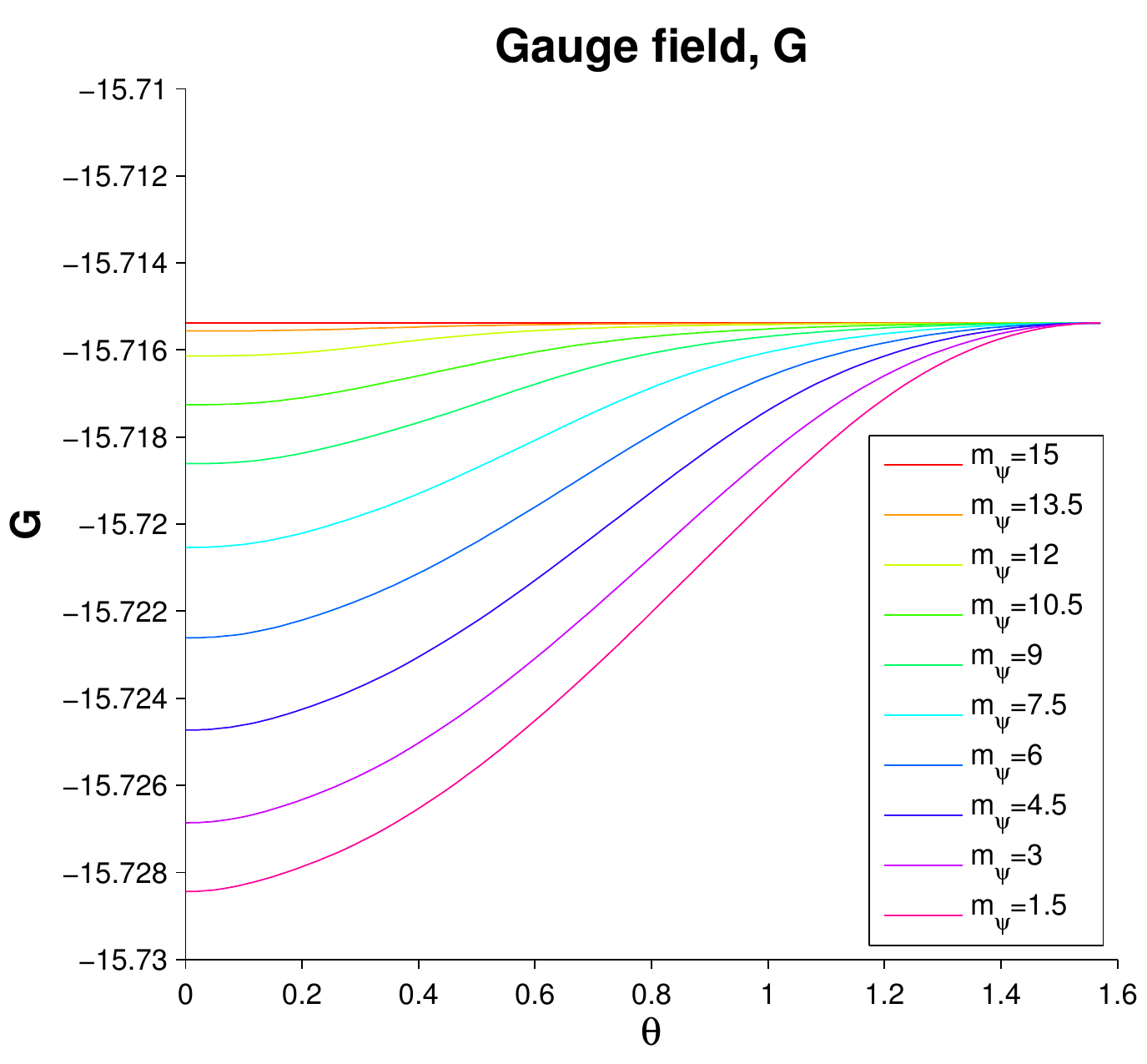}
   \includegraphics[width=4cm]{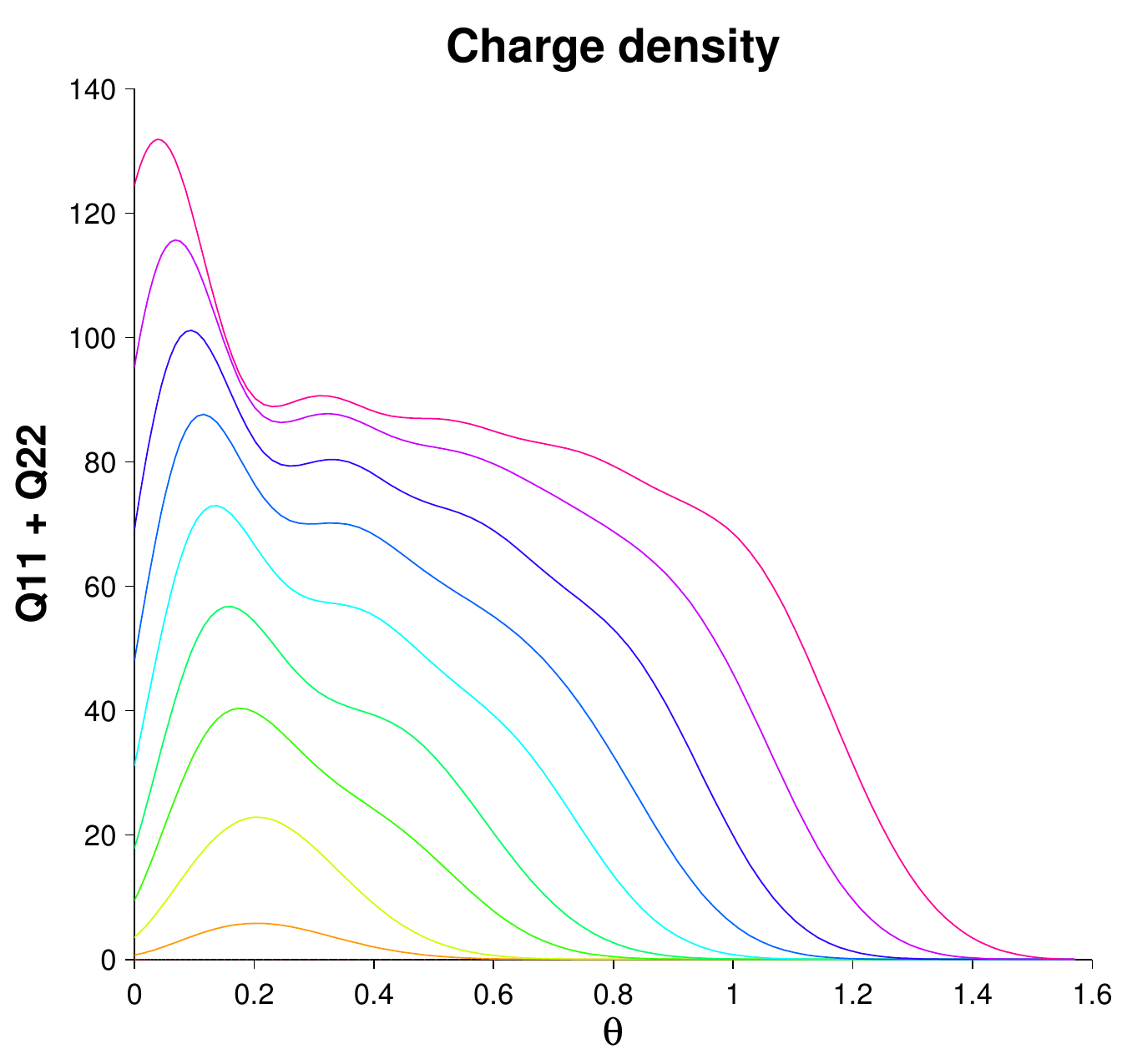}
   \includegraphics[width=4cm]{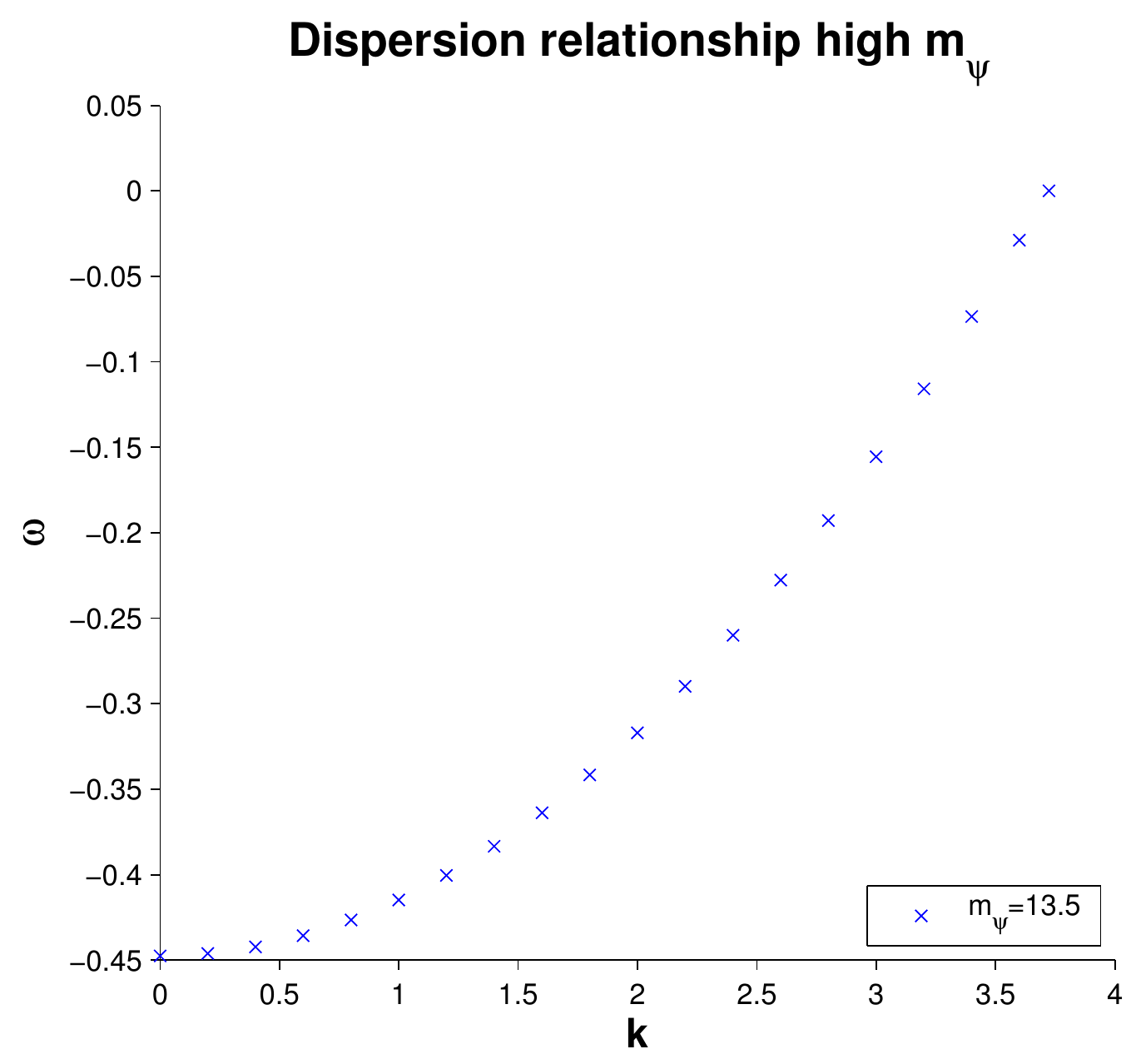}
   \includegraphics[width=4cm]{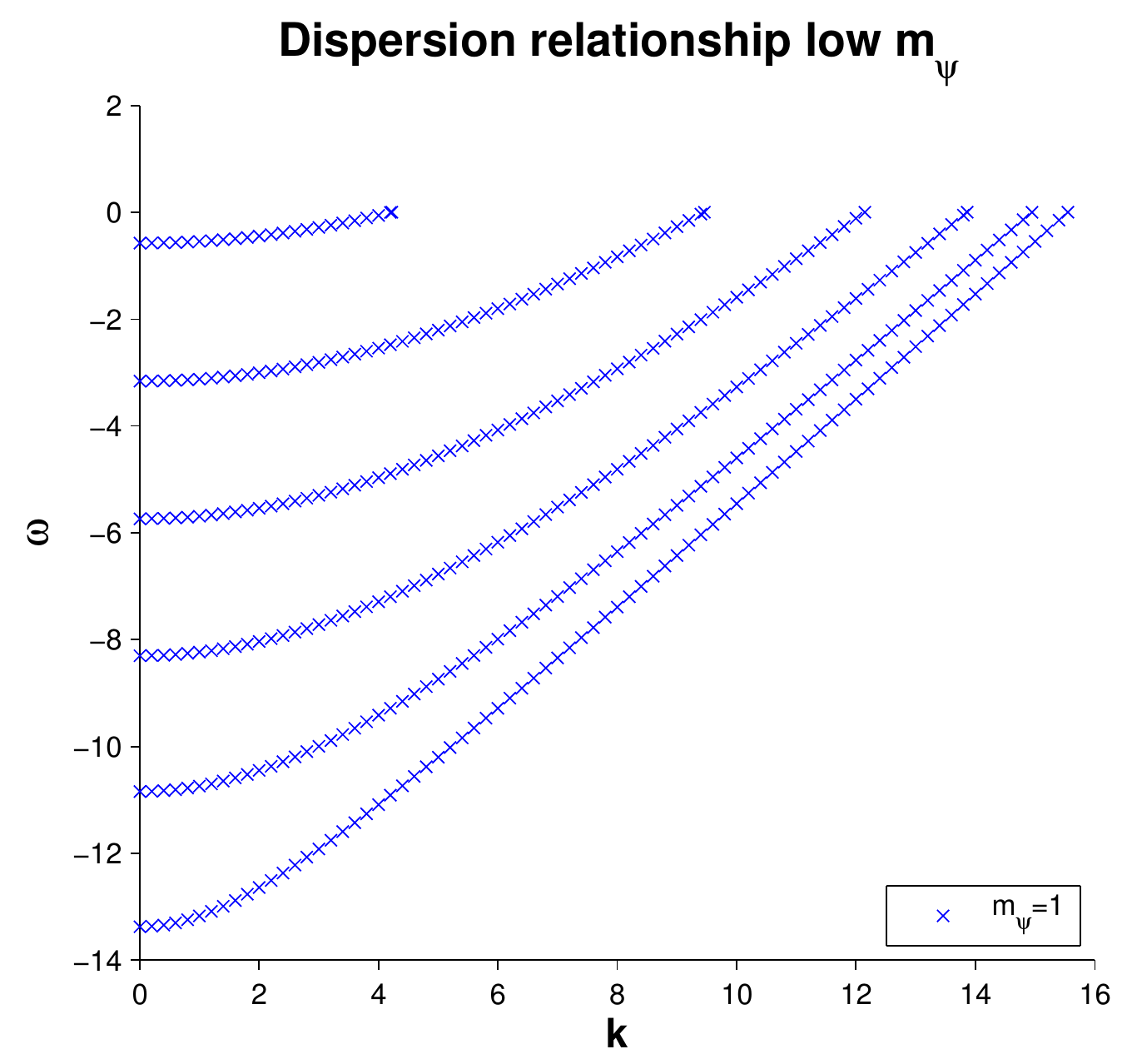}
    \caption[]{Bulk field profiles resulting from varying $m_{\psi}$
with $m_0=1$, $\mu=-15.7154$, and $\beta=-0.001$. We
see that increasing the bare mass has the effect of reducing
the number of filled states, the amplitude of the gauge field
and the contributions to the bulk charge. We also note that as
$m_{\psi}$ decreases the ratio of width to amplitude of the bulk
charge density increases. This results in amplification of the charge densities in the dual QFT.}
\label{fig:mb_probe}
\end{figure}

        We now examine the effects of moving within our parameter space of theories, defined by $(m_{\psi}, m_0, \beta)$. The results for variations of the bare mass term, $m_{\psi}$, are shown in figure \eqref{fig:mb_probe}. We see that decreasing the bare mass has the effect of pushing us towards a regime of higher fermion density.   A reduction of $m_{\psi}$ from an initial high value results in a rapid increase in the peak amplitude of the bulk charge density. However for sufficiently low values of $m_{\psi}$ the charge density profile is seen to broaden and shift  towards the UV geometry. This  results in an enhancement of the charge density of the dual QFT at low $m_{\psi}$.

\begin{figure}
\vspace{-0.5cm}
\center
   \includegraphics[width=4cm]{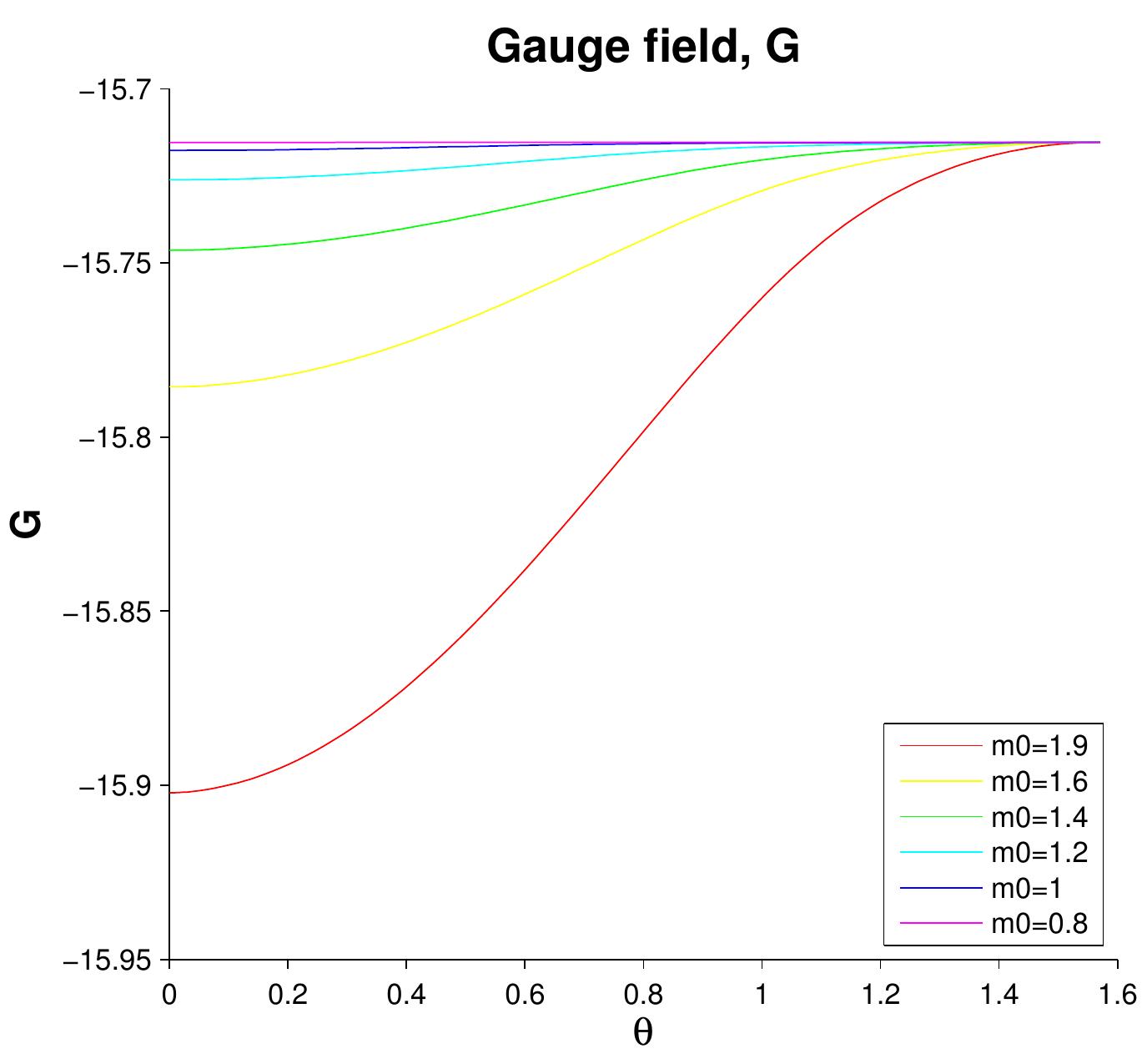}
     \includegraphics[width=4cm]{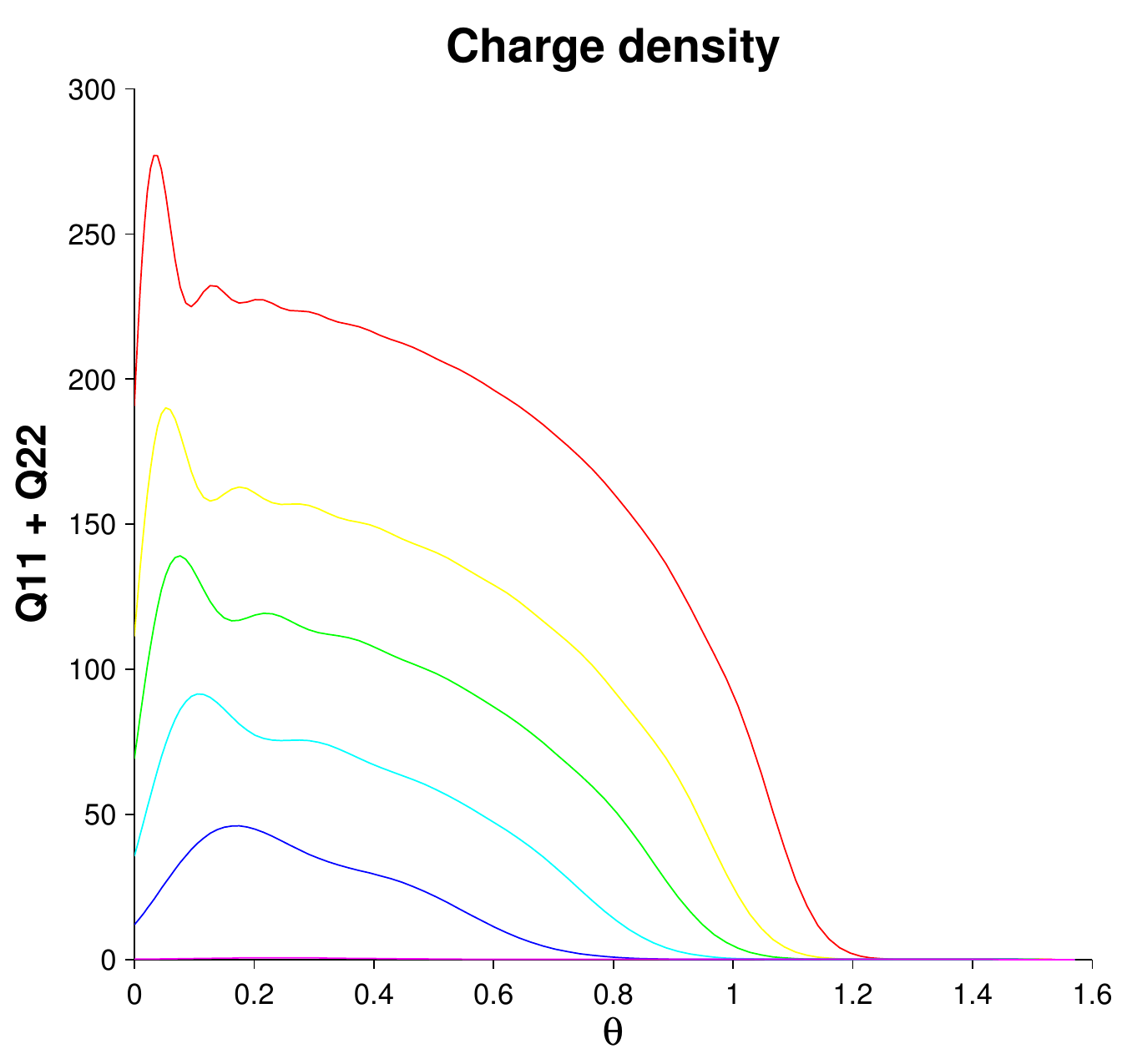}
      \includegraphics[width=4cm]{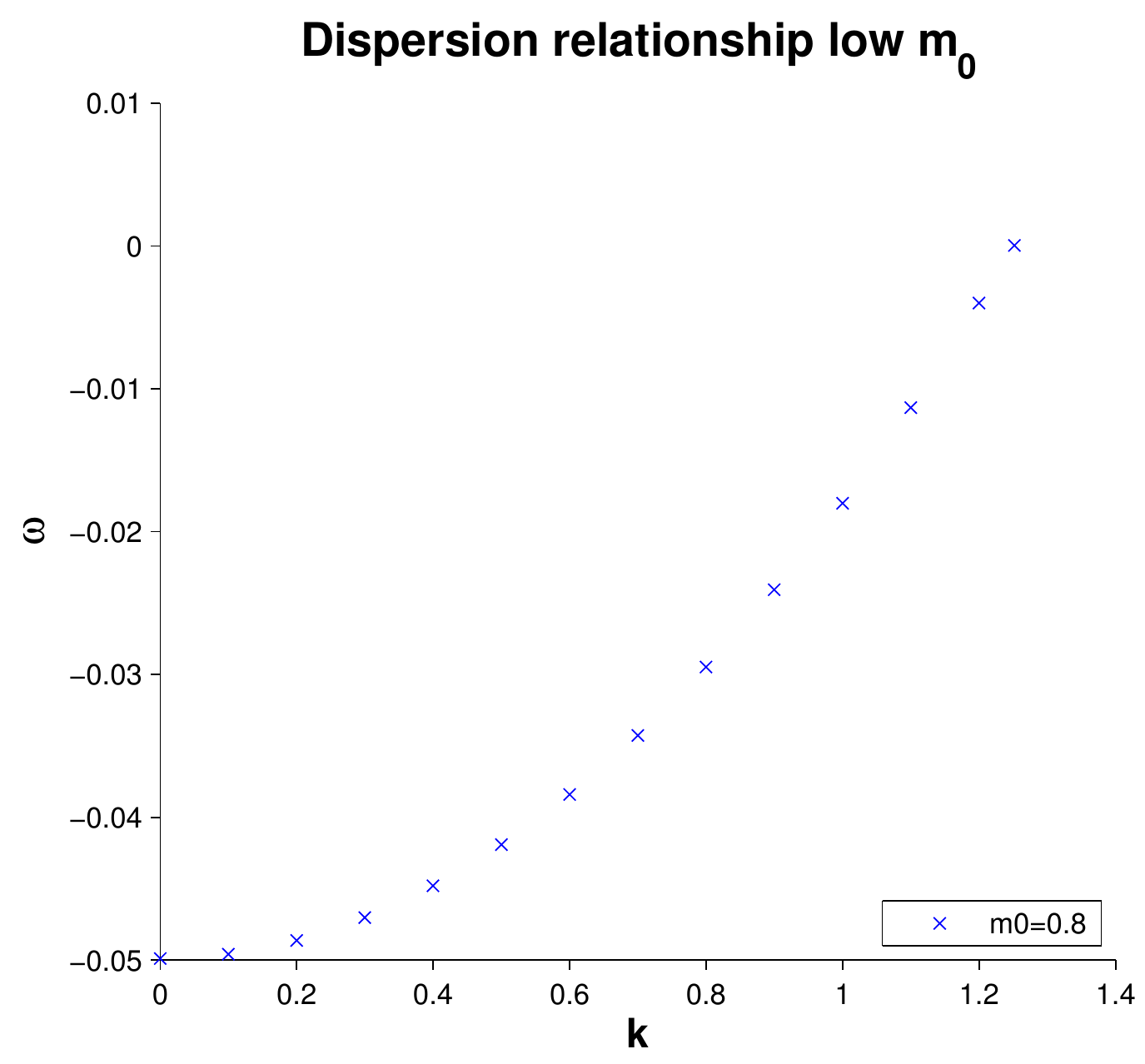}
      \includegraphics[width=4cm]{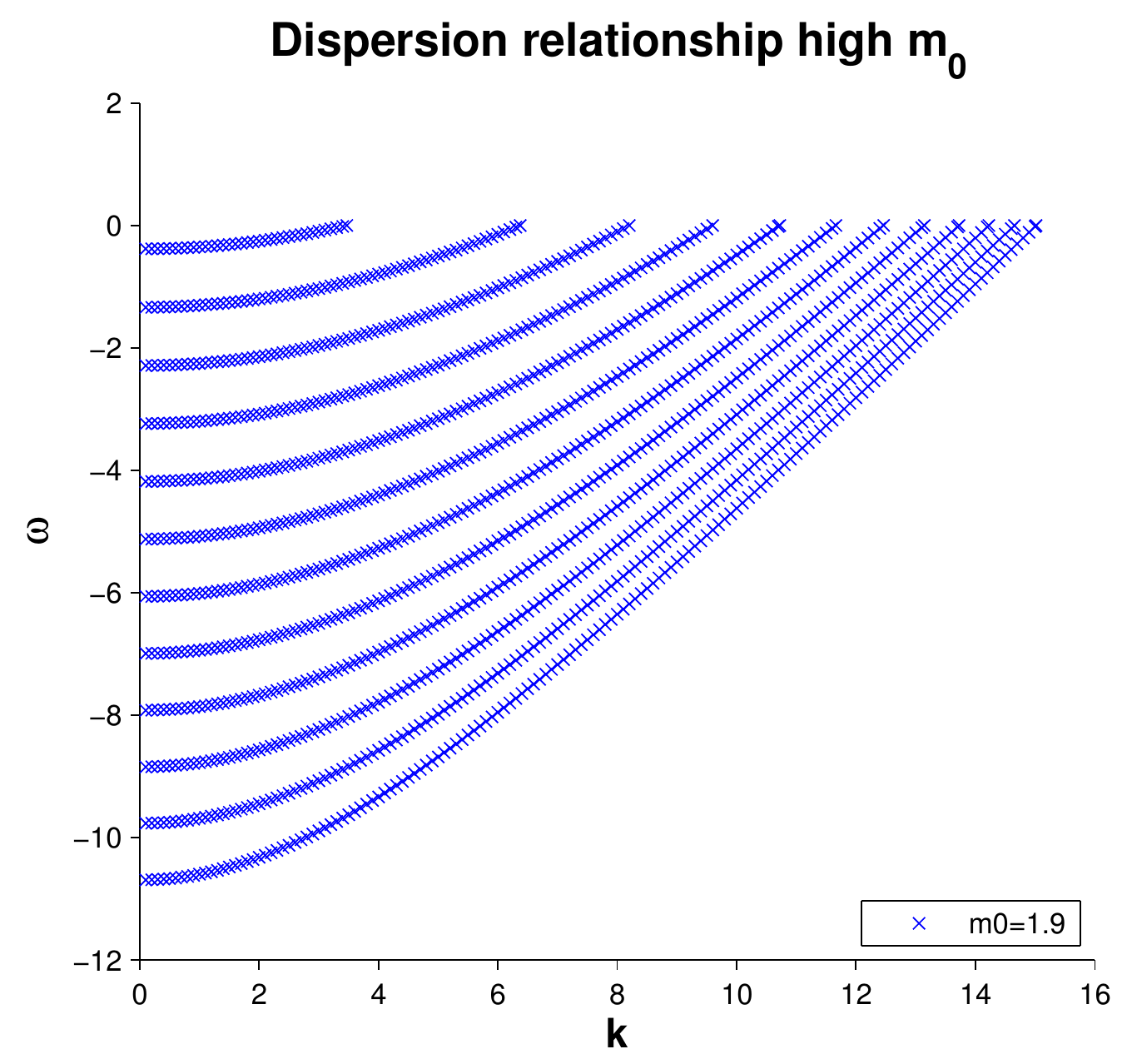}
    \caption[]{The bulk field profiles resulting from varying $m_0$ with $\beta=-0.001$, $\mu=-15.7154$ and $m_{\psi}=10$. We note that increasing $m_0$ has the opposite effect to increasing $m_{\phi}$ as it pushes the solution to regimes of higher fermion density. It can be seen the behaviour of the bulk gauge field and charge density under the variation of $\mu$ and $m_0$ is qualitatively similar. This  is reflected in the behaviour of QFT quantities.}
\label{fig:m0_probe}
\end{figure}

        We can look at the effects of adjusting our (frozen)  embedding cap off point by changing $m_0$.  This simulates the  expected adjustment
in the position of the embedding cap-off  in response to the
finite charge density, once backreaction is included. From figure\eqref{fig:m0_probe} we see that larger values of $m_0$ correspond to larger fermion densities, and electric field strengths. 

\begin{figure}
\vspace{0.5cm}
\center
   \includegraphics[width=4cm]{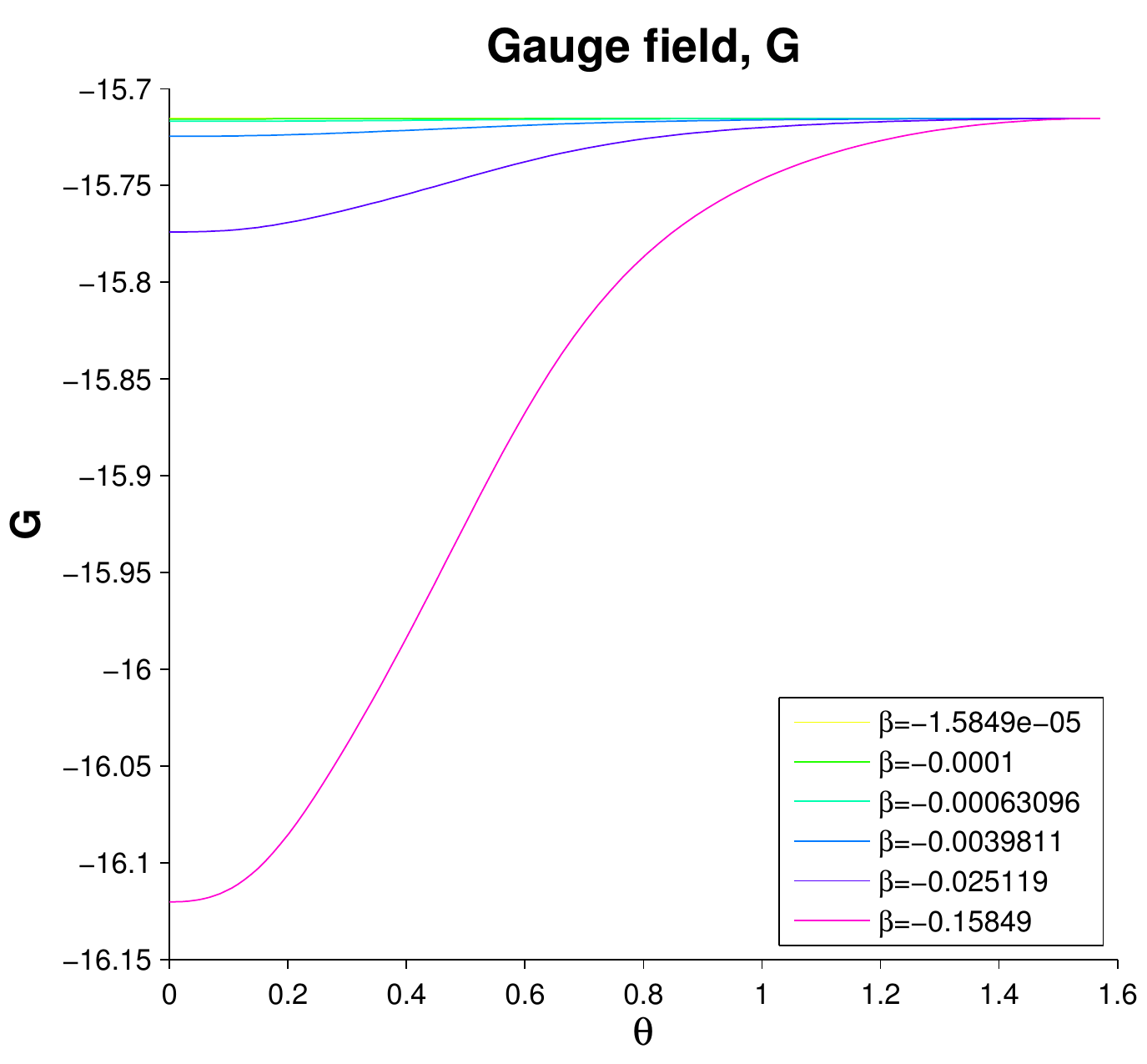}
     \includegraphics[width=4cm]{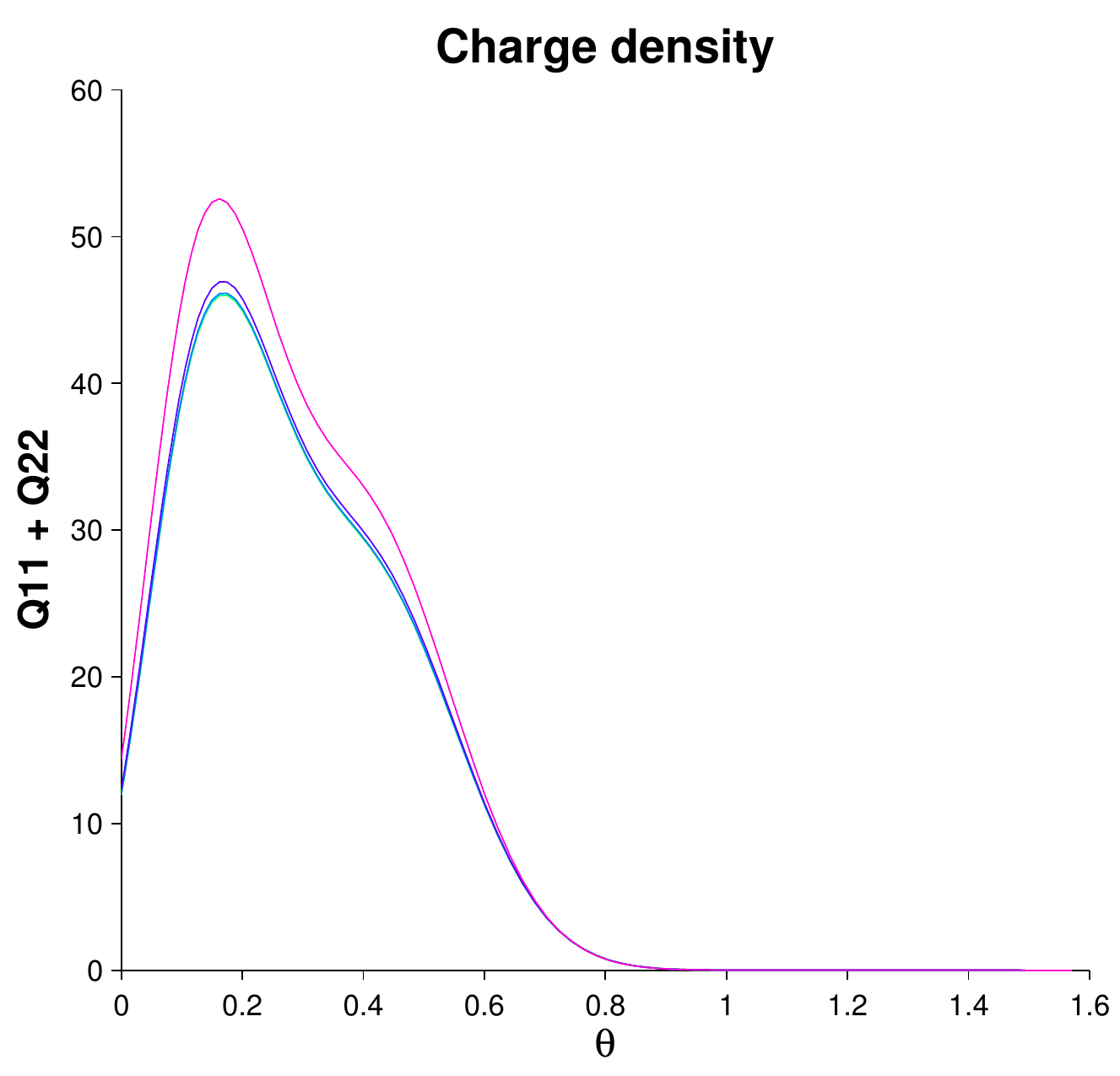}
      \includegraphics[width=4cm]{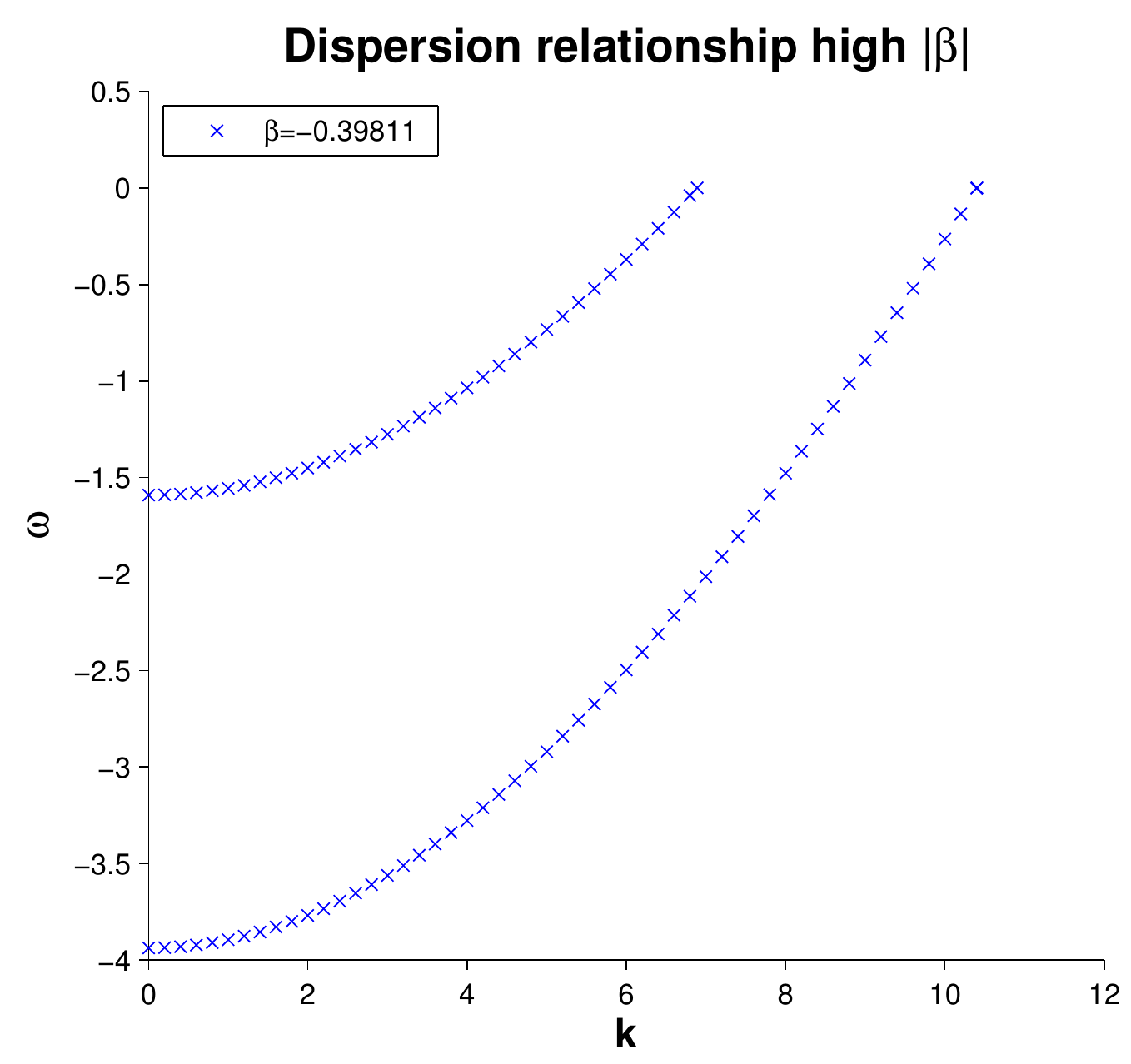}
        \includegraphics[width=4cm]{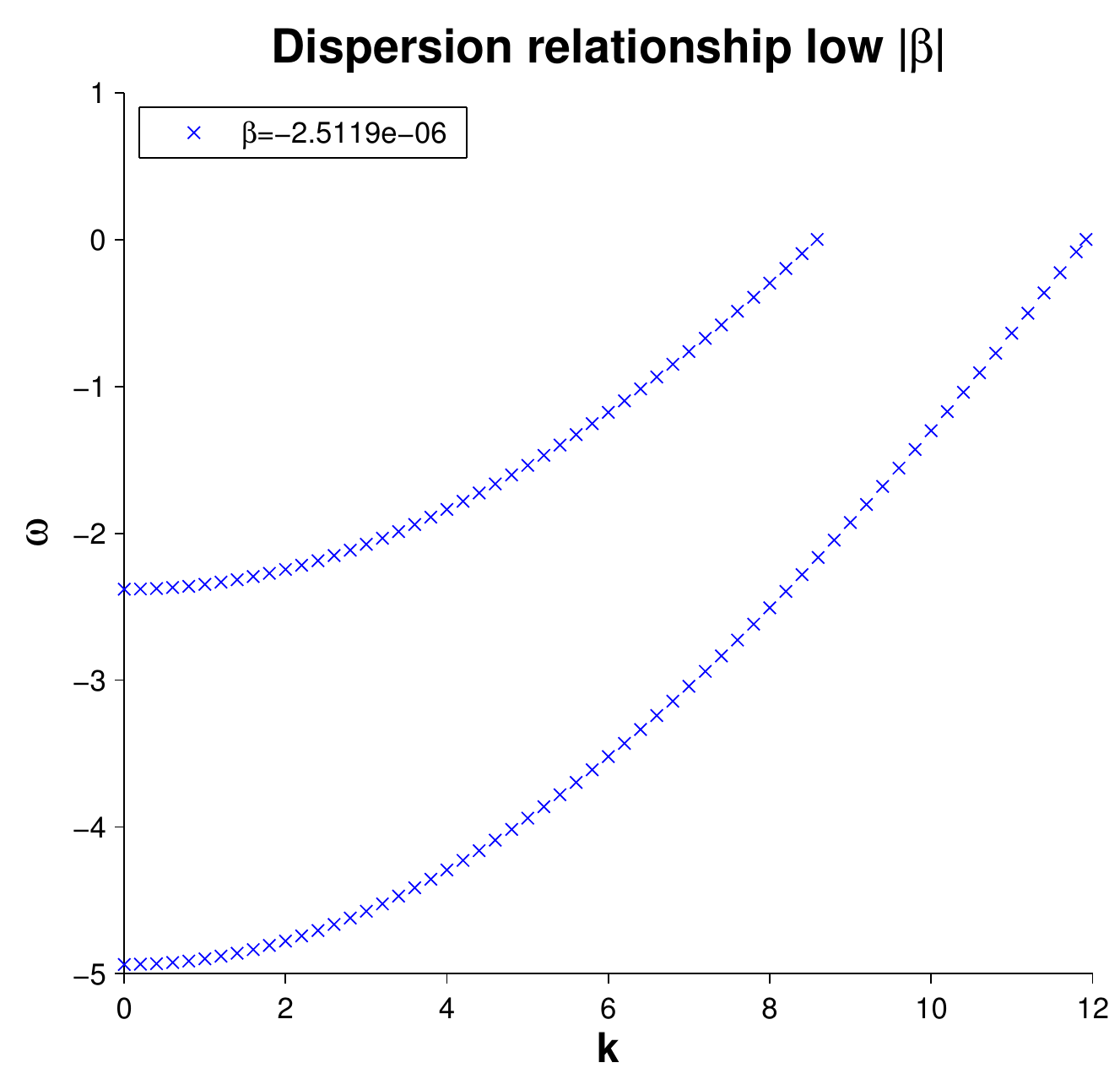}
    \caption[]{Varying $\beta$ with $m_0=1$, $m_{\psi}=10$, $\mu=-15.7154$. We see that increasing the magnitude of $\beta$ has the effect of the increasing the bulk charge density and correspondingly the strength of the bulk electric field.}
\label{fig:beta_probe}
\end{figure}

Finally, turning our attention to $\beta$ we examine the influence of the strength of the fermion-boson coupling in figure \eqref{fig:beta_probe}. Increasing the magnitude of $\beta$ produces larger bulk charge densities and electric fields. 
        
\subsection{Including Backreaction}

\begin{figure}
\center
\vspace{-1cm}   
   \includegraphics[width=4cm]{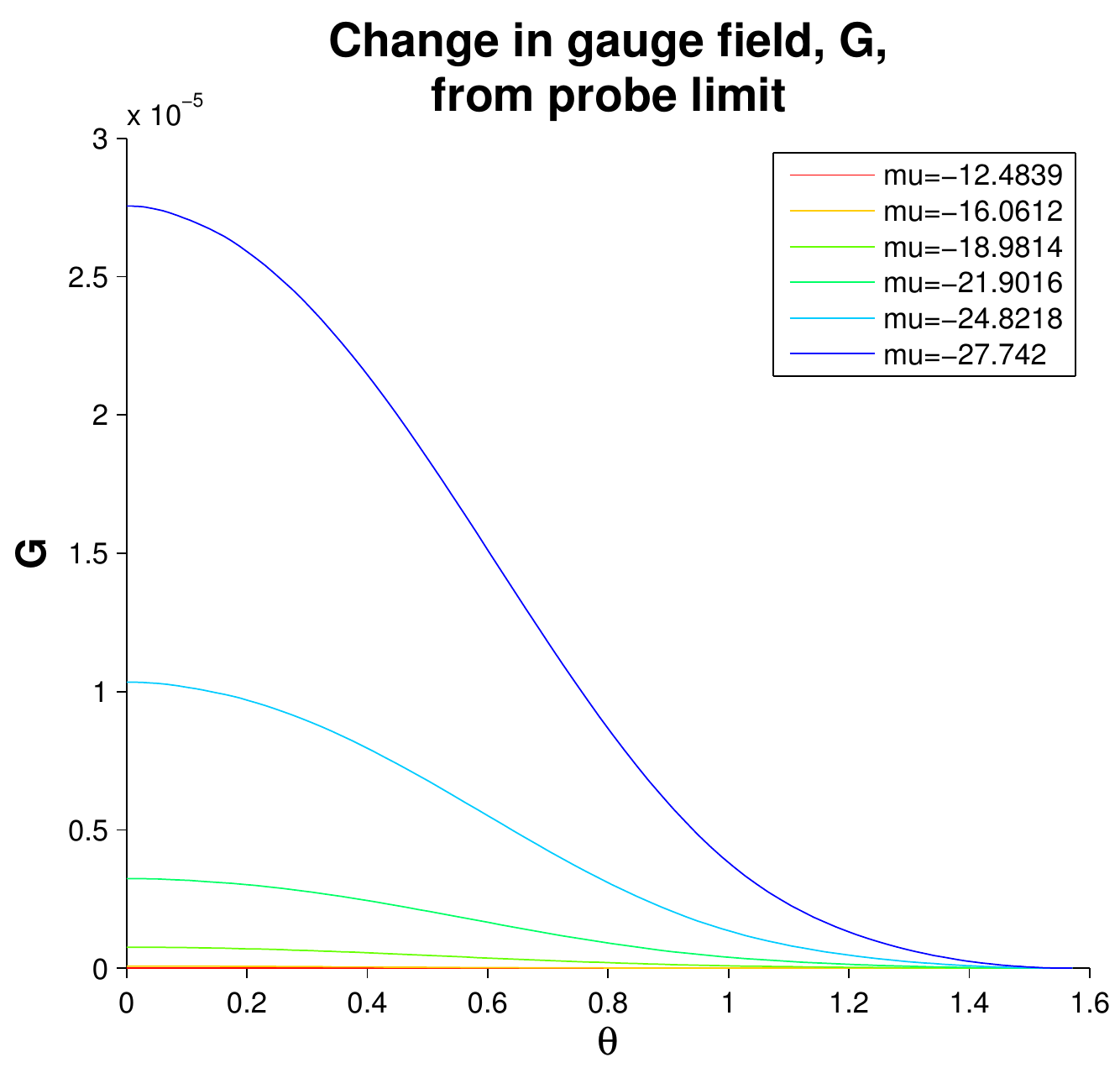}s
   \includegraphics[width=4cm]{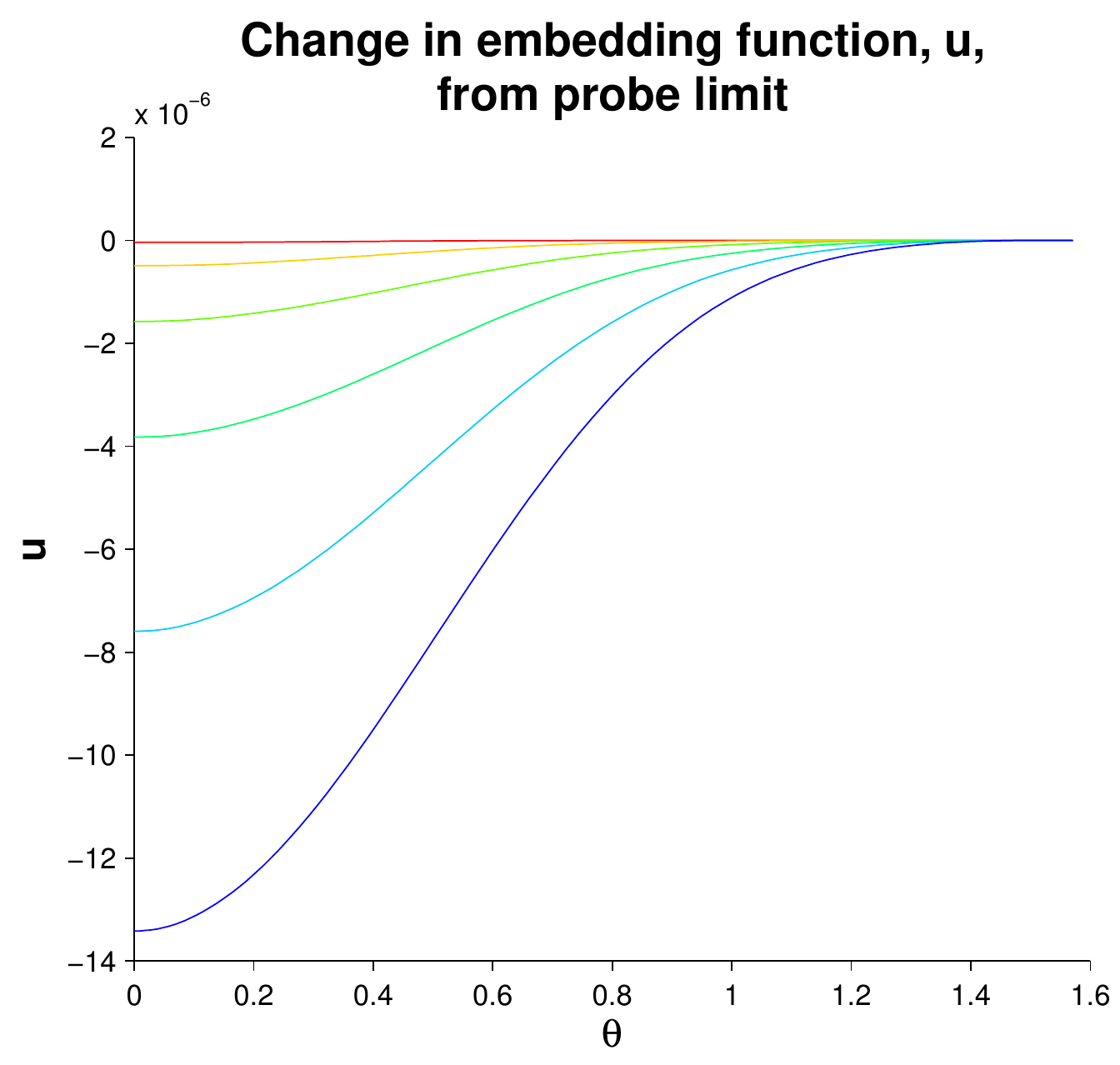}
   \includegraphics[width=4cm]{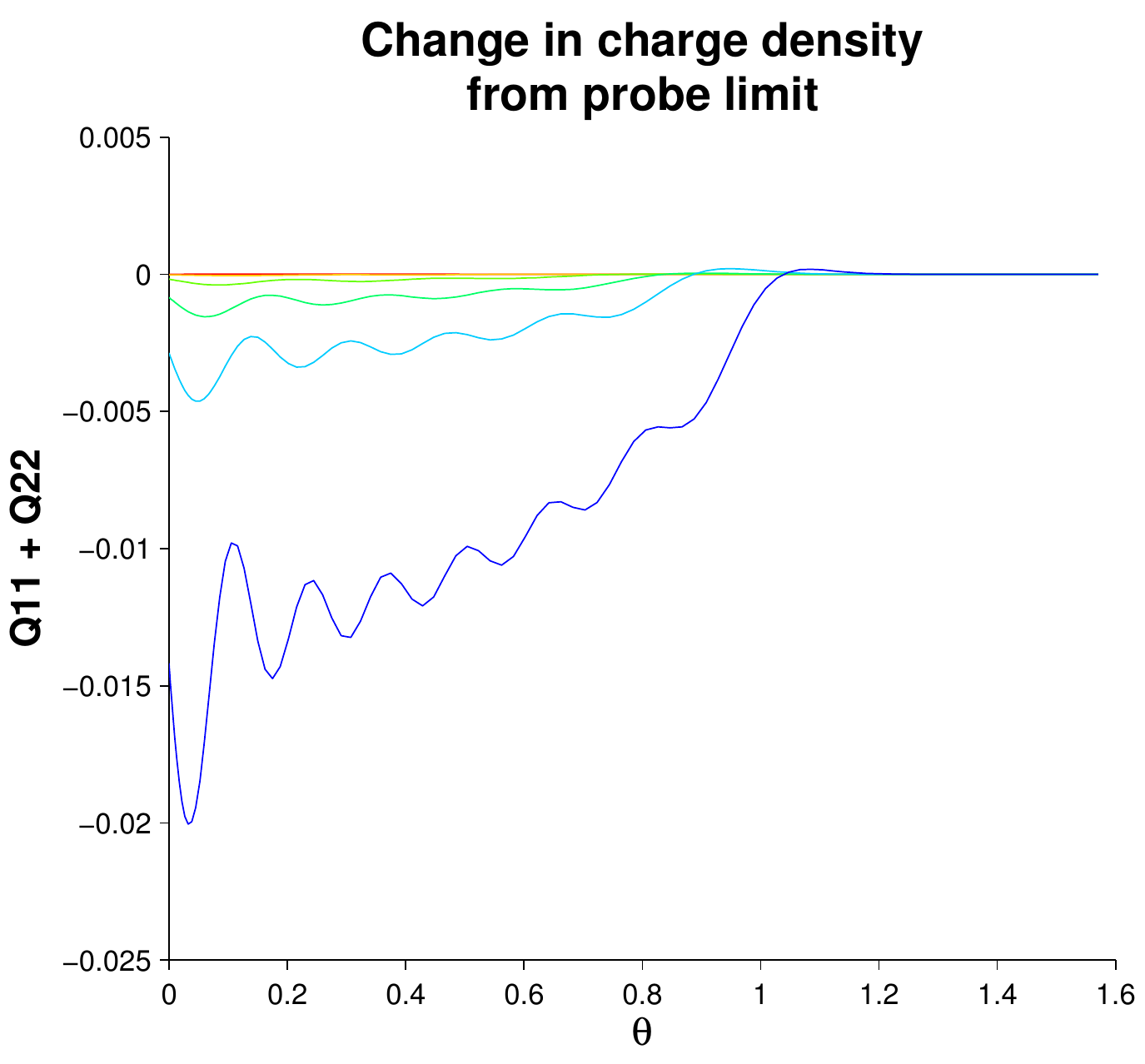}
   \includegraphics[width=4cm]{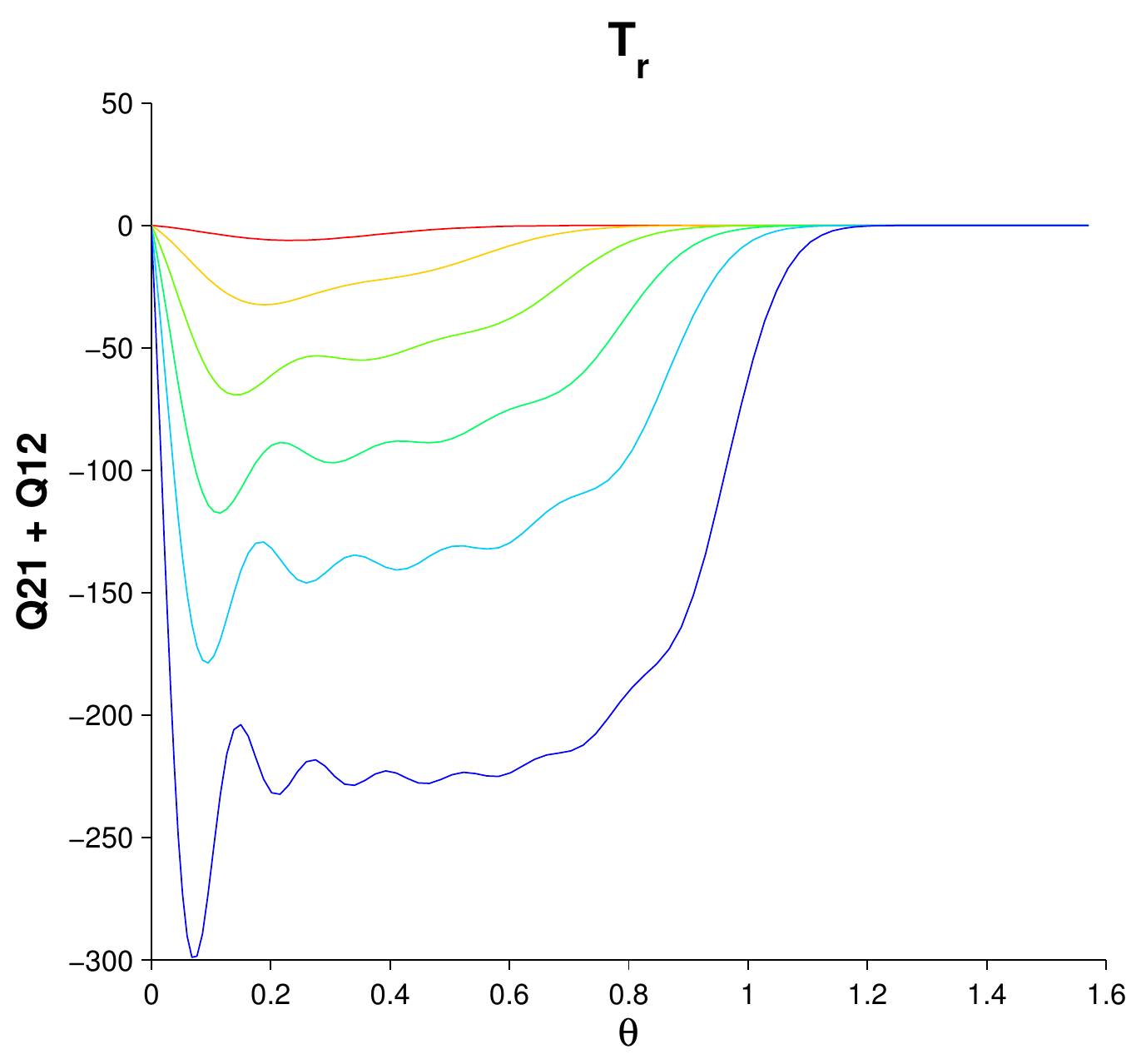}
   \includegraphics[width=4cm]{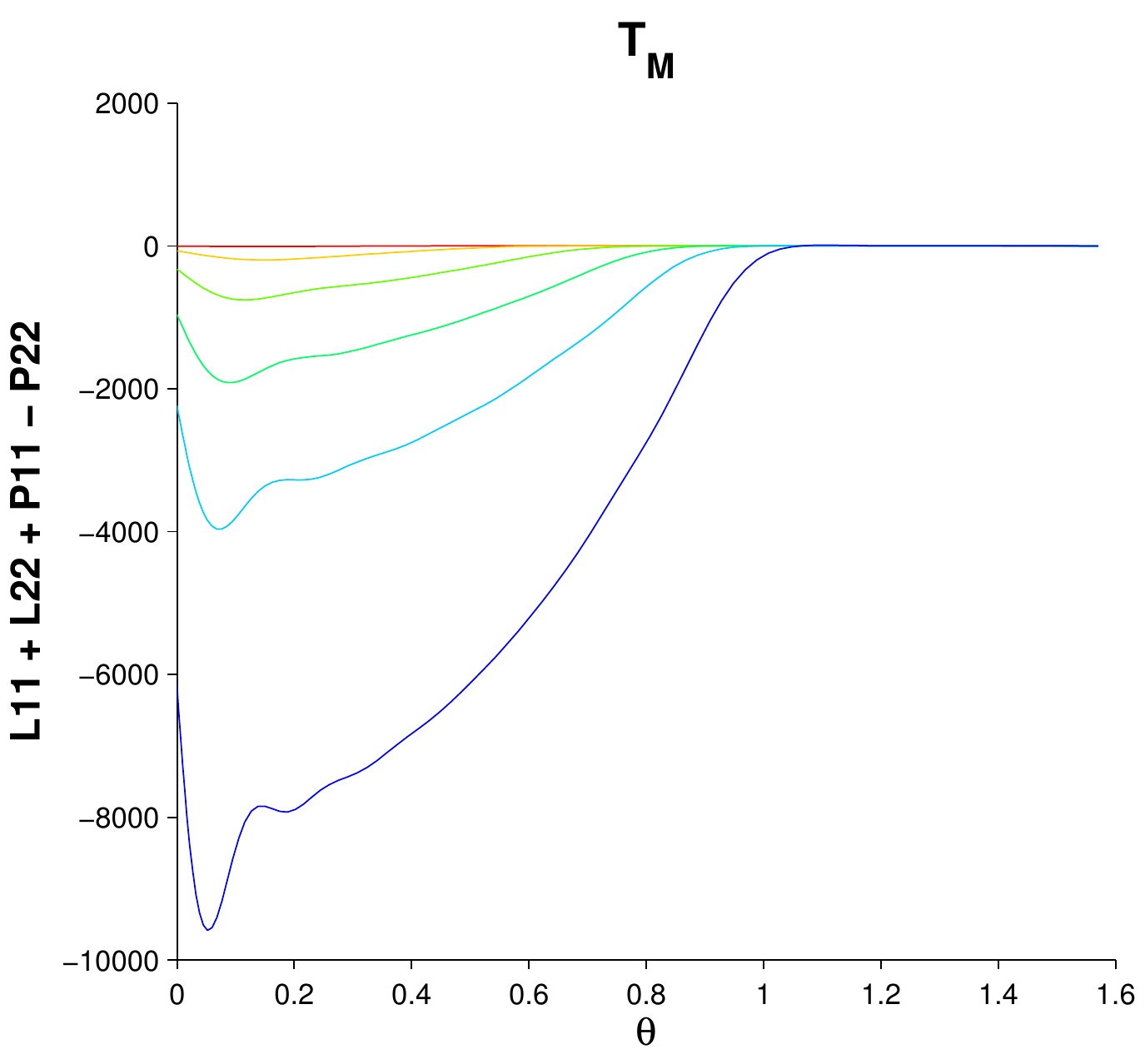}
   \includegraphics[width=4cm]{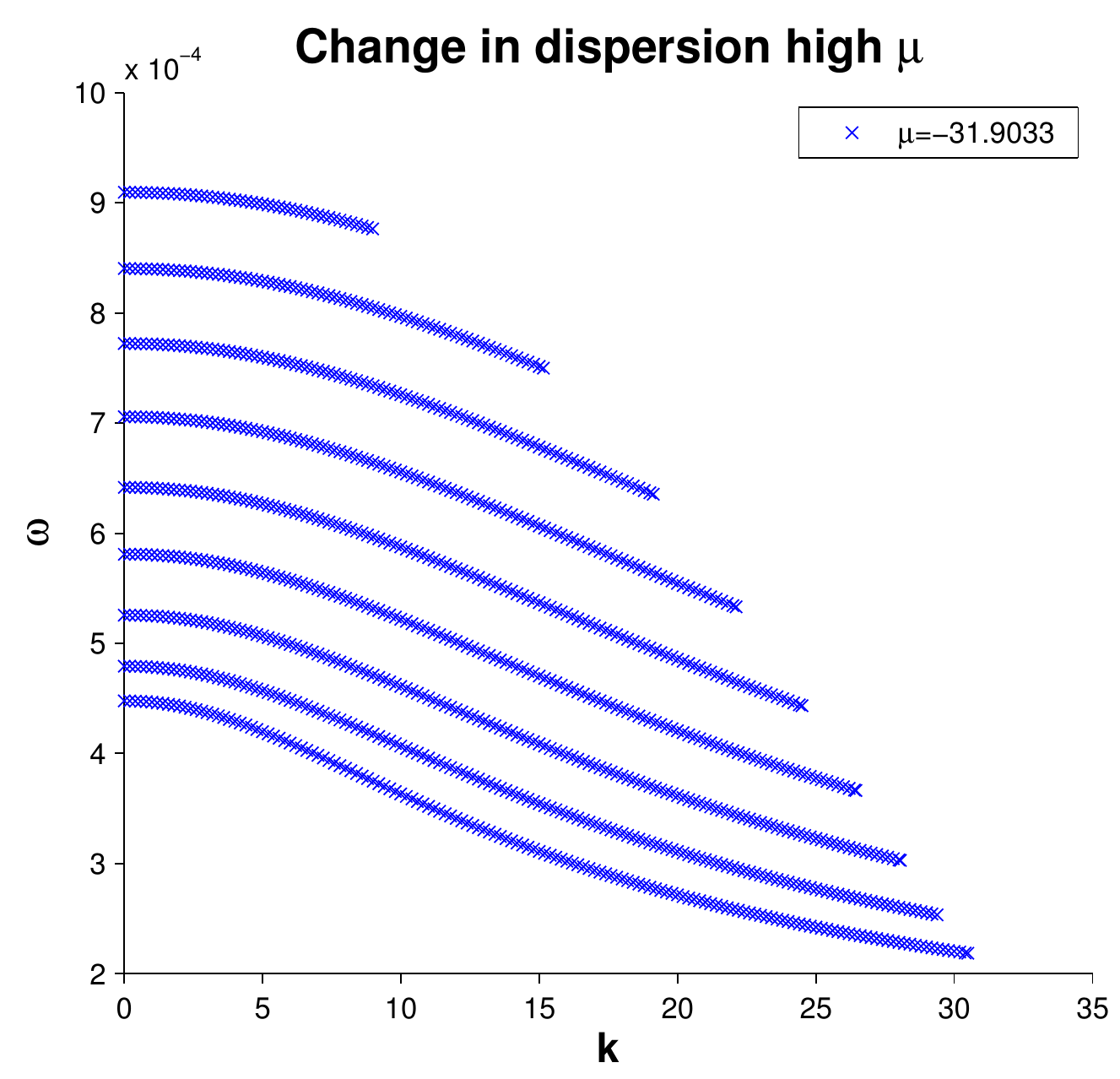}
    \caption[]{Varying $\mu$ with backreaction included, and normalizing
our solutions with respect to the probe limit. The other parameters
were set as $\beta=-0.01$, $\epsilon=0.1$, $m_0=1$, $m_{\psi}=10$.
Electric field strength and charge density are shifted further
away from the embedding cap-off relative to the probe limit.
We see that the sources to the embedding have a profile whose relative importance away from the cap-off
grows with increasing $\mu$.}
\label{fig:mu_full}
\end{figure}
We now consider the effects of backreaction in our system via tuning $\epsilon$ to non-zero values. We now have a five parameter family of solutions characterized by $(\mu,m_0, m_{\psi},\beta, \epsilon)$. In analogy to the previous section we investigate the influence of each individual parameter on the bulk solutions when the others are kept fixed. In order to determine the significance of backreaction we normalize our solutions where appropriate by subtracting off the probe limit background.

 We are particularly interested in examining whether the embedding tends to evolve, when changing parameters, in a direction where the volume in the IR increases significantly. If, in a limiting case, the IR geometry were to tend towards becoming non-compact, it may signify the breakdown of bulk perturbation theory. In the dual QFT this phenomena would signal the breakdown of the large $\mathcal{N}$ expansion. We will illustrate in section \eqref{Greenfunc_sec} below that such a breakdown is necessary for large quasiparticle scattering rates. 
\begin{figure}
\center
 \vspace{-1.5cm}
   \includegraphics[width=4cm]{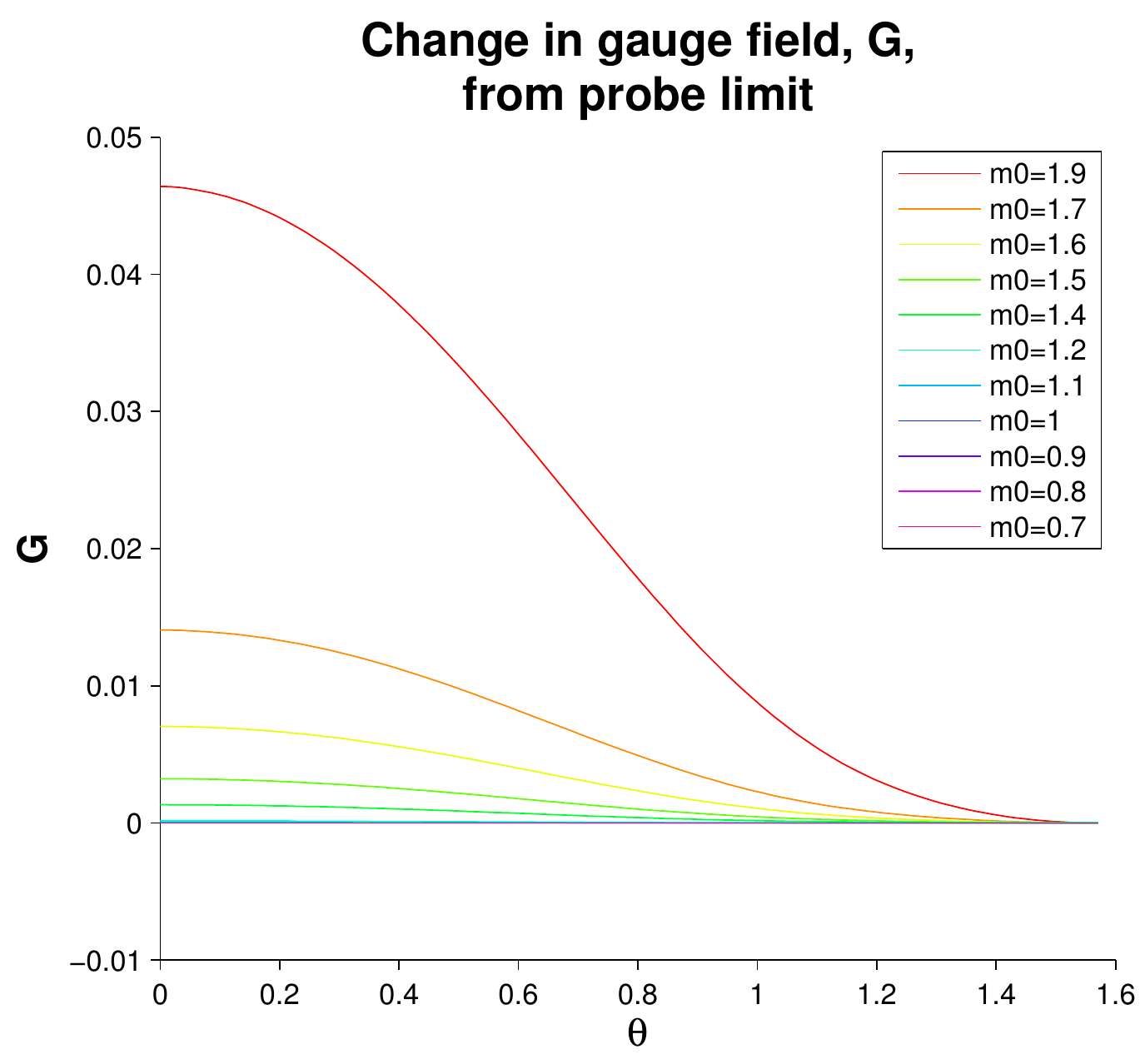}
   \includegraphics[width=4cm]{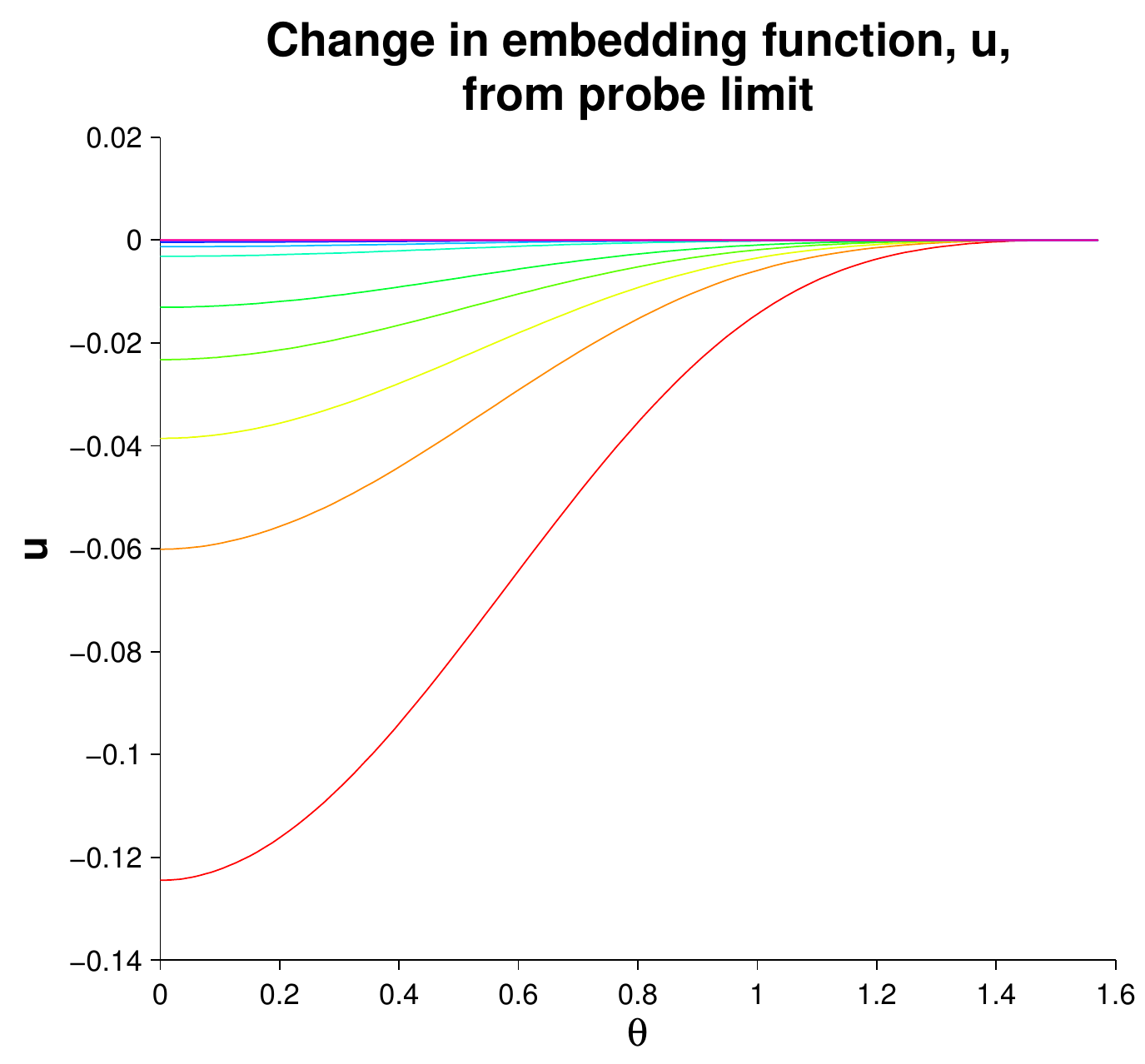}
     \includegraphics[width=4cm]{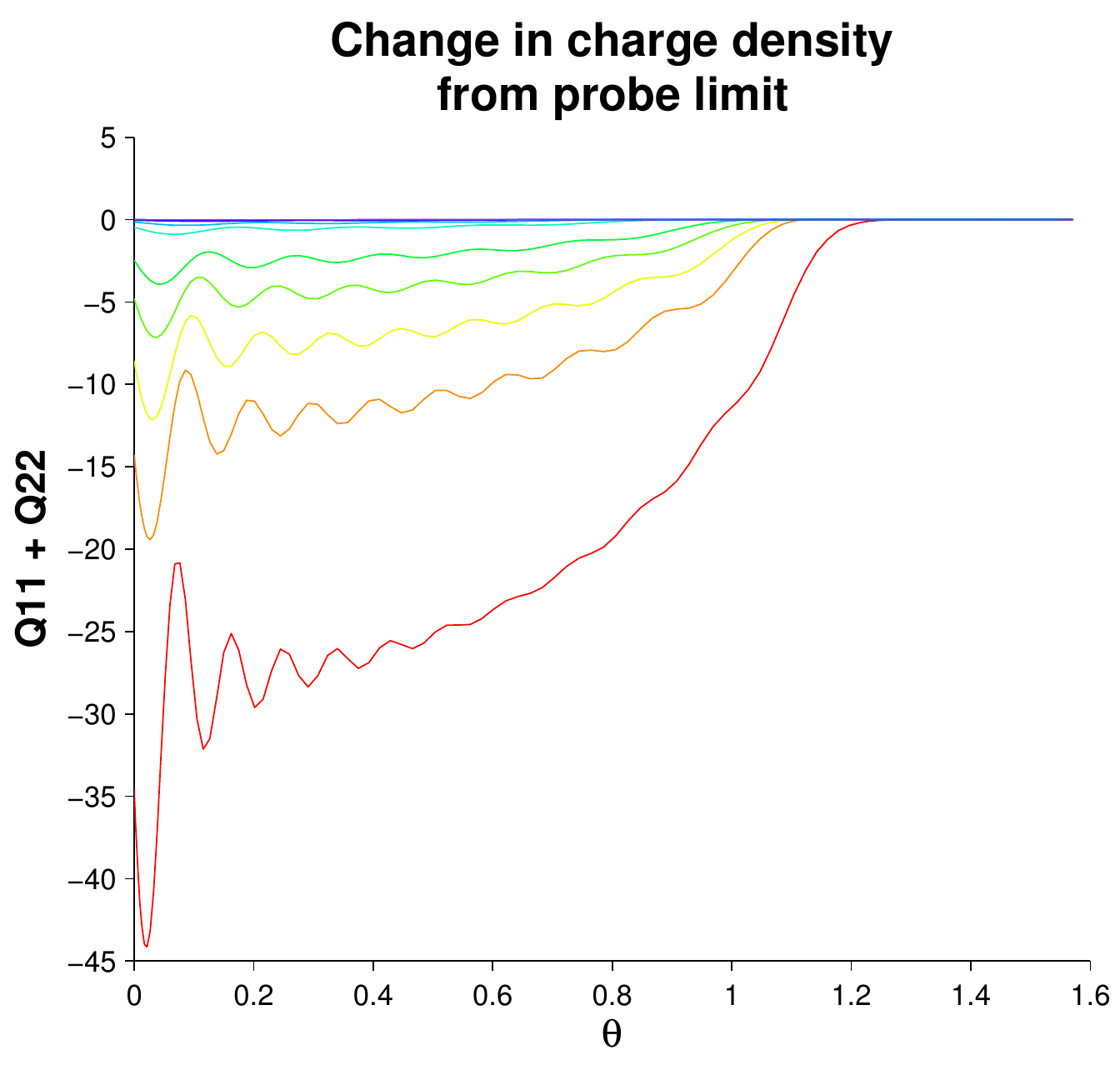}
     \includegraphics[width=4cm]{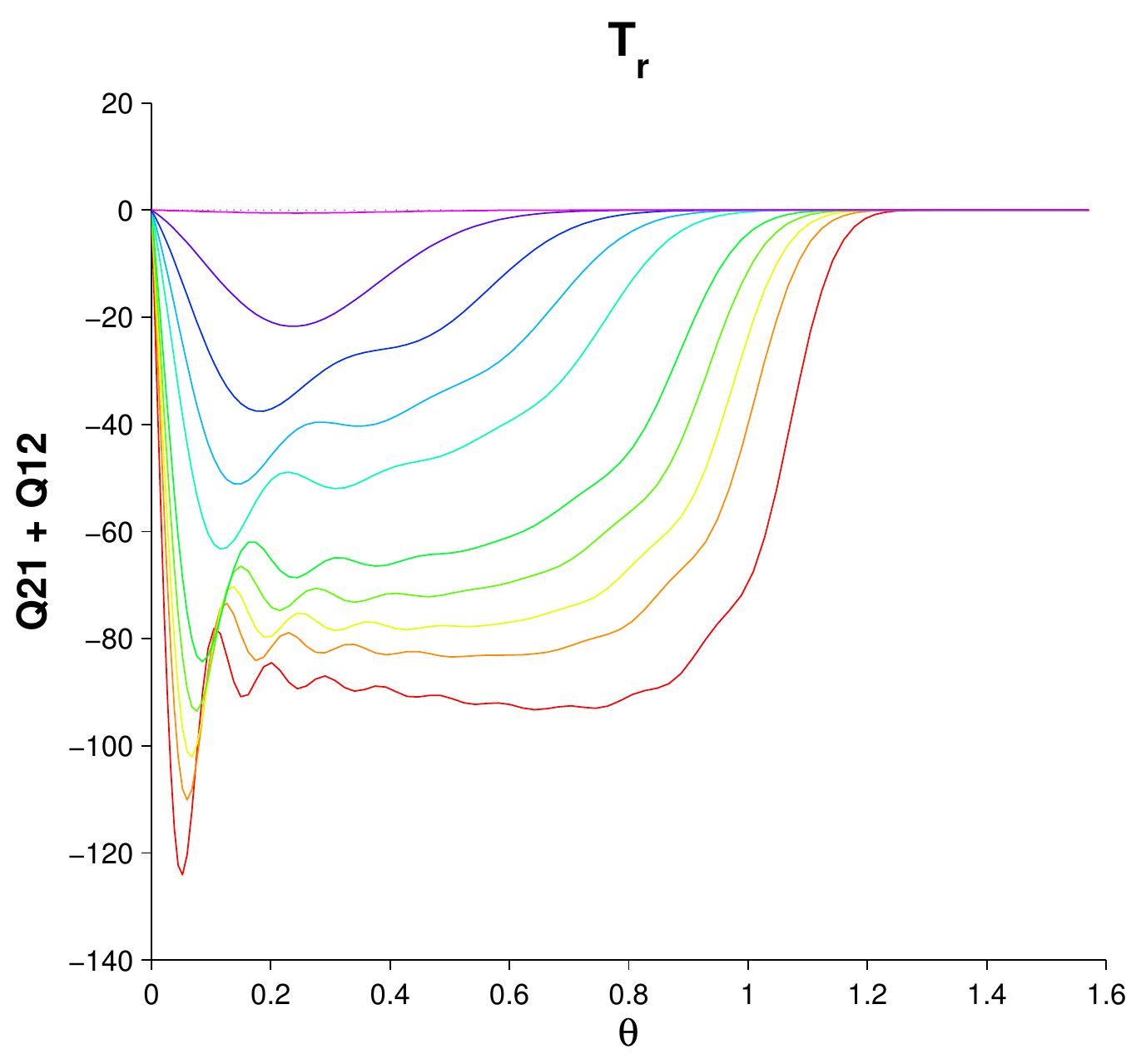}
      \includegraphics[width=4cm]{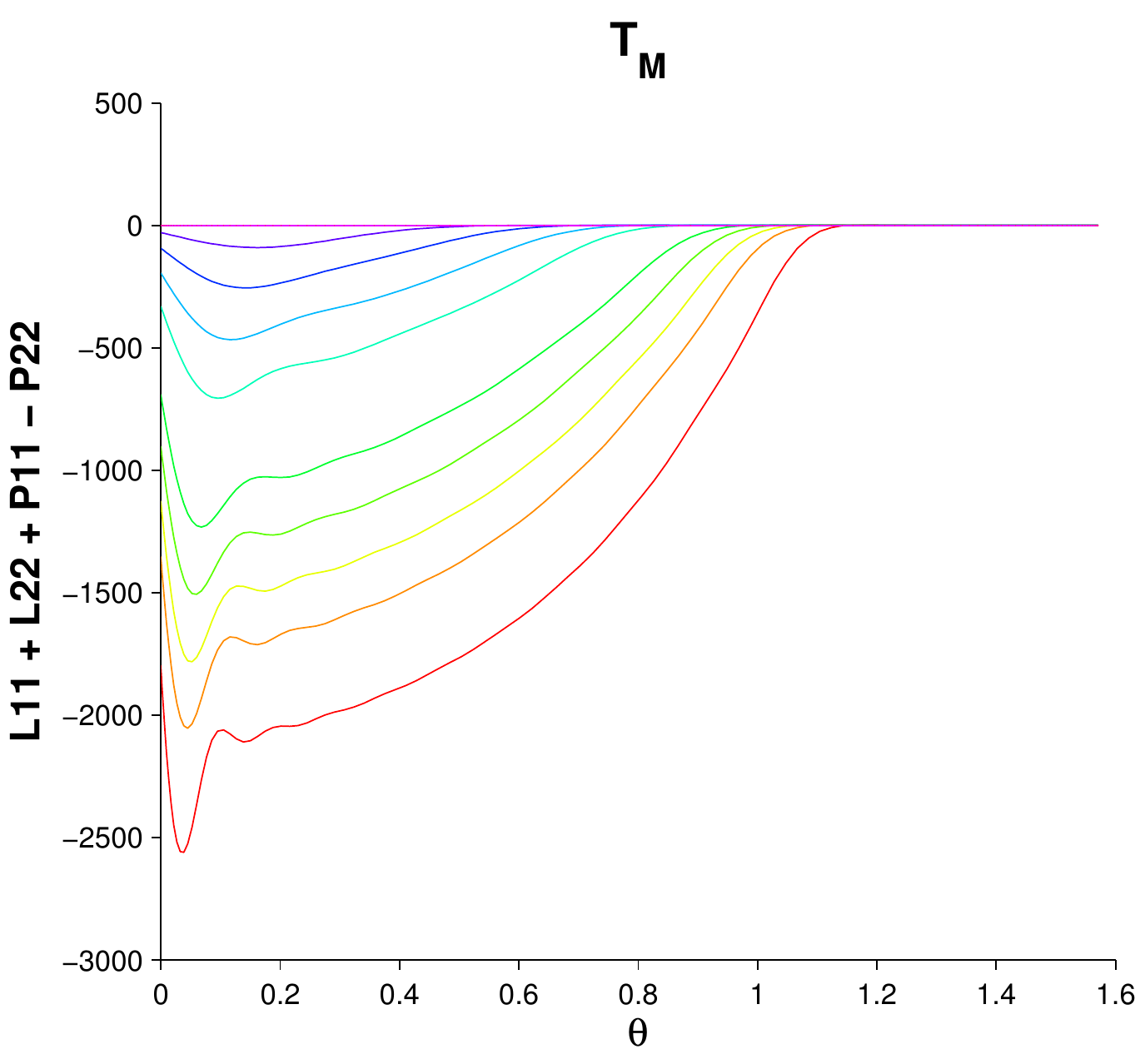}
      \includegraphics[width=4cm]{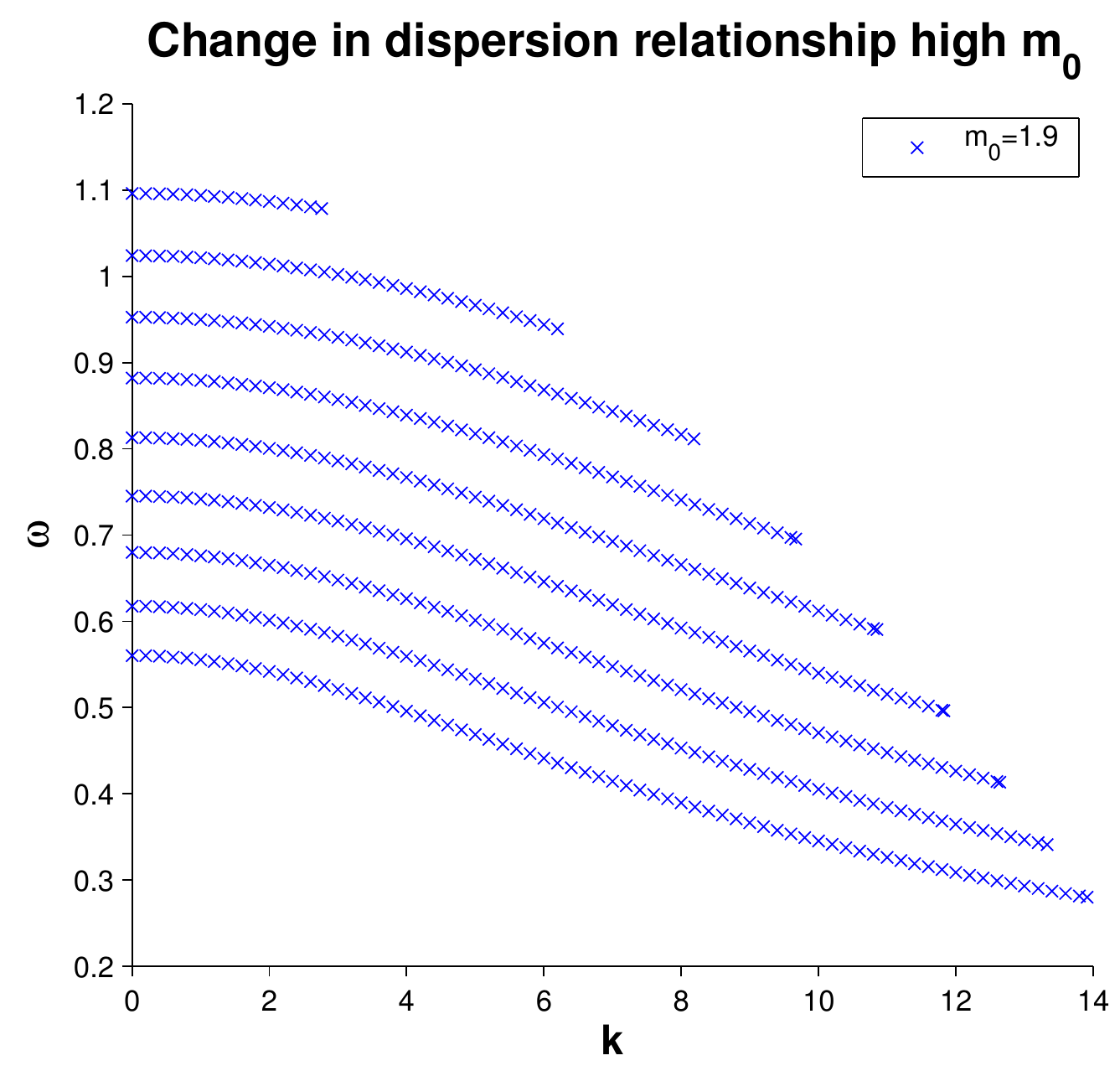}
    \caption[]{Varying $m_0$ with backreaction included, and normalizing our solutions with respect to the probe limit. The
couplings are set as  $\mu= -15.7154, \epsilon=0.1, \beta =-0.001,
m_{\psi}=10$. We note the qualitative similarity to the
changes observed in the profiles at fixed $\mu$ as seen in figure\eqref{fig:mu_full}.}
\label{fig:m0_full}
\end{figure}

        As in the previous section, we first consider changes in our thermodynamic variables. Our ensemble is now defined via $(\mu, m_0)$, as encoded in the boundary conditions for the relevant bulk fields. In figure \eqref{fig:mu_full} we plot the change in the bulk fields displayed in figure \eqref{fig:mu_probe} once backreaction is included. We note that the charge density decreases slightly in the vicinity of the cap-off once backreaction is included. This has the effect of decreasing the electric field strength in the cap-off region and allowing it to retract slightly towards the UV. This change is sourced by $T_{r} $ and $T_{M}$ which we also display below. These effects become more pronounced as $|\mu|$ increases. The net result is that the transverse sphere collapses to a point slightly sooner then in the probe limit case, but still does so at a finite value as a result of support provided by the Fermi pressure. We also note that the energy of the filled states is decreased relative to the probe limit. A similar story emerges as we follow the branch of solutions parameterized by $m_0$, as seen in figure \eqref{fig:m0_full}. Electric field and charge density profiles are shifted slightly towards the mid-region of the geometry while the cap off itself retracts towards the conformal boundary.

\begin{figure}
 \vspace{-1cm}
\center
   \includegraphics[width=4cm]{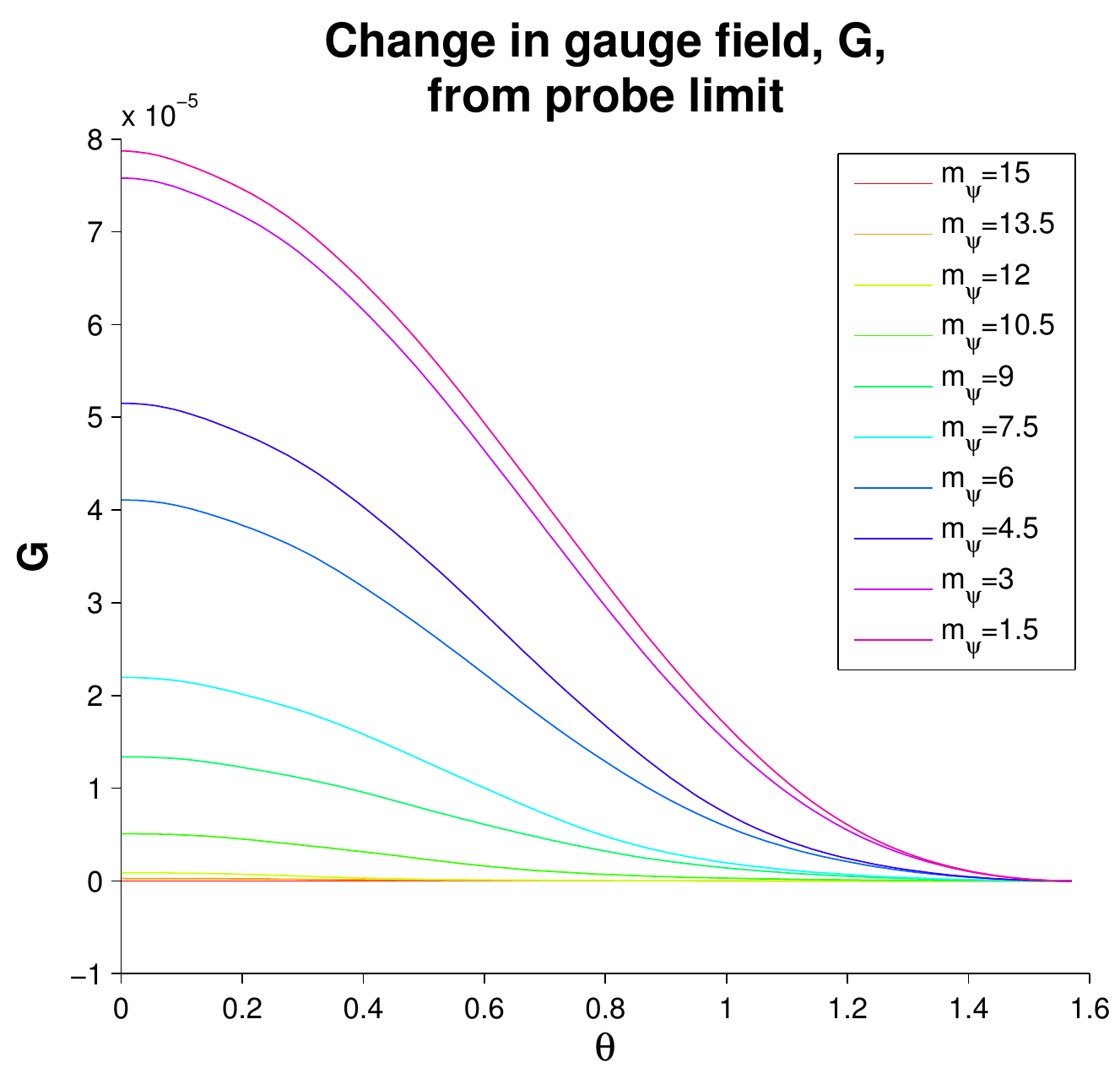}
      \includegraphics[width=4cm]{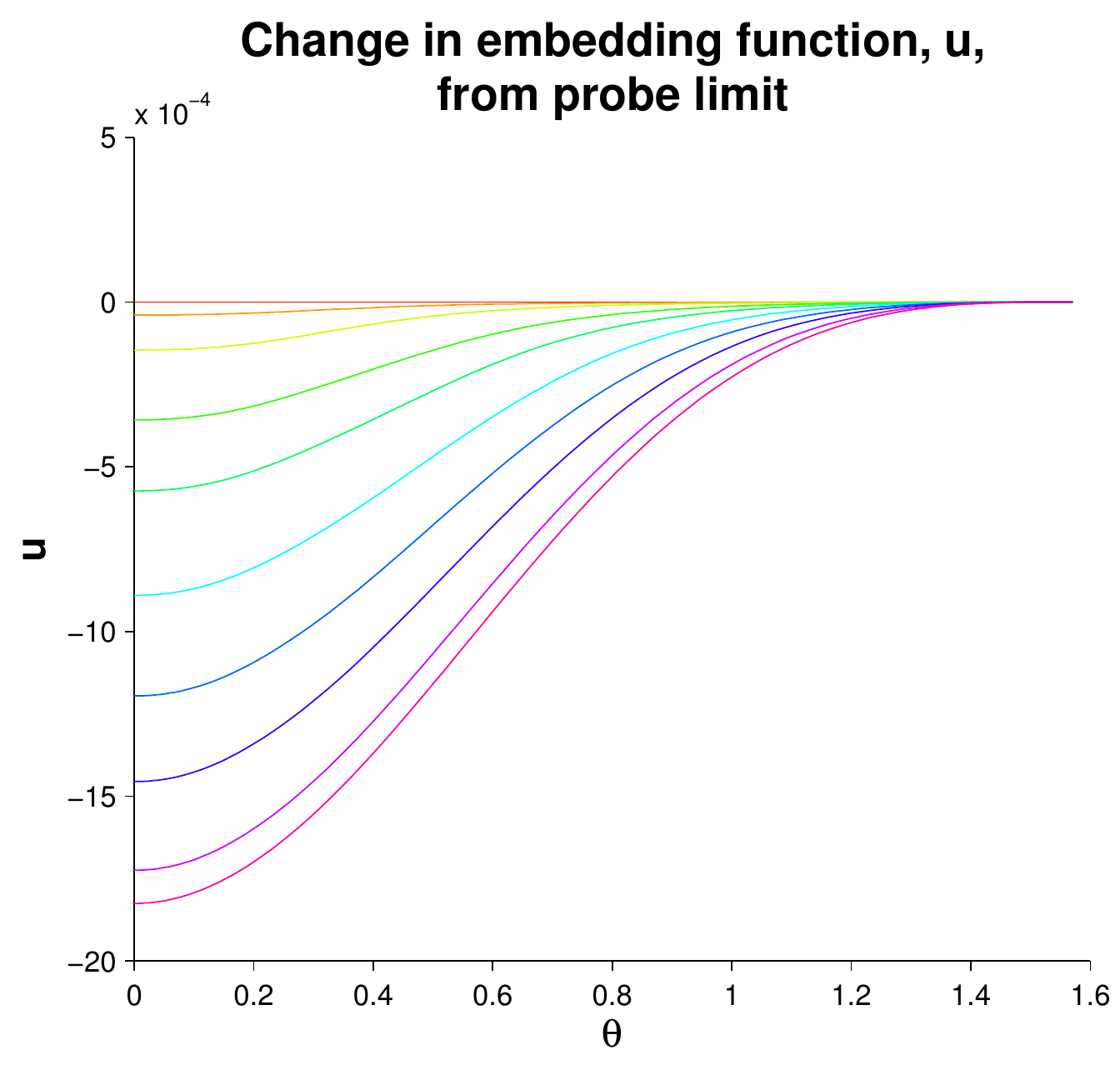}
     \includegraphics[width=4cm]{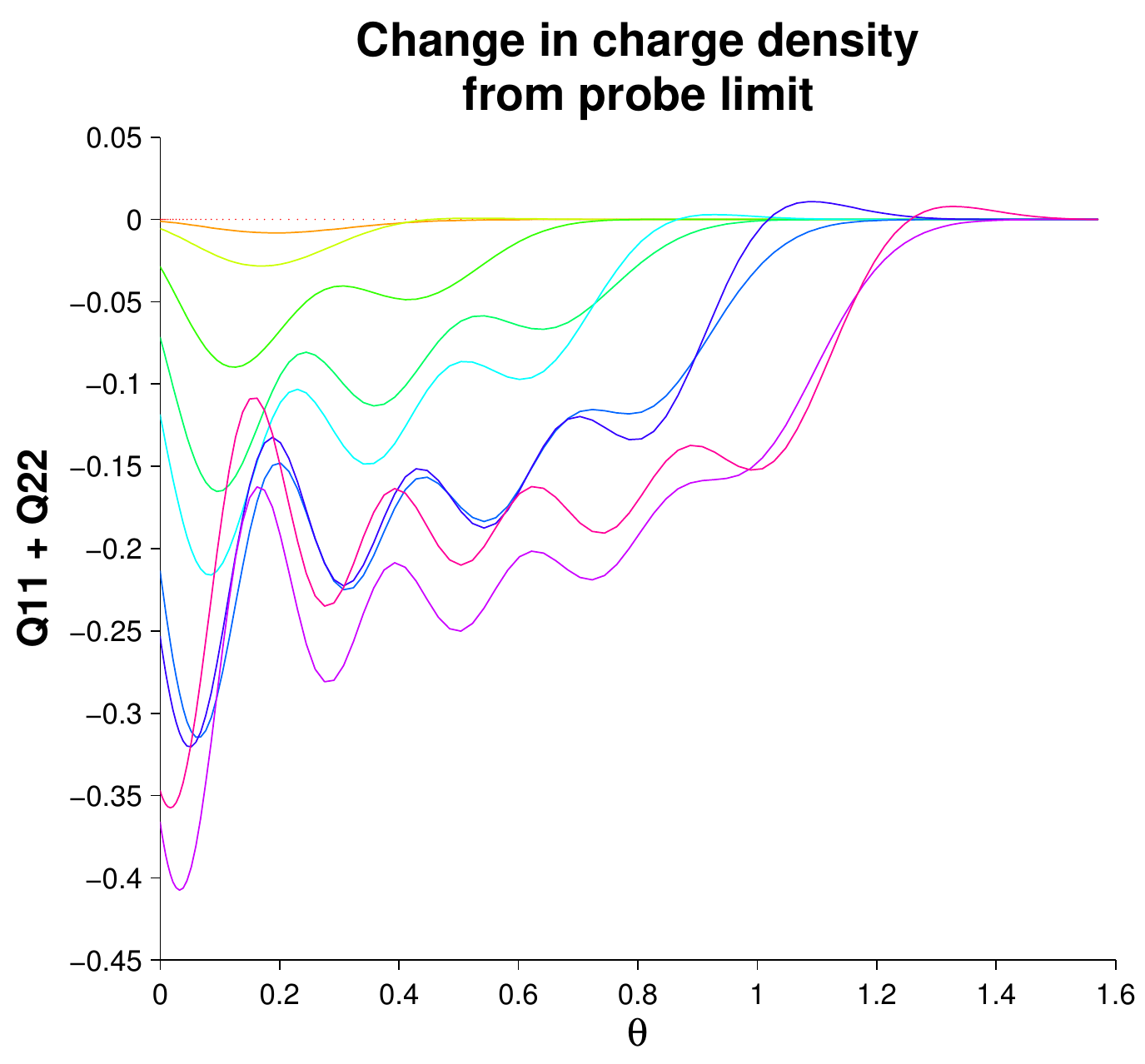}
     \includegraphics[width=4cm]{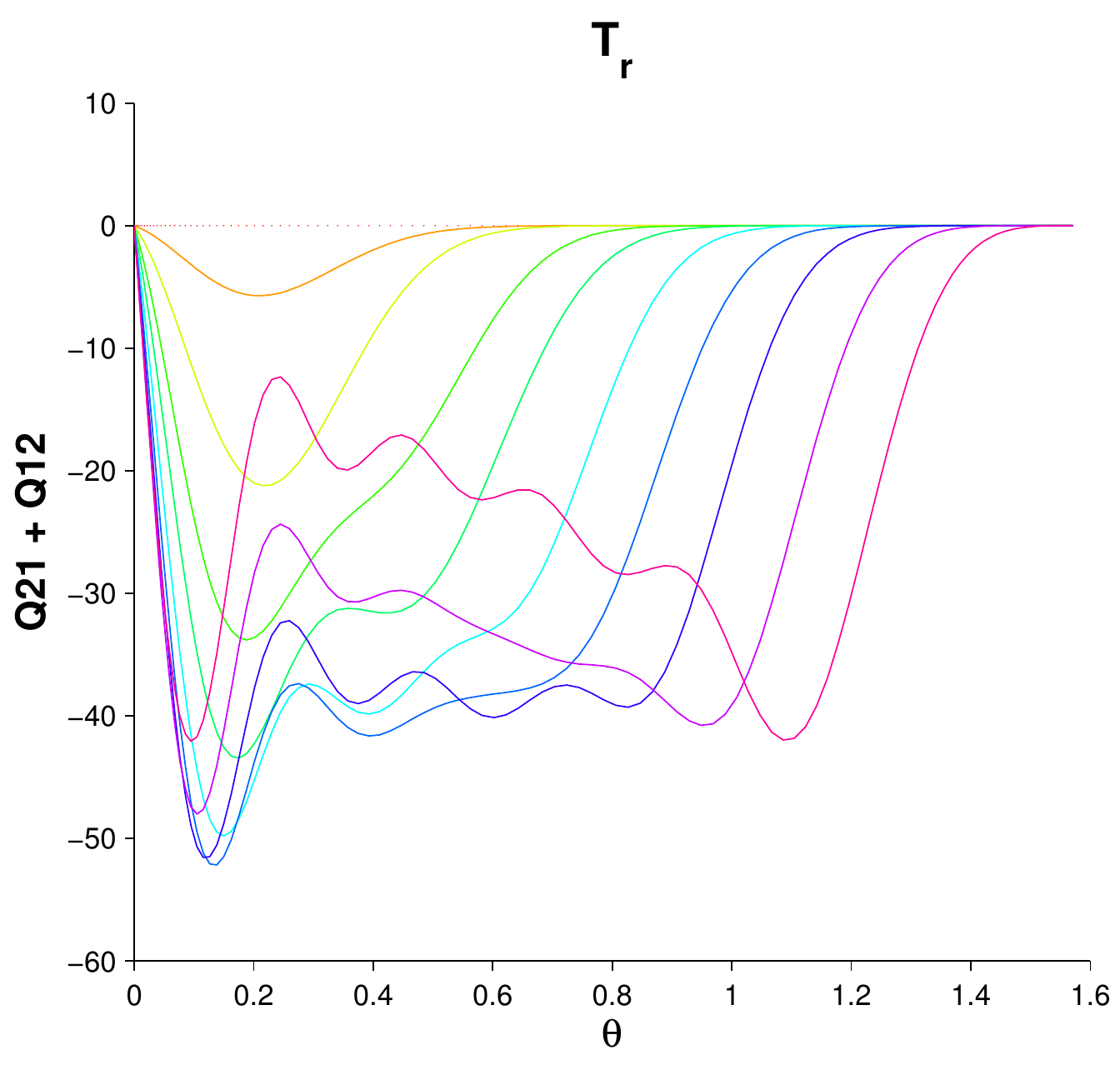}
      \includegraphics[width=4cm]{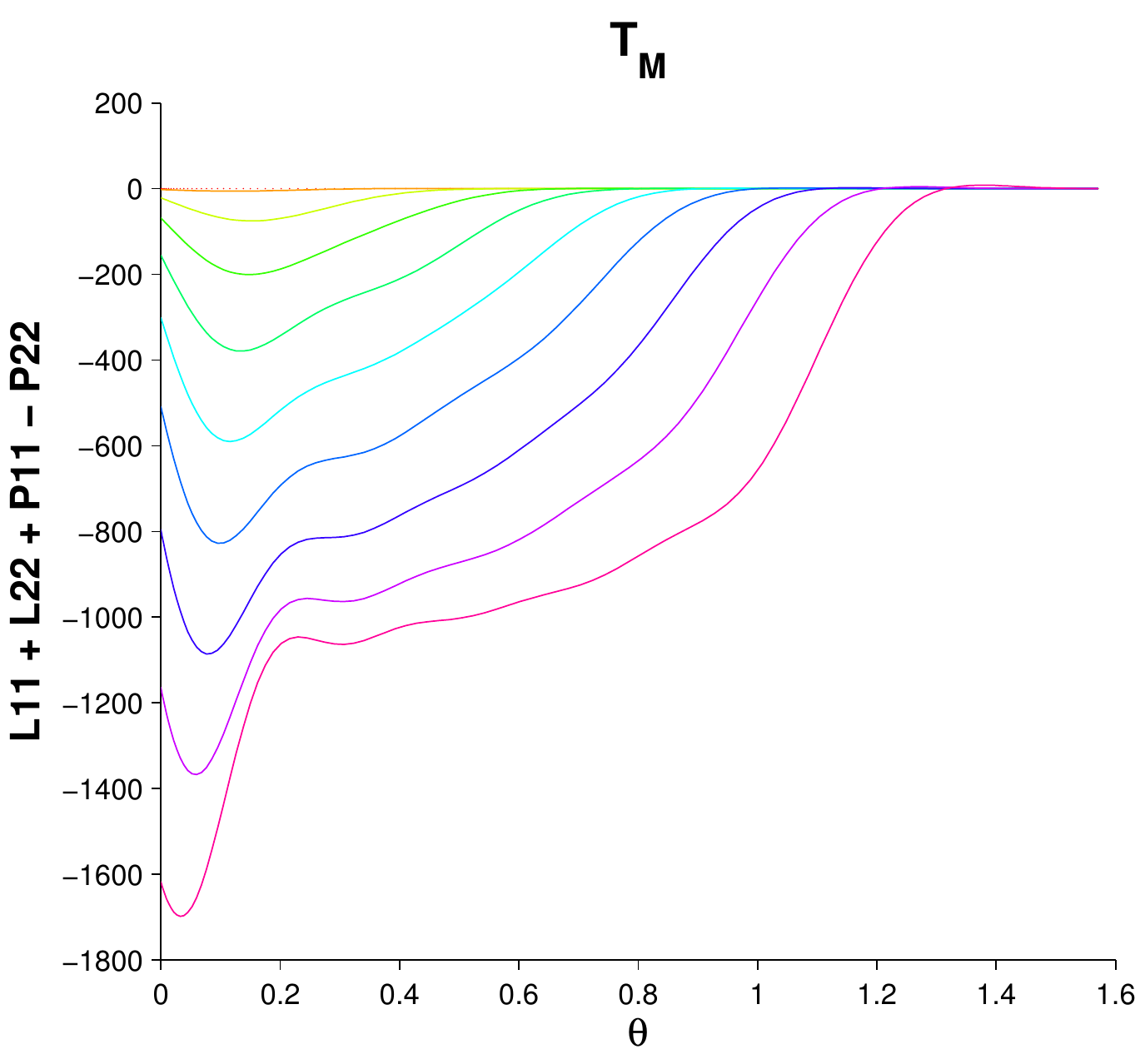}
       \includegraphics[width=4cm]{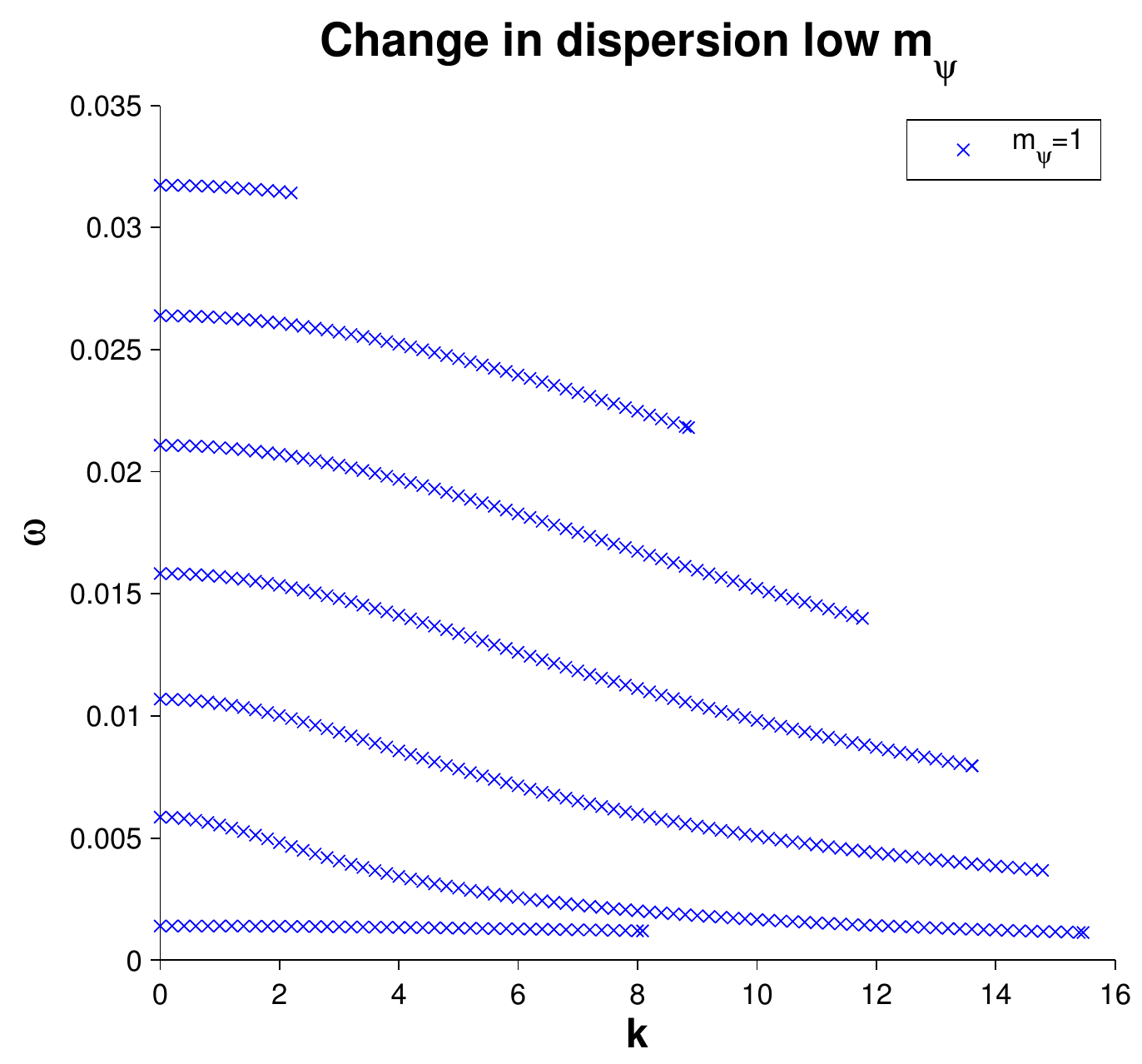}
    \caption[]{Varying $m_{\psi}$ with $\mu=-15.7154$, $m_0=1$,
$\beta=-0.001$, $\epsilon=0.1$. Here we plot the change in the
embedding function, gauge field and charge density relative to
the probe limit. We also plot the the un-subtracted components
of the sources for the embedding function, $T_r$ and $ T_M$. It can be seen
that the importance of backreaction is suppressed for both sufficiently
small and large value of $m_{\psi}$}.
\label{fig:mb_full}
\end{figure}

         Turning to $m_{\psi}$ we see that, for sufficiently large values of this parameter, changes to the bulk fields due to backreaction are suppressed. Initially, decreasing $m_{\psi}$ has a similar effect on the sources for the embedding function, charge density and electric fields as increasing $m_0$ or $\mu$. However at a critical value (for the choice of parameters in figure  \eqref{fig:mb_full} this occurs at $m_{\psi} \simeq 6$) a qualitative change in behaviour occurs. The rate of change of the peak modulation relative to the probe limit slows down and saturates. Instead the effects of backreaction are seen to extend further and further towards the UV boundary.

\begin{figure}

\center
   \includegraphics[width=4cm]{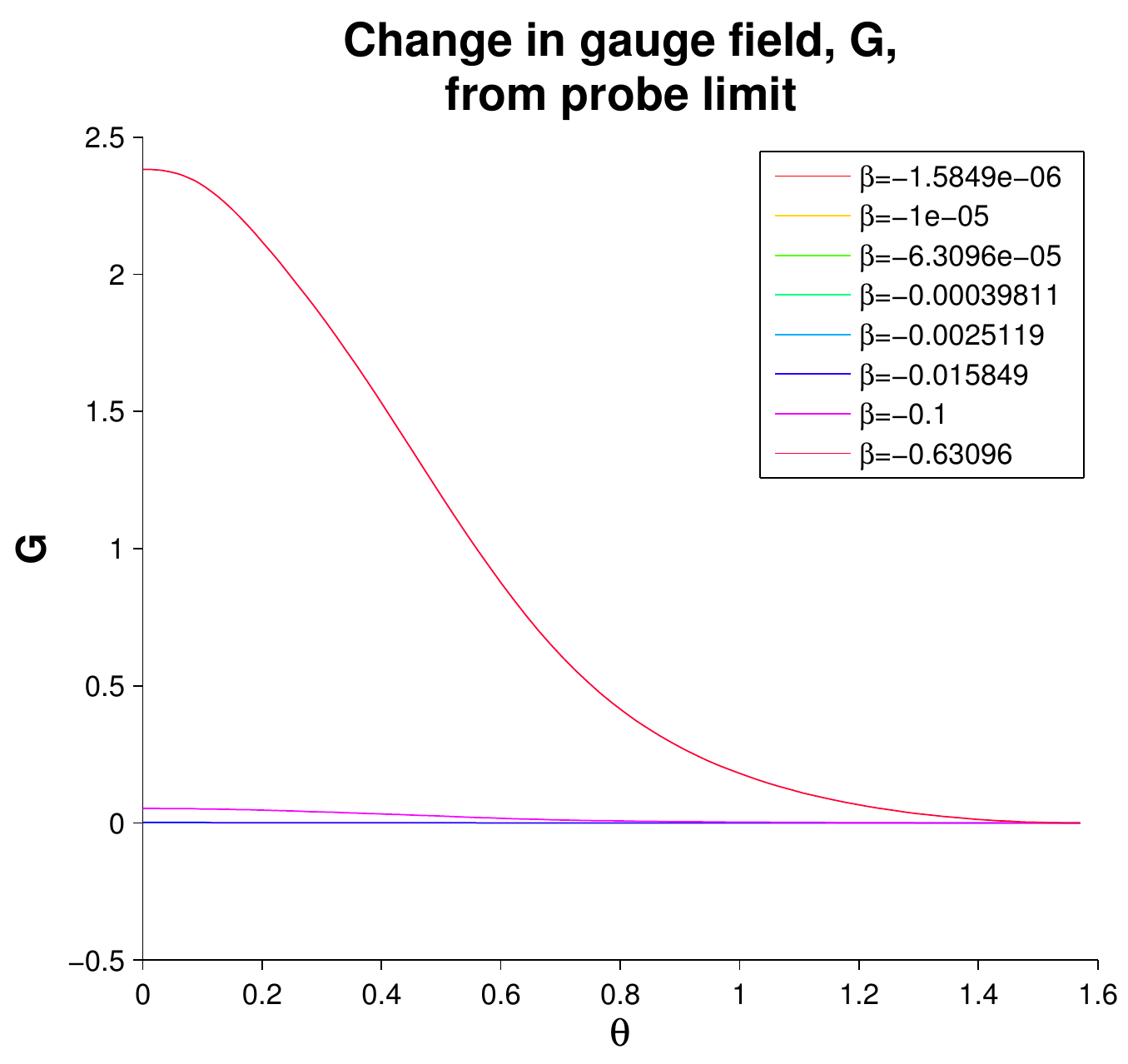}
     \includegraphics[width=4cm]{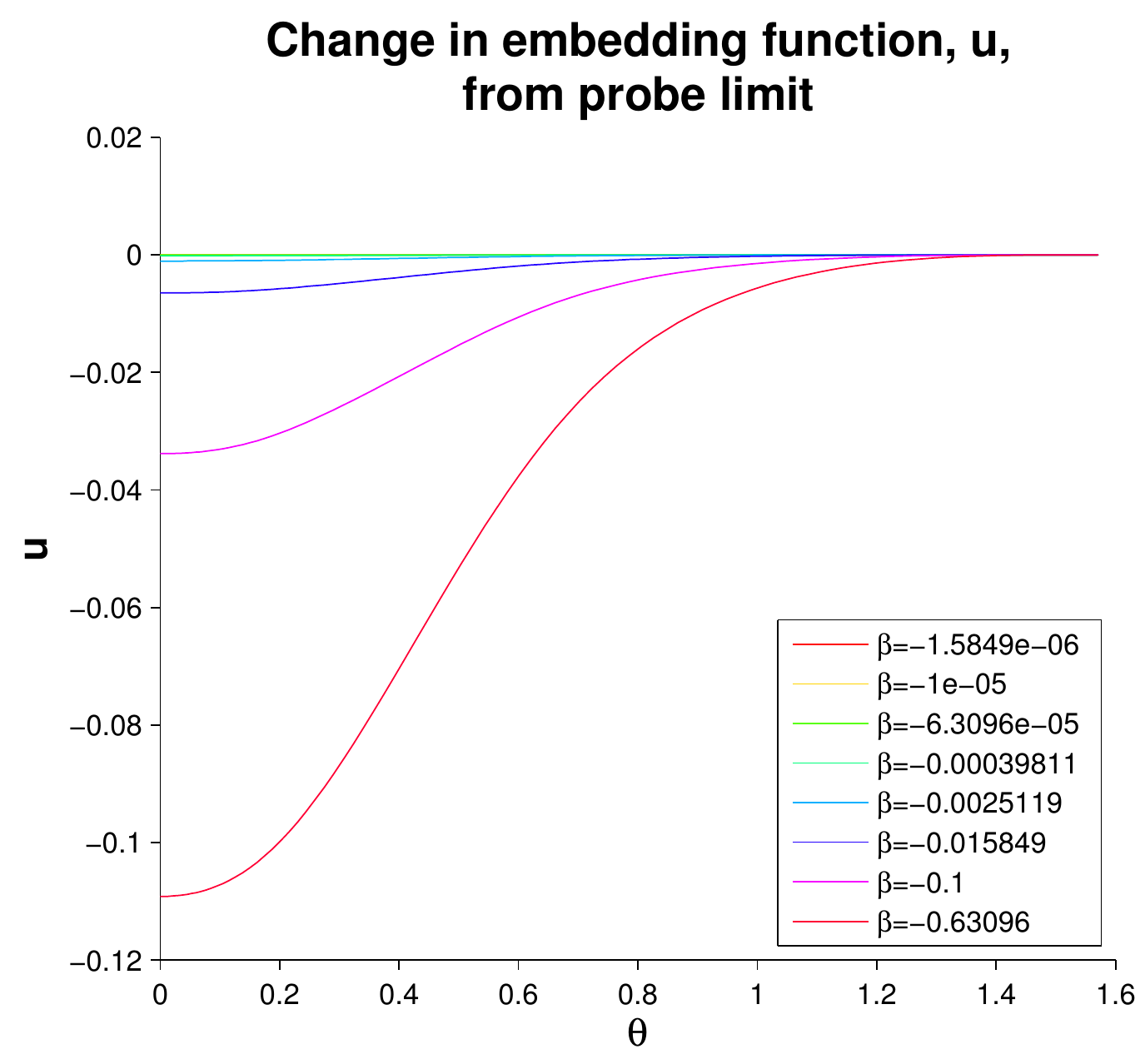}
     \includegraphics[width=4cm]{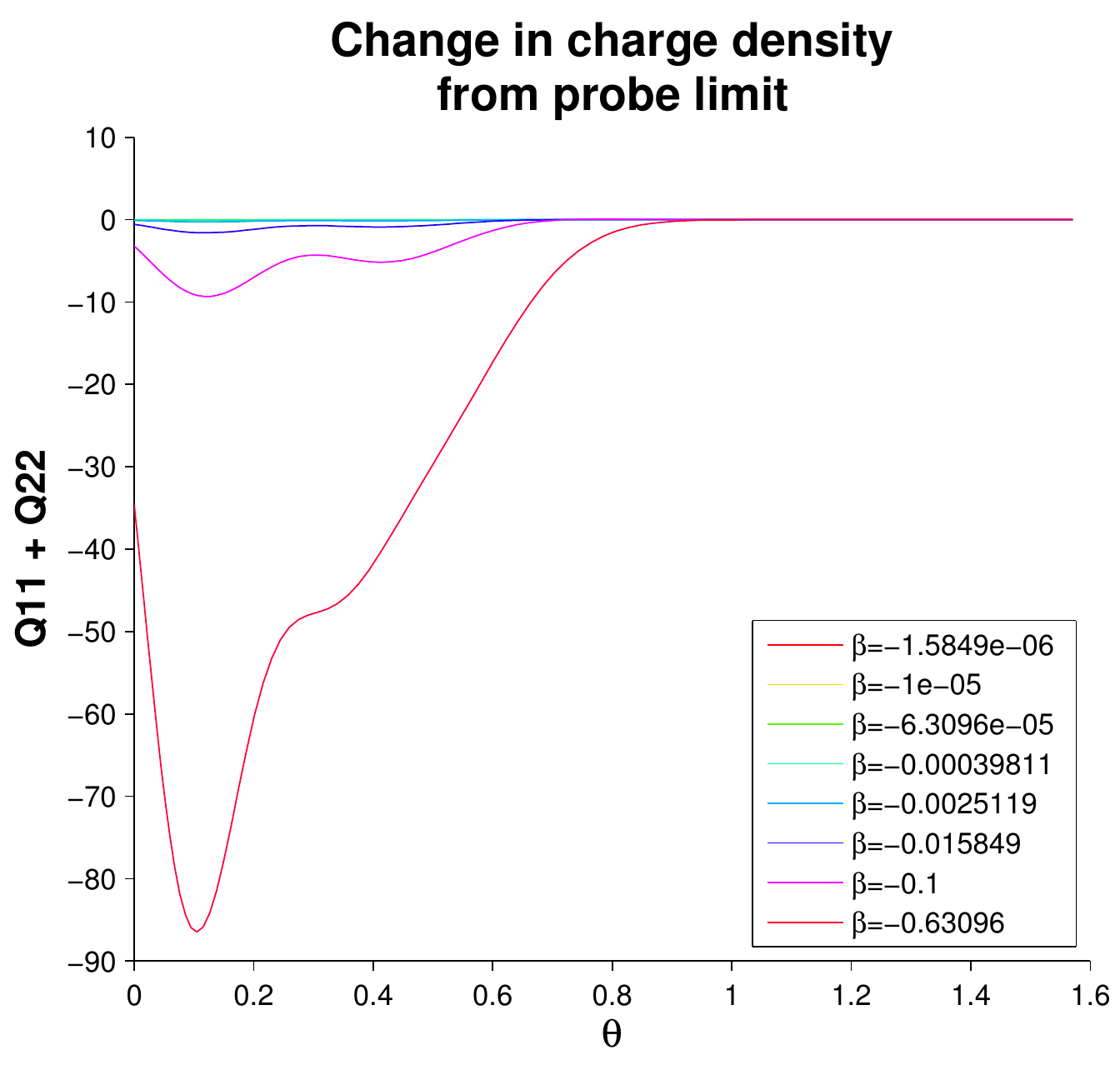}
     \includegraphics[width=4cm]{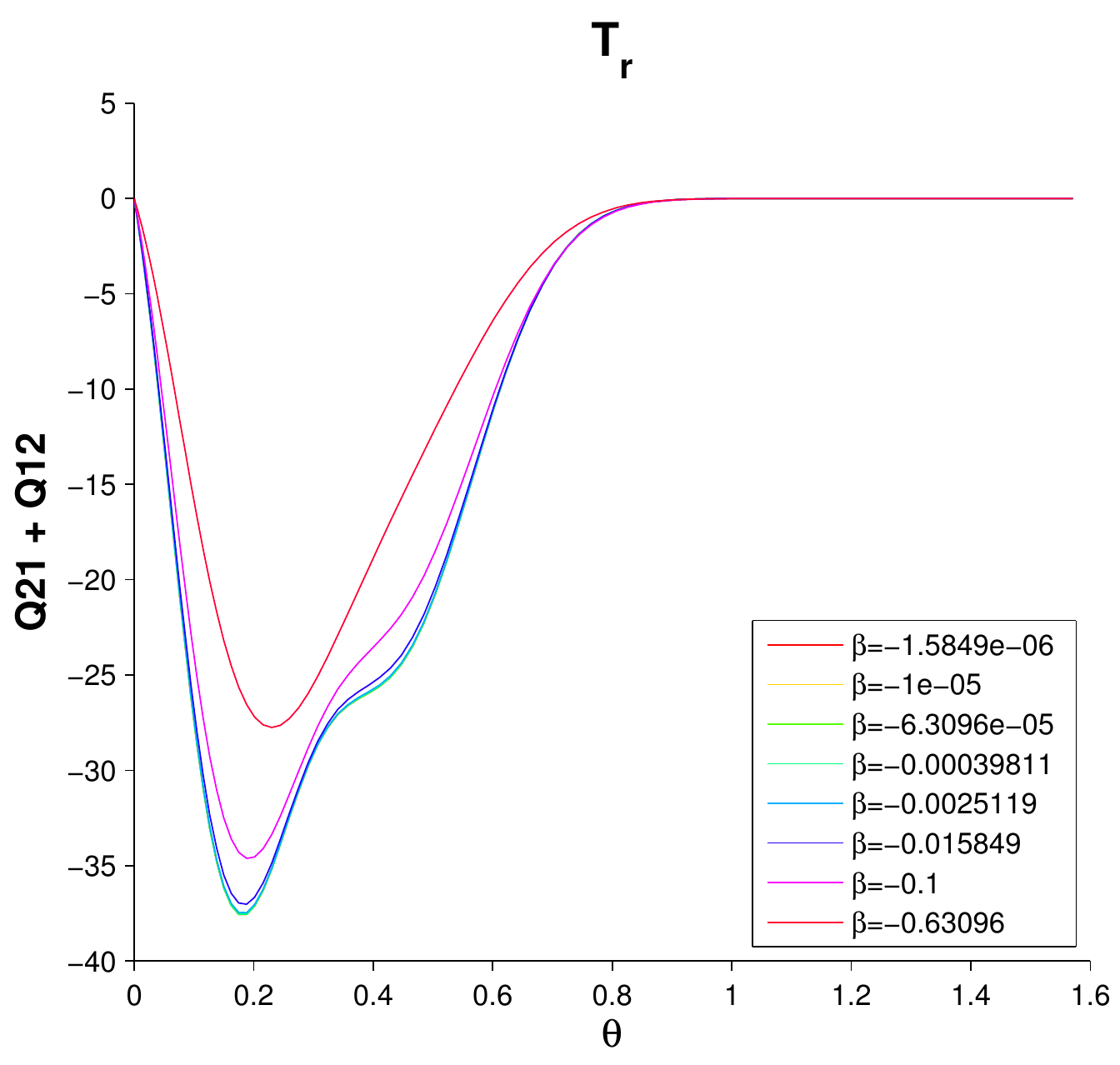}
      \includegraphics[width=4cm]{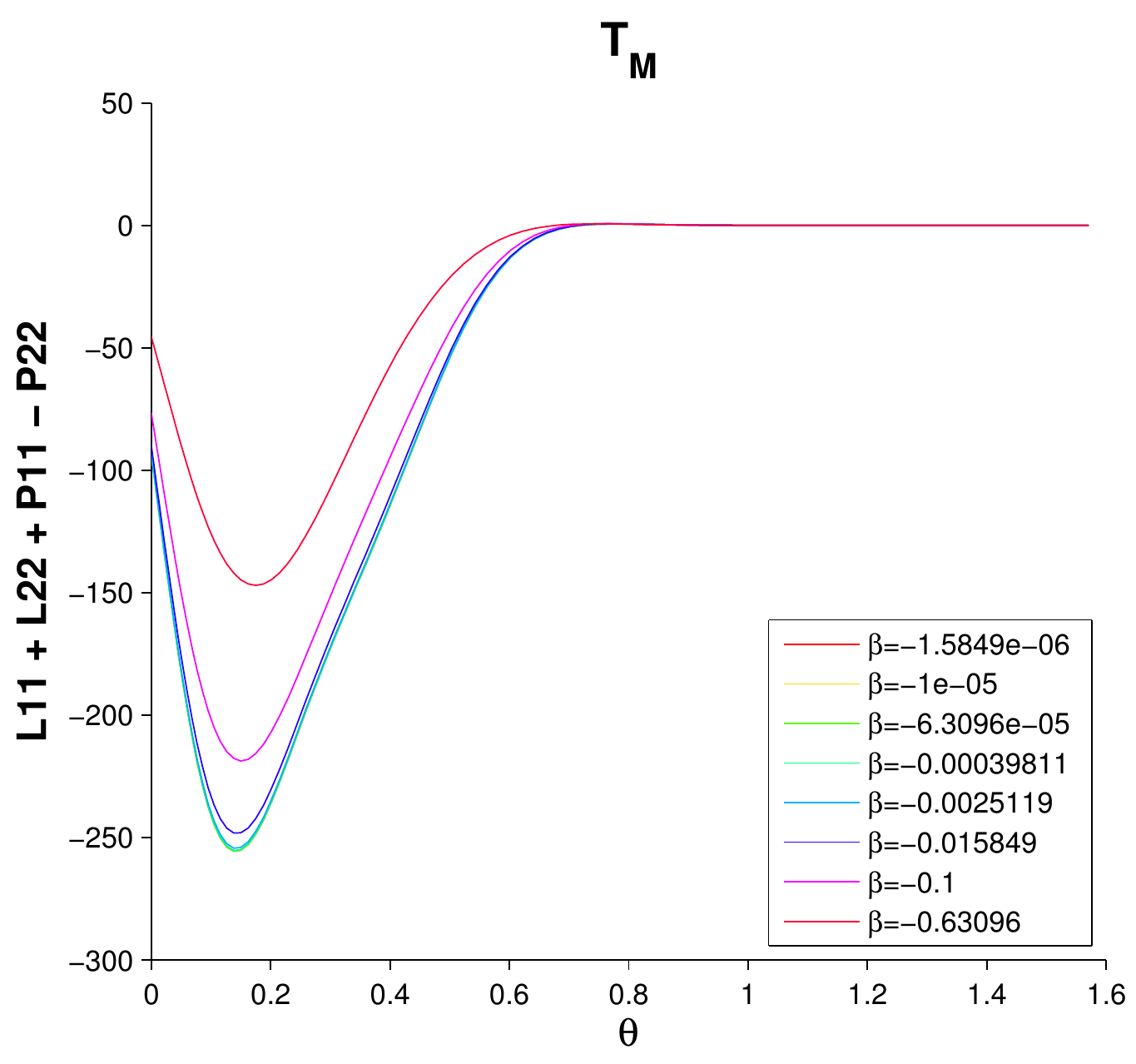}
      \includegraphics[width=4cm]{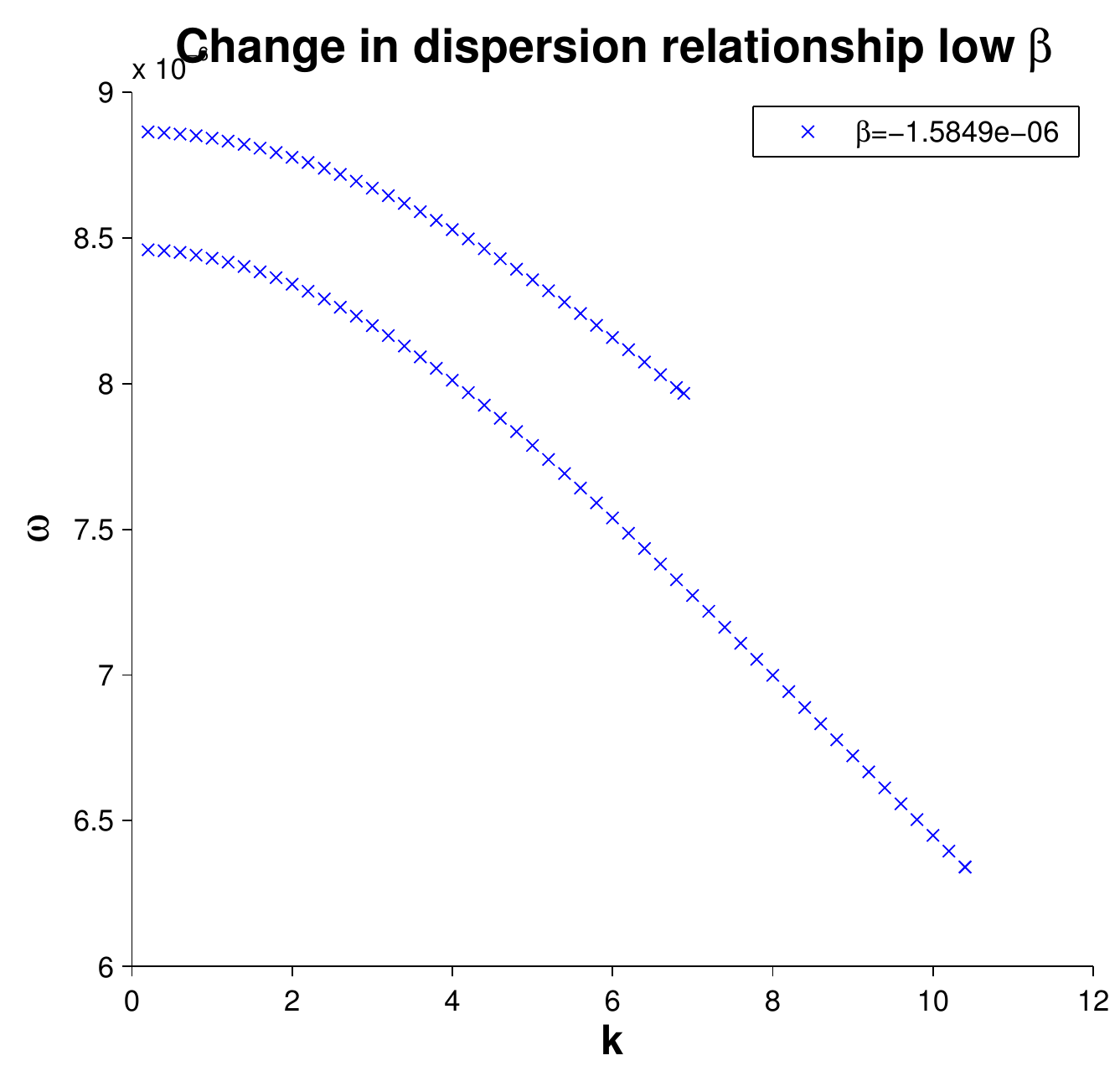}
    \caption[]{Varying $\beta$, with $\mu=-15.7154$, $m_0=1$, $m_{\psi}=10$, $\epsilon=0.1$. It is interesting to note the relative lack of modulation of the profiles of the embedding sources or the charge density, in comparison to variations involving $m_0$, $\mu$ or  $m_{\psi}$. This may be attributed to the relatively minor changes which occur in the spectrum of filled states. The energy of the filled bands changes as $\beta$ is varied however their number and qualitative shape does not. A similar pattern may be observed for $\epsilon$ variations in figure (9).}
\label{fig:beta_full}
\end{figure}

Next, in figures \eqref{fig:beta_full} and (9) we examine the effects of backreaction for variations of $\beta$ and $\epsilon,$ respectively. As these control the coupling of the matter sector fields to the embedding function,  increasing the magnitude of either parameter serves to accentuate the effects of backreaction. The qualitative nature of this behaviour is similar to that that observed for the other parameter variations described previously. The details are described in the captions to figures \eqref{fig:beta_full} and  (9).

From the above analysis we conclude that the incorporation of
backreaction in the model has two principle consequences:
\begin{itemize}
\item Due to the finite charge density at the cap-off and the
associated Fermi pressure Minkowski embedding are now possible.
The effect of backreaction is modify the geometry in the region
near where the geometry caps off.
\item While backreaction remains small its effect on the bulk
fields is to shift the charge density distribution and associated
electric field slightly in the direction of the conformal boundary.
The fermions are less tightly bound then in the probe limit.
\end{itemize}

\begin{figure}
\center
   \includegraphics[width=4cm]{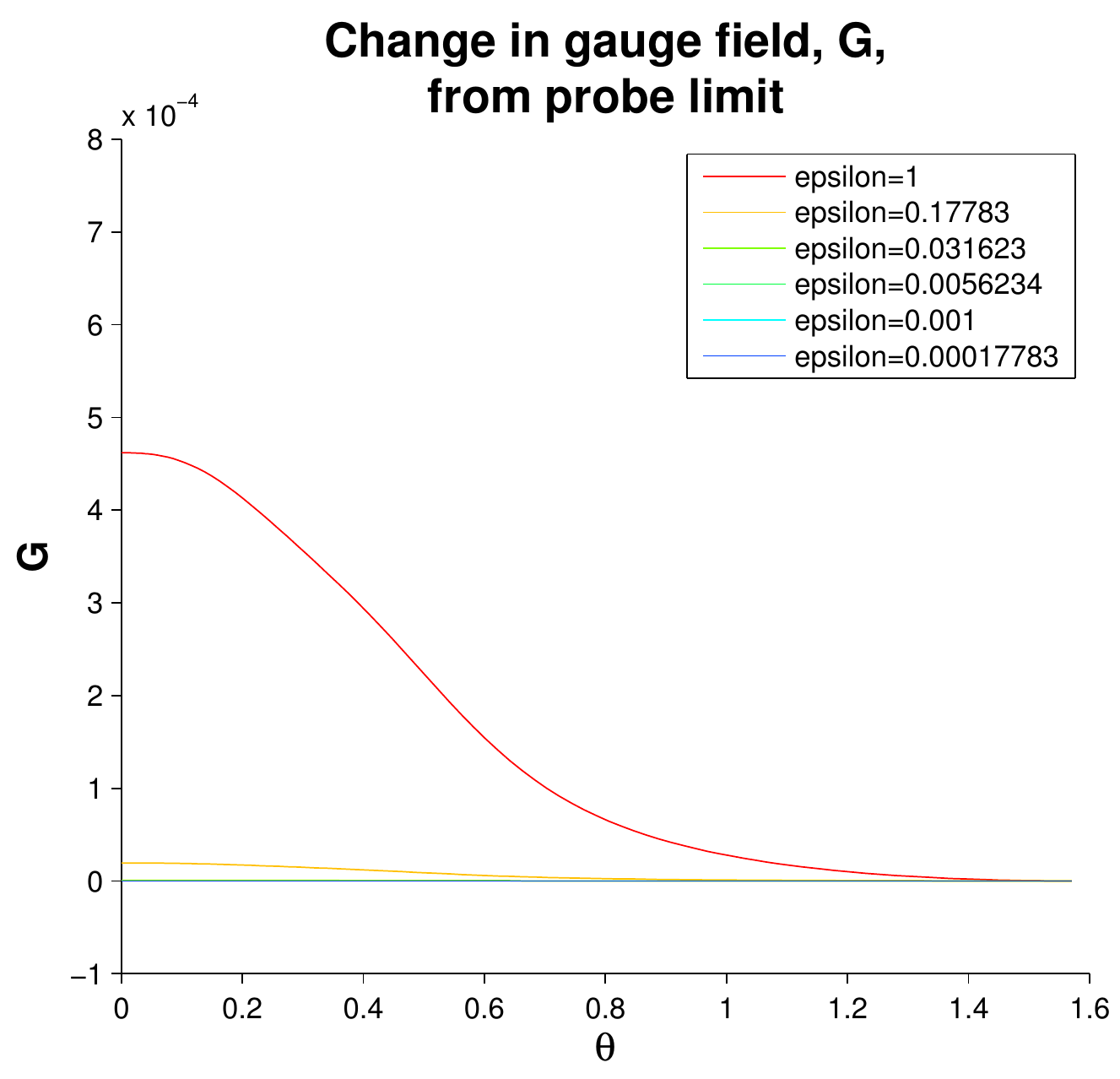}
    \includegraphics[width=4cm]{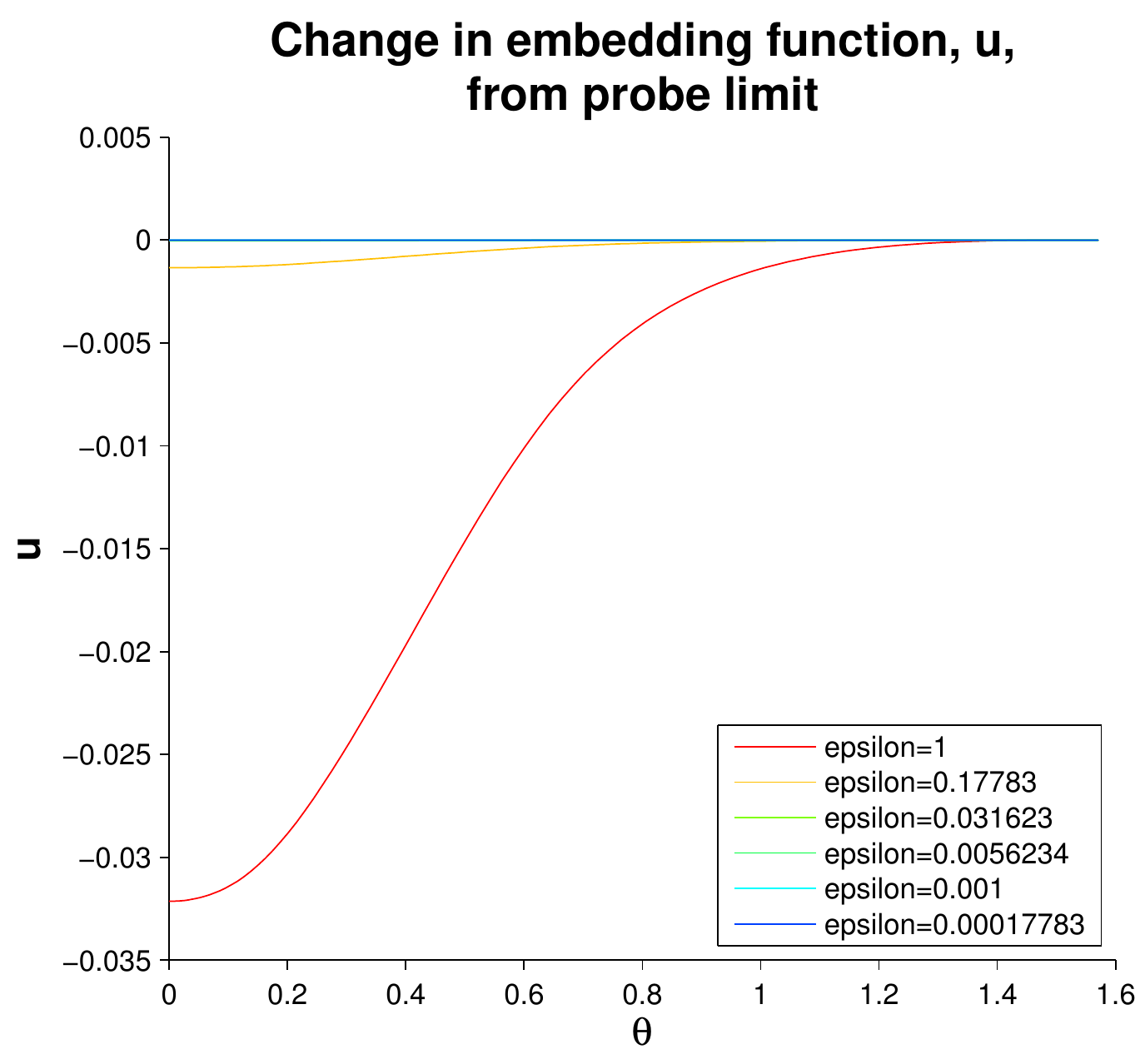}
     \includegraphics[width=4cm]{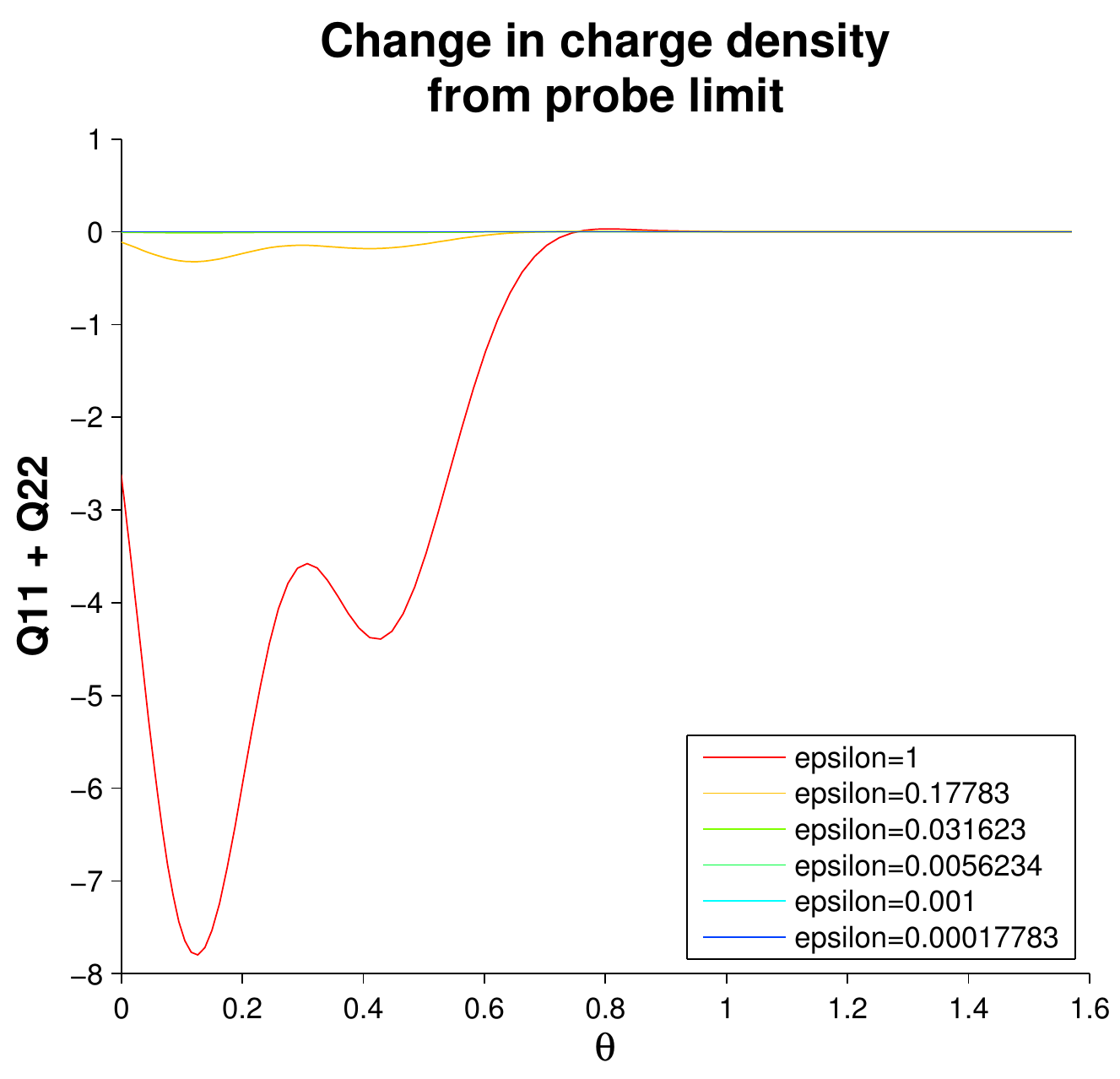}
    \includegraphics[width=4cm]{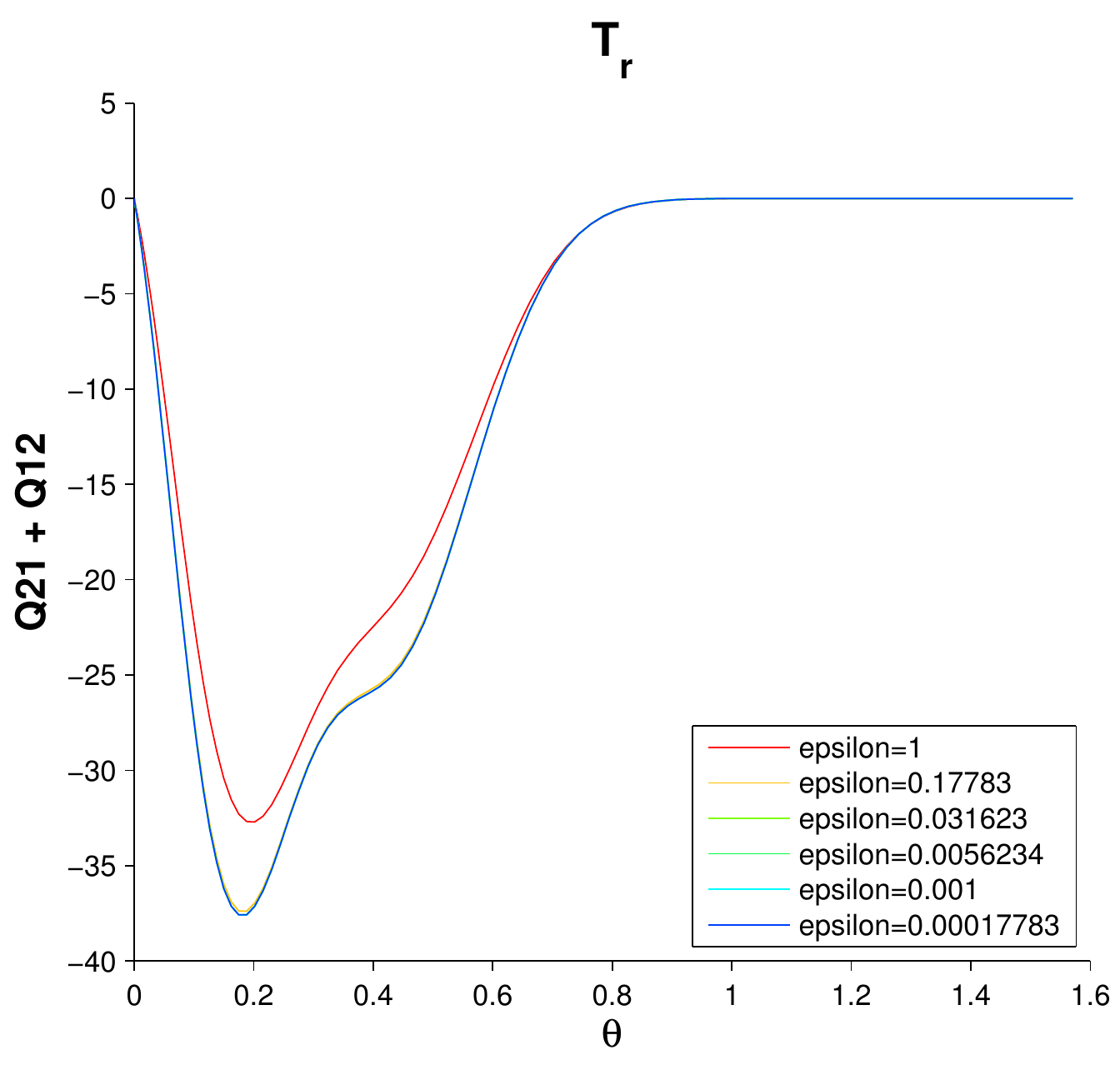}
     \includegraphics[width=4cm]{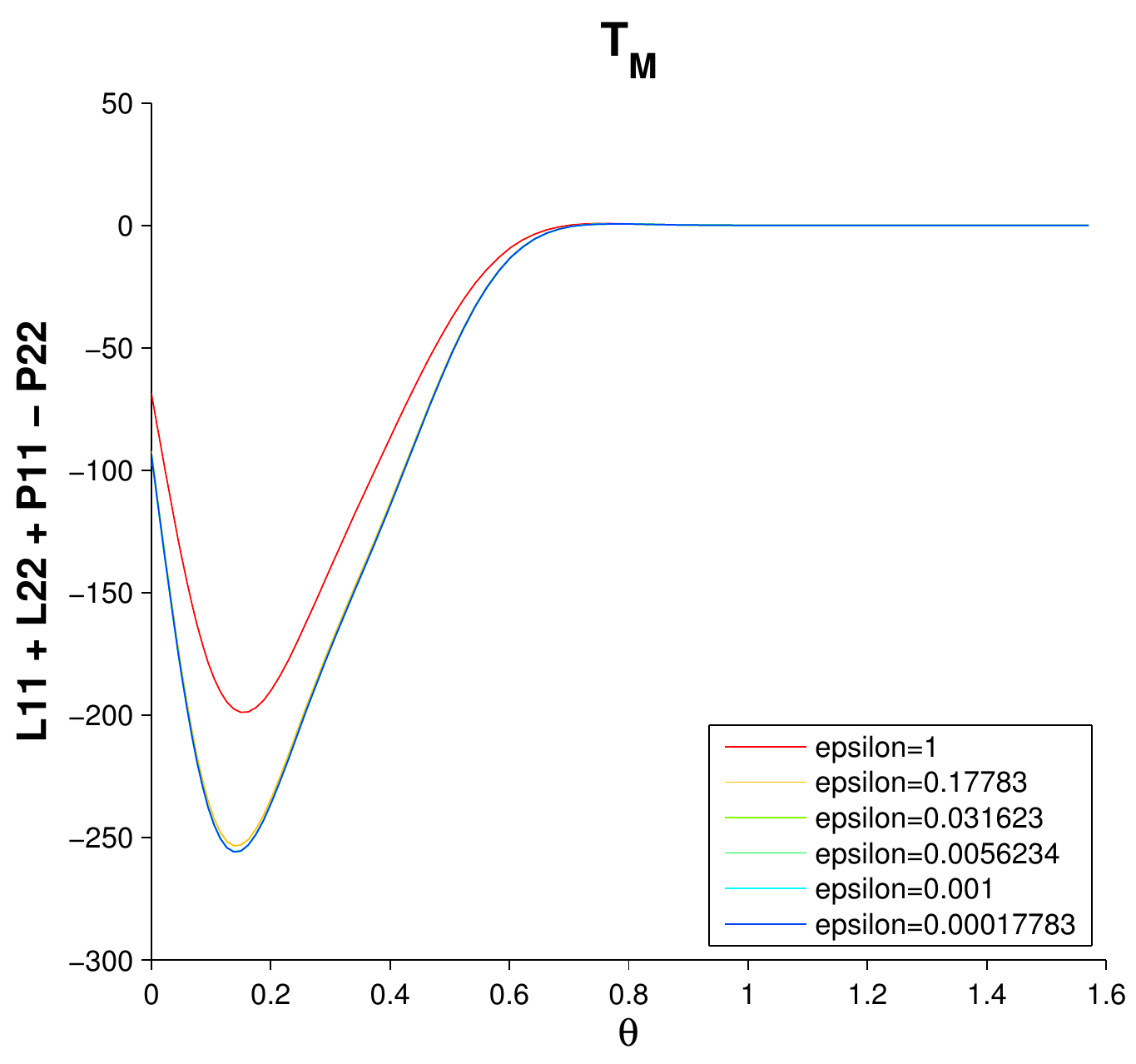}
      \includegraphics[width=4cm]{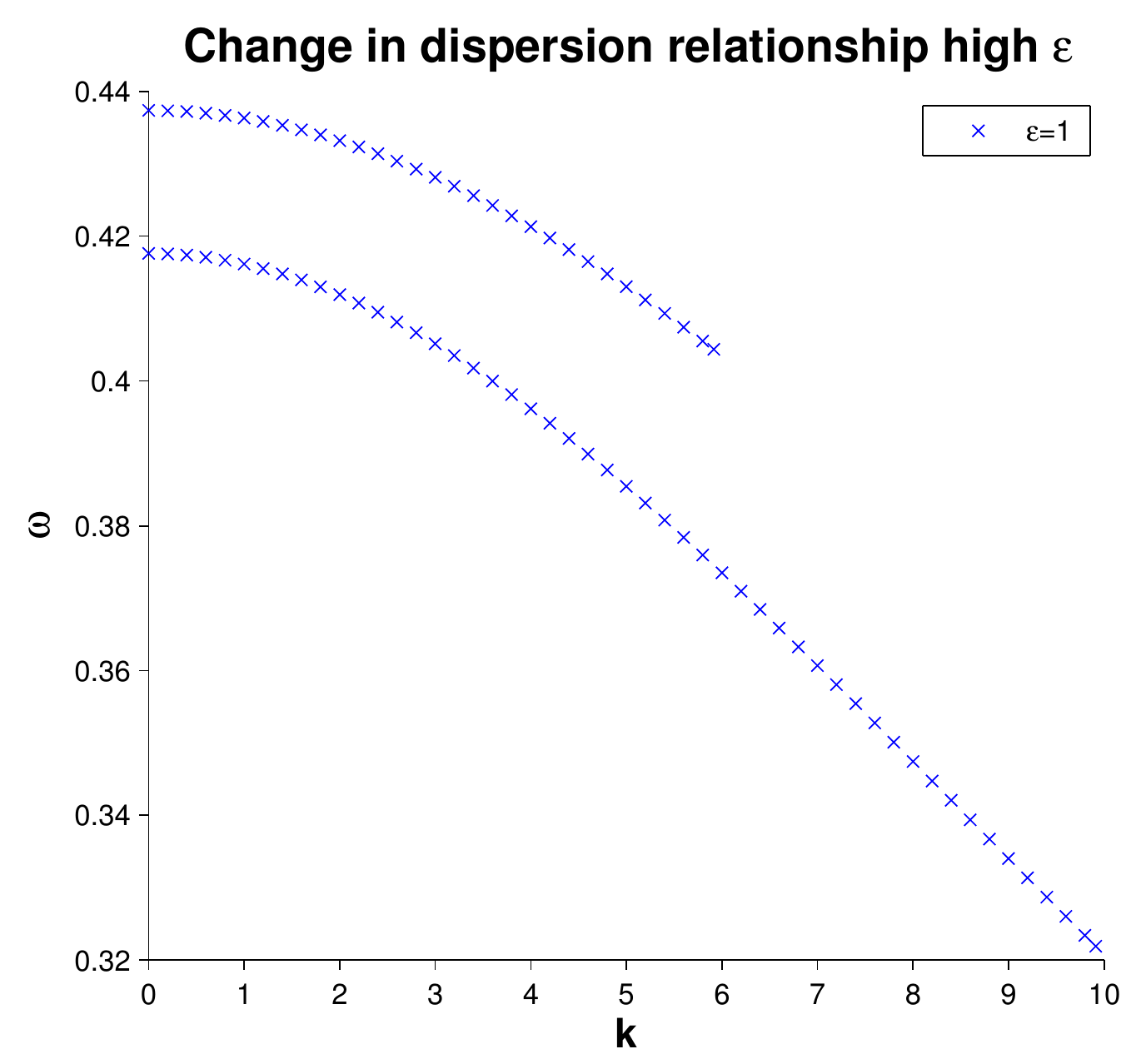}
    \caption[]{Varying $\epsilon$ with $\beta=-0.001$, $m_0=1$, $m_{\psi}=10$ and $\mu=-15.7154$. Increasing the value of $\epsilon$  has a similar effect to increasing the magnitude of $m_0$ or $\mu,$ resulting in lower charge densities near the cap off and a slight movement of this cap-off towards the IR.}
\label{fig:eps_full}
\end{figure}

\begin{figure}[h]
\vspace{-0.5cm}
\center
      \includegraphics[width=3.5cm]{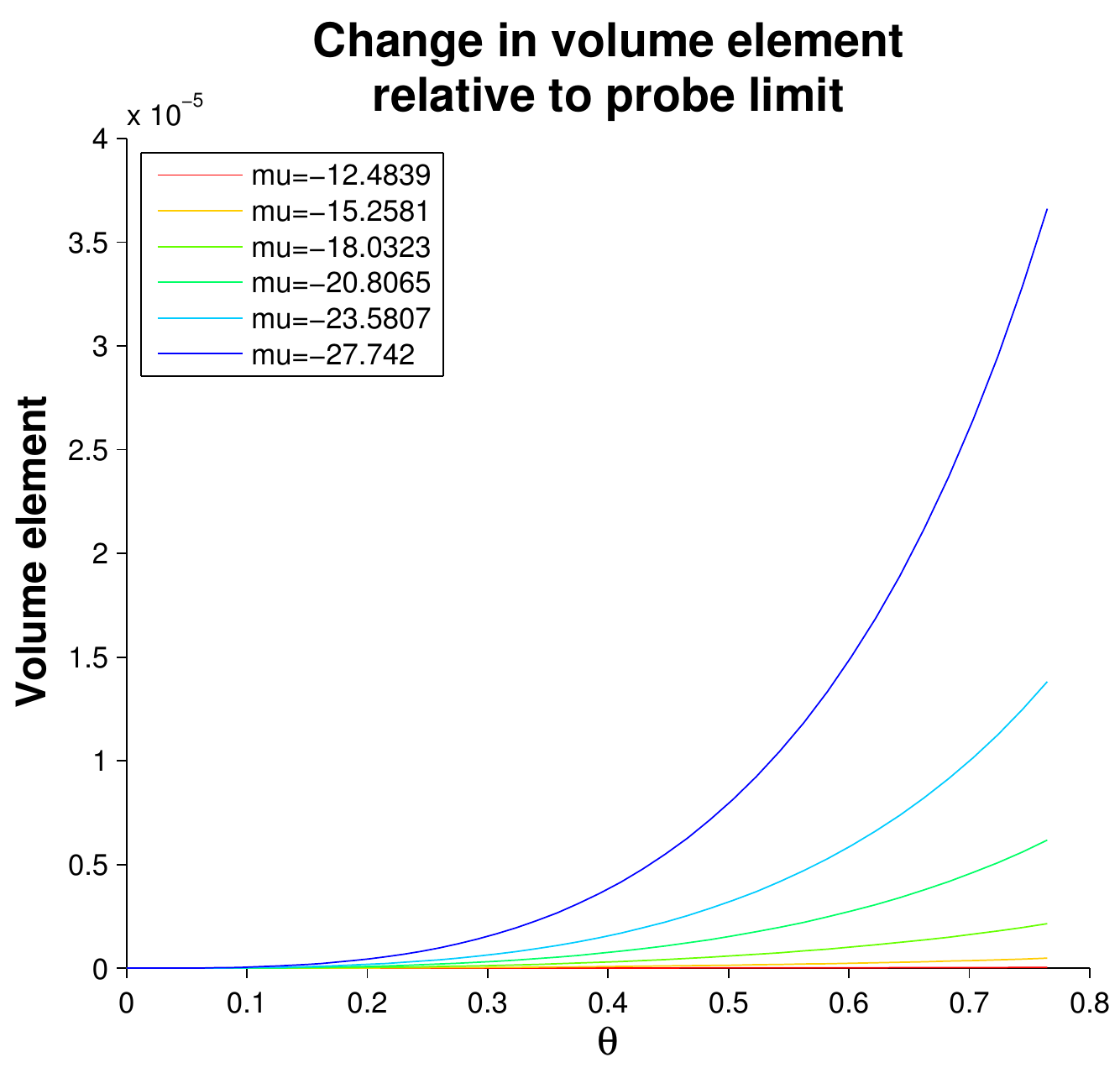}
      \includegraphics[width=3.5cm]{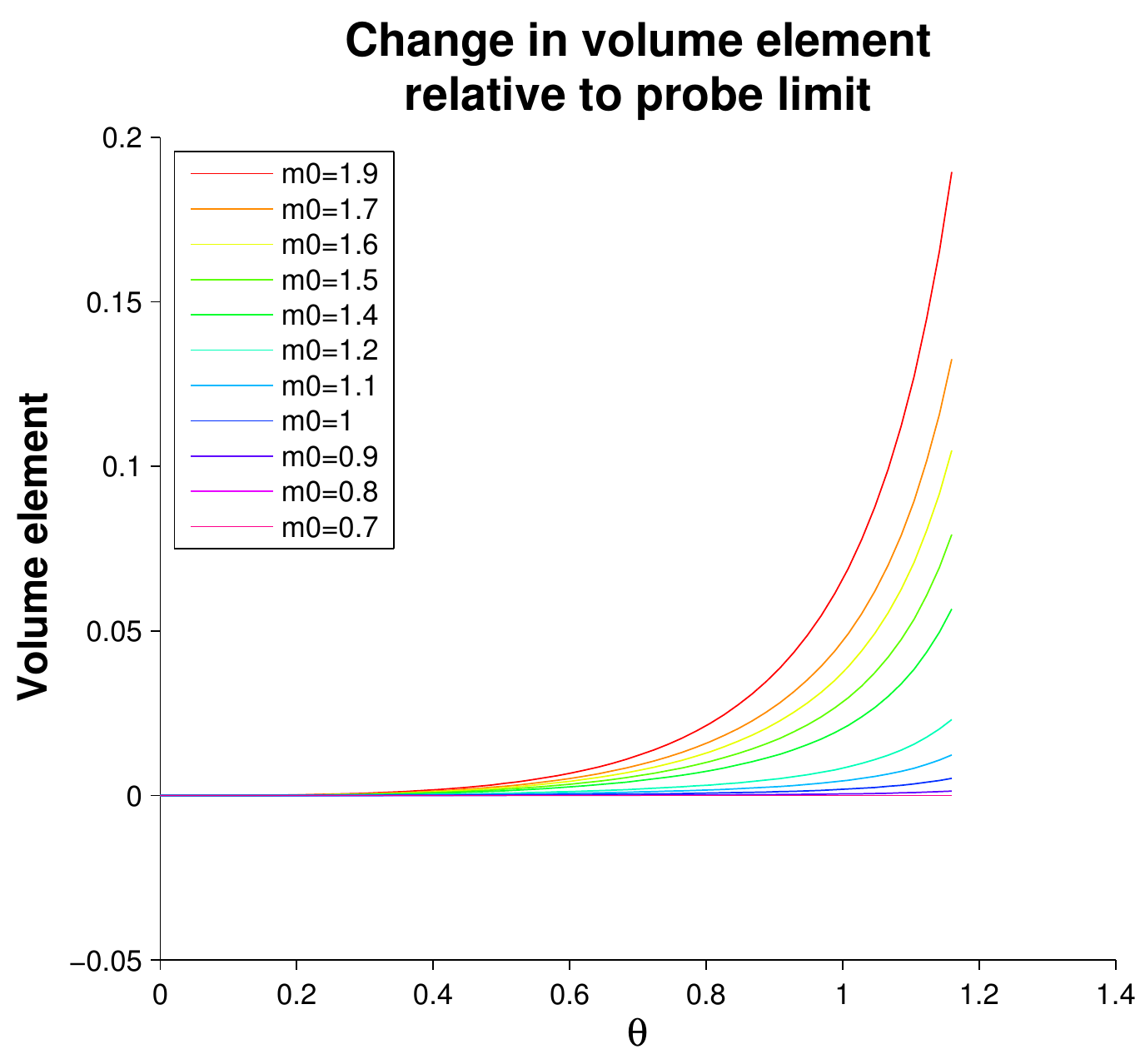}
      \includegraphics[width=3.5cm]{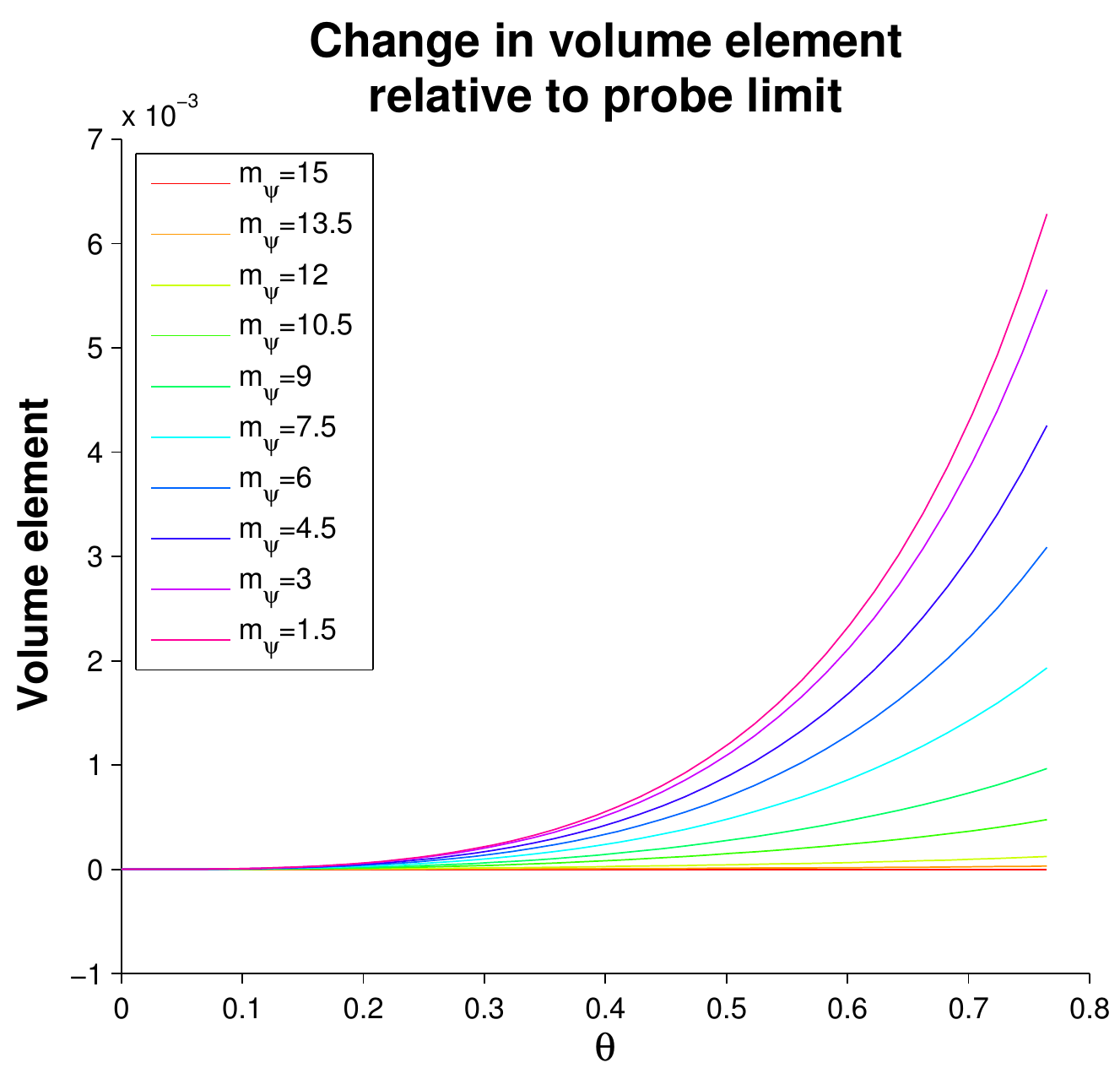}
      \includegraphics[width=3.5cm]{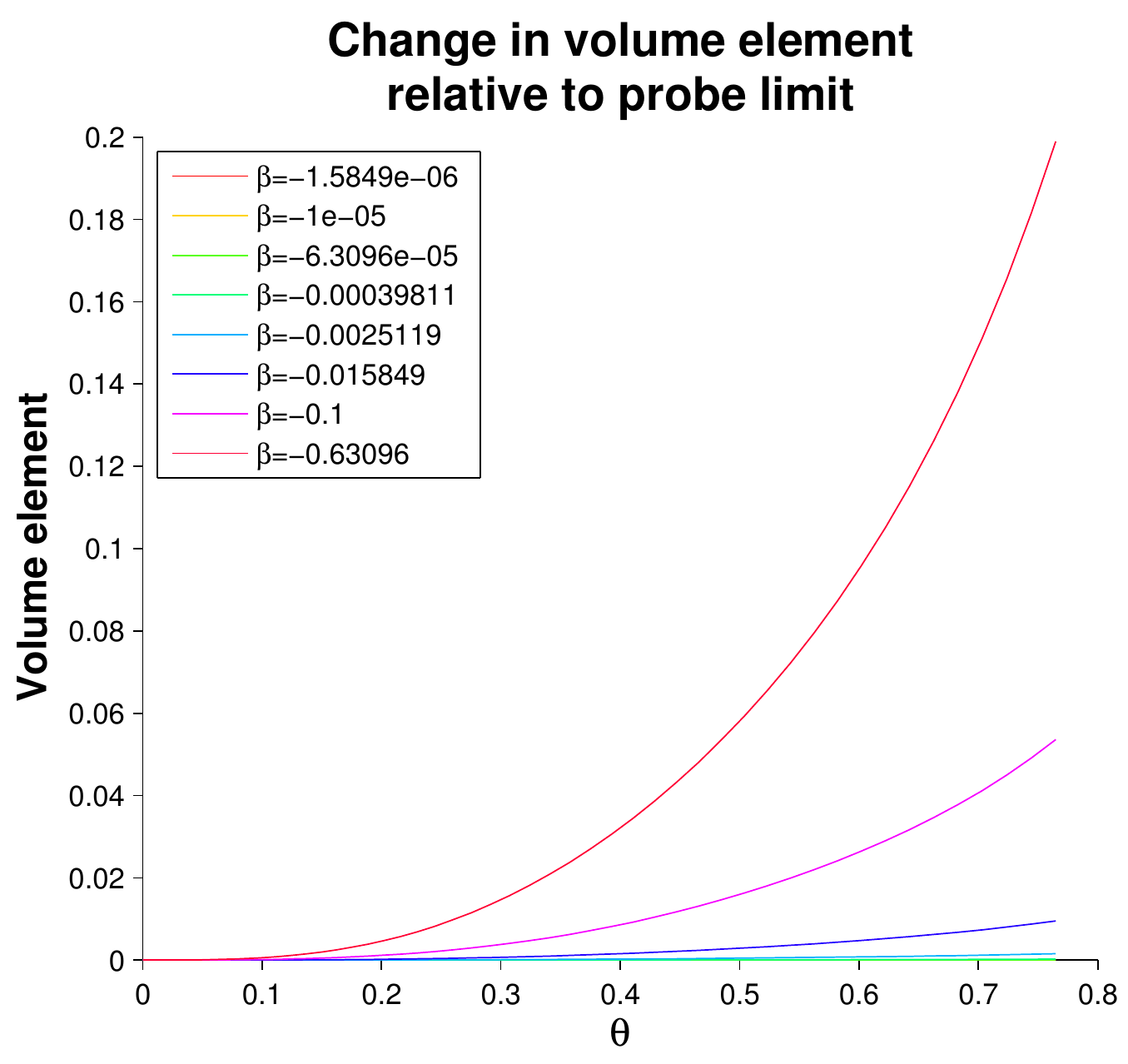}
      \includegraphics[width=3.5cm]{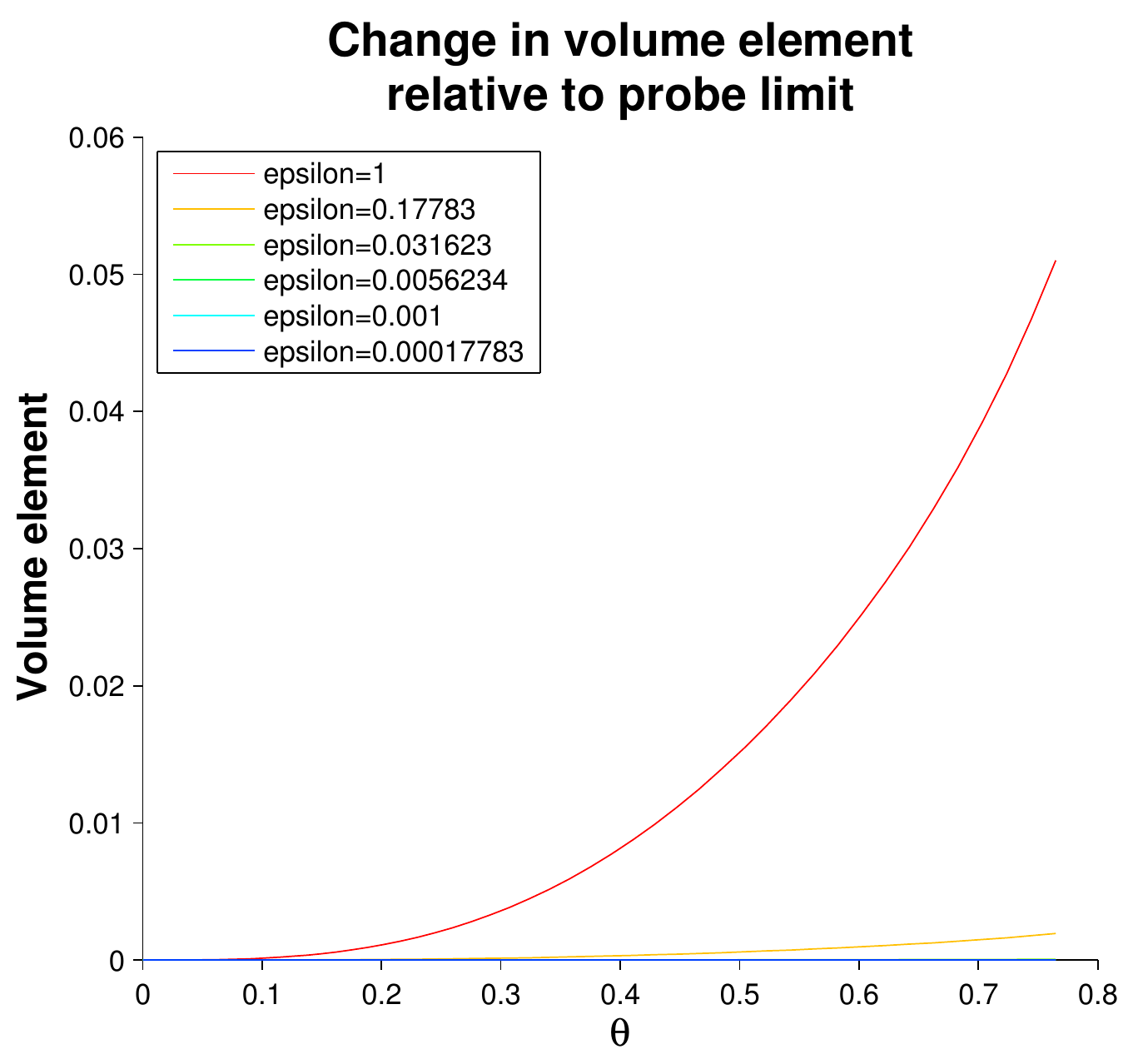}
    \caption[]{The change in volume near the embedding cap off. Higher fermion densities tend to  increase the volume element, relative to the probe limit.}
\label{fig:pd_graph}
\end{figure}

Finally, in figure \eqref{fig:pd_graph} we examine the change in the volume element from its the probe limit value, for the parameter variations described previously. We see that generically the inclusion of backreaction leads to an increase in the volume element near the cap-off. This difference increases for increasing $|\mu|$, $m_0$, $|\beta| $ and $|\epsilon|$ corresponding to states of larger charge density. For variations of $m_{\psi}$ the change of the volume element is maximized when the change in the embedding itself is greatest (low $m_{\psi}$). These results provide a tentative indication that the high density limit may be an interesting regime to probe for possible non-Fermi liquid behaviour.  

\section{Boundary Fermions}\label{section4}
With the above constructed bulk solutions, we can proceed to probe the behaviour of the dual QFT in the state we constructed. Our principle aim in this section is to demonstrate that the states we are considering are those of a Fermi liquid. We also wish to keep in mind future generalizations which could produce non-Fermi liquid behaviour. To this end we study the equation of state as a function of various parameters and examine the analytic properties of the fermion Green's function.

\subsection{Fermion Density}
         
\begin{figure}
\center
\vspace{-1cm}
       \includegraphics[width=3.5cm]{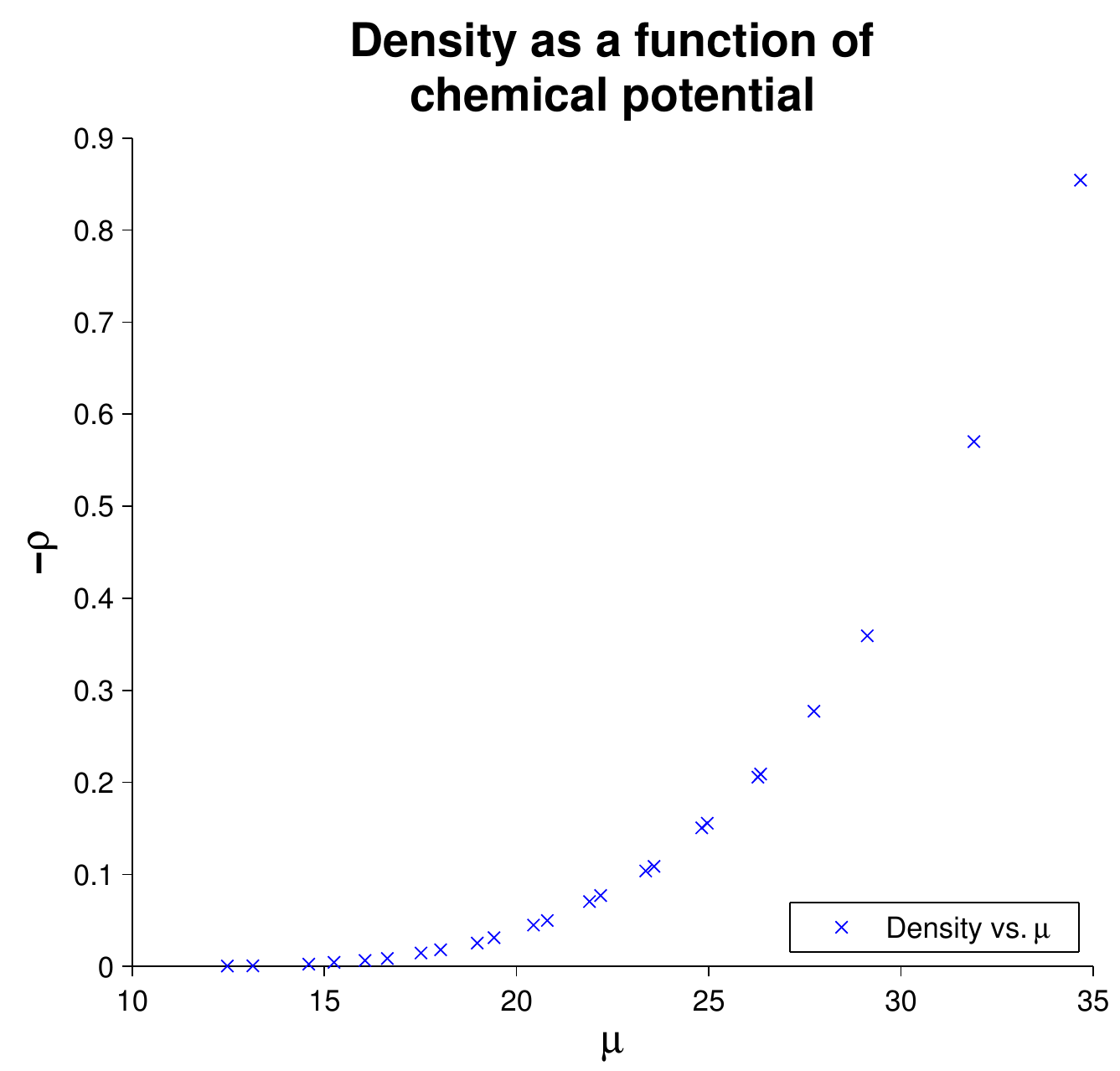}
       \includegraphics[width=3.5cm]{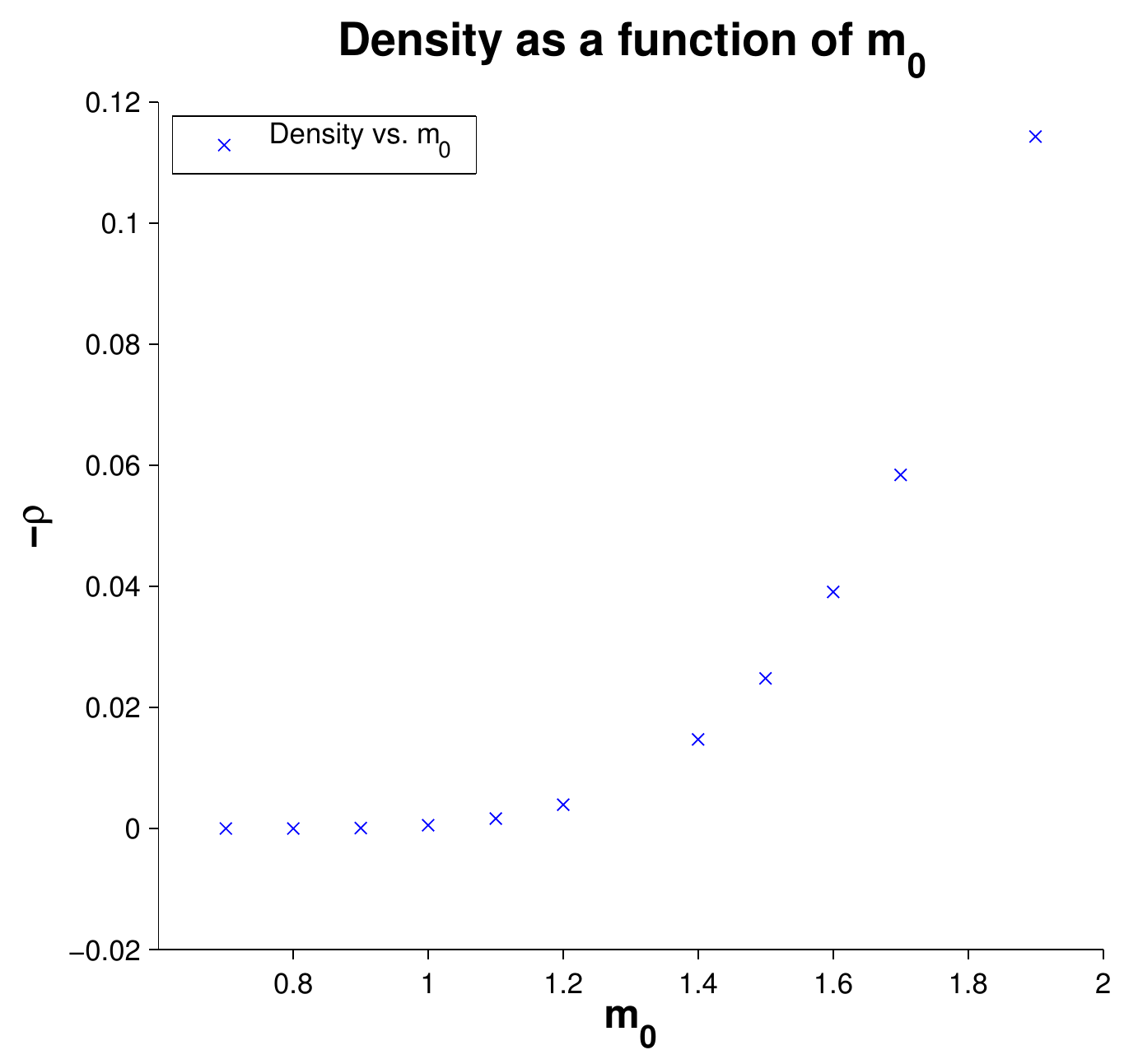}
        \includegraphics[width=3.5cm]{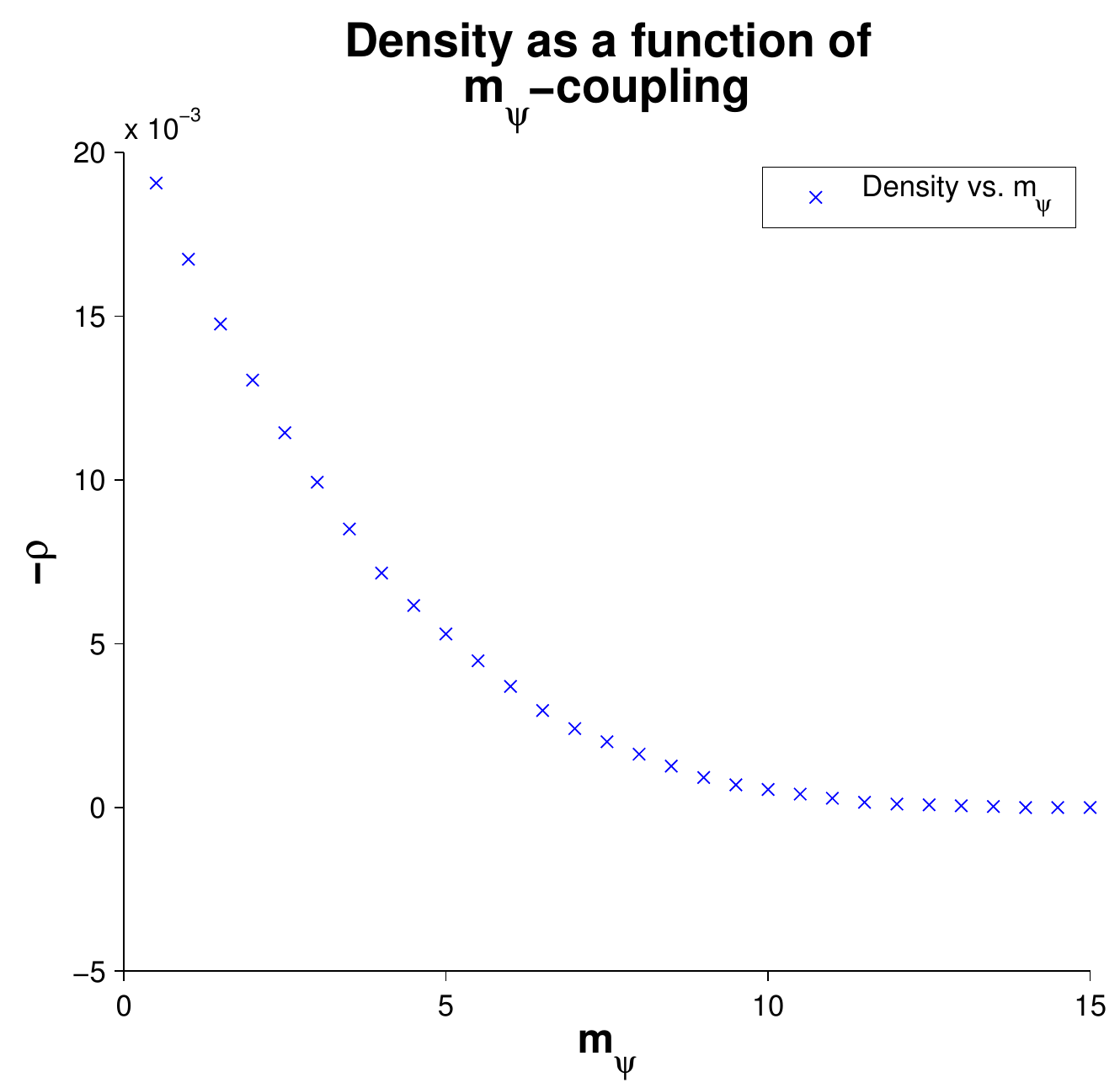}
        \includegraphics[width=3.5cm]{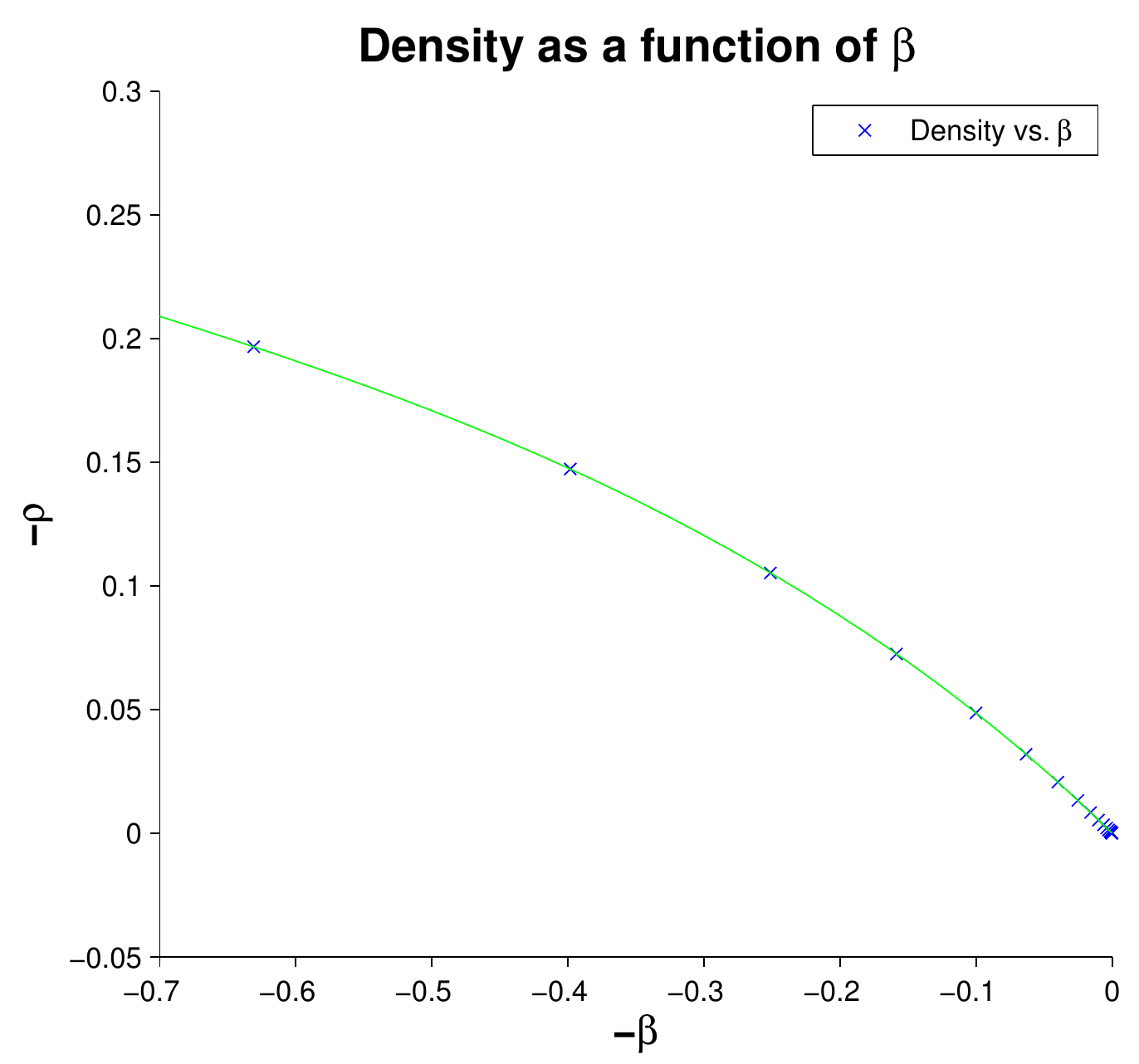}
         \includegraphics[width=3.5cm]{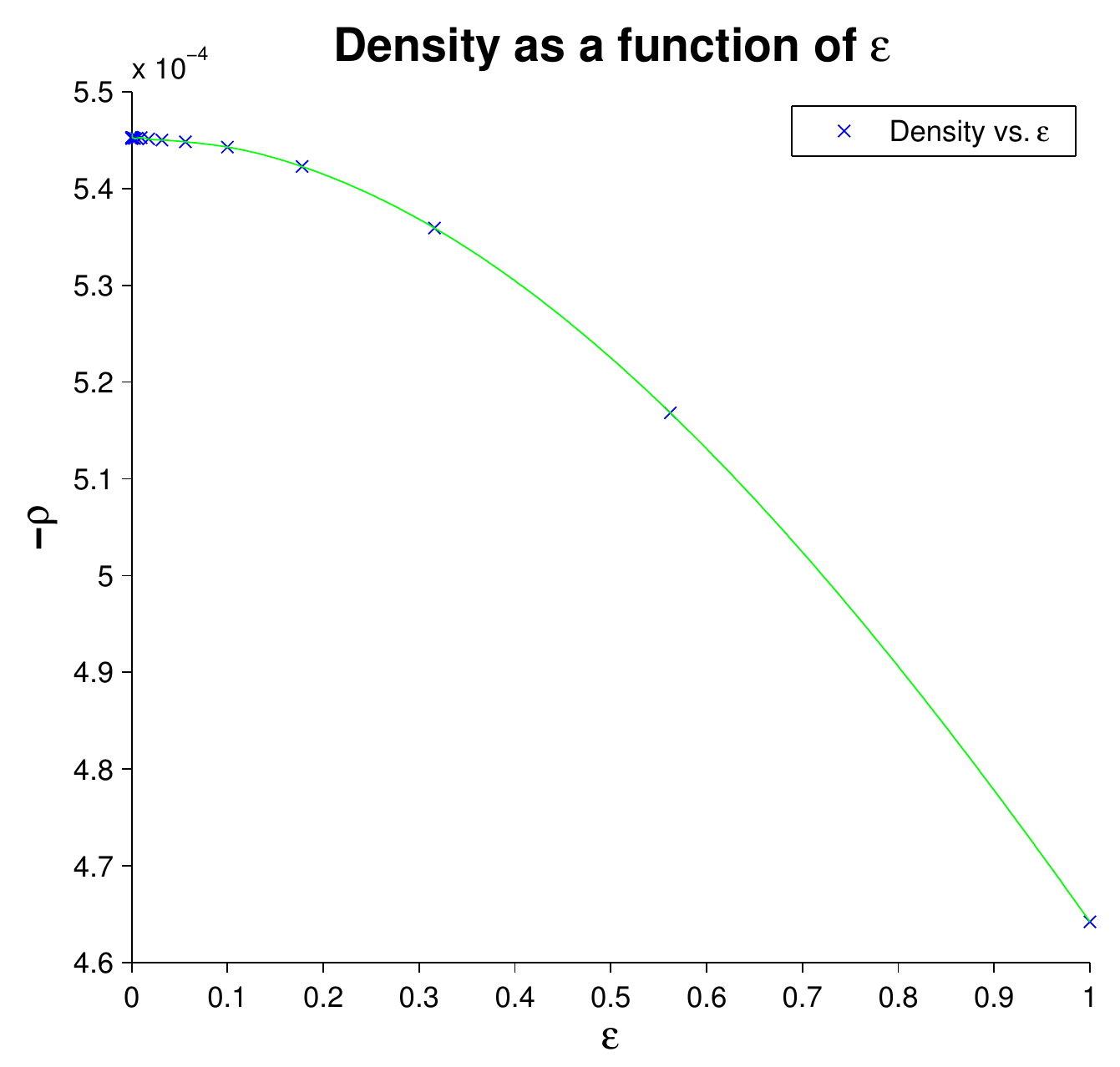}
    \caption[]{The boundary fermion density for the
various parameter ranges discussed in the text.
The plots are calculated in the backreacted case, though they
are not greatly influenced
by the backreaction.}
\label{fig:eqn_state}
\end{figure}     

In figure \eqref{fig:eqn_state} we examine the fermion density of the boundary theory as a function of the various parameters discussed previously. We start by discussing the equation of state: the dependence of the density on the chemical potential. Plotting $\rho$ versus $\mu$ we see that the system is   gapped,  with the mass of boundary fermions determined by the intercept of that plot. For the parameters chosen in figure\eqref{fig:eqn_state} $m_{QFT} \simeq 12.2173$. 

One of the basic signals of a Fermi surface is that the state
is \textit{compressible}, i.e that $\frac{d\rho}{d\mu}\neq 0$
for all $\mu$ for which the Fermi surface exists (i.e. for chemical potential above the gap). Our plot demonstrate
that our model indeed has  this feature.
We also fit the asymptotic behaviour of the curve to the  form $\rho \propto b \mu^{a}$ , with $b=5.945 e^{-5}$ and $a=3.064$  for the choice of parameters plotted. Note that for a single species of fermions in 2+1 dimensions, the exponent $a$ ranges between one (non-relativistic fermion)  to two (relativistic fermion), in the regime of asymptotically large $\mu$. The asymptotic scaling we find is a result  of the multiple species of fermions which exist in our model due to the presence of multiple bulk bands.
At asymptotically large chemical potential large number of these bands are occupied.

Examining the density as a function of other parameters is also interesting.  We find that the density decreases with $m_{\psi}$ and increases with $m_{0}$. This is in line with our expectations from the observed behaviour of the bulk fields in figures\eqref{fig:mb_probe} and \eqref{fig:m0_probe}. Zero density was found to occur at a finite value of $m_0$ which we label as $m_{0}^{critical}$ (in the case plotted in figure\eqref{fig:eqn_state} $m_{0}^{critical} \simeq 0.701$). This indicates that compact embedding solutions only exist for solutions which penetrate sufficiently far into the bulk. Fitting the density to the form $d (m_0-m_{0}^{critical})^c$ we find $c=0.05791$, $d=3.77$. Similarly $m_{\psi}^{critical}$  was found to be $\simeq 13.9981$ in our current case. 

The last two graphs indicate that $\rho$ increases with $|\beta|$ and $|\epsilon|$ in some nontrivial manner. Again this is to be expected from the plots of the bulk fields in figures\eqref{fig:beta_probe} and \eqref{fig:eps_full}.

\bigskip

\subsection{Retarded Green's function}\label{Greenfunc_sec}
The retarded Green's function in the vicinity of the Fermi surface has the form 
\begin{align}
G_R (\omega,k)= \frac{Z}{\omega - v_F (k-k_F)+\Sigma (\omega,k)}
\end{align}
The quasiparticle decay rate is determined by the self-energy $\Sigma$. In the case of  a Landau-Fermi liquid it is known to universally scale as $\Sigma = \frac{i \Gamma}{2} \sim i w^2$, and therefore the quasiparticle lifetime diverges as the Fermi surface is approached. In the case of non-Fermi liquids $\Sigma$  scales faster than $\omega $\footnote{There will be an additional logarithmic term in the case of a Marginal Fermi Liquid.} at low frequencies. This indicates that a quasiparticles are not a good description of the physics close to the Fermi surface.

\begin{figure}
\center
\vspace{-1cm}
    \hspace*{1cm}
    \includegraphics[width=6cm]{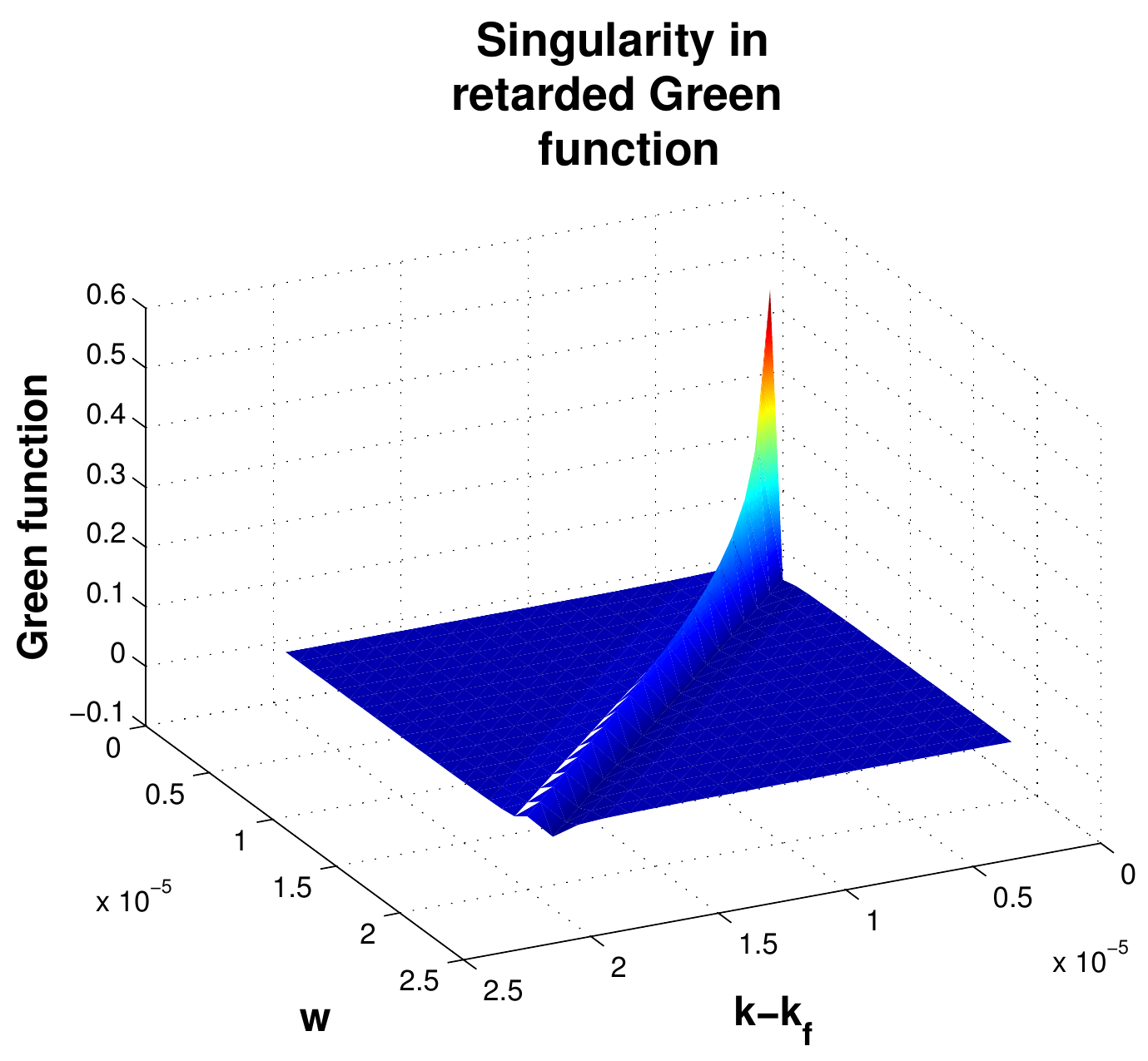}
    \caption[]{The multiple filled bands for bulk fermions translate
into  multiple Fermi surfaces in the boundary theory.
Each of these Fermi surfaces is associated with a pole in the
retarded Green's function. Here we plot one illustrative example
of the structure of the Green's function in the $(k,w)$ plane
in the region near one Fermi surface at $k_f=15.3676$.}
\label{fig:Green_func}
\end{figure}
\bigskip

The first step is to identity the presence of a Fermi surface in the dual QFT.  As we noted previously there are two distinct fermion modes as $u \rightarrow 0$:
 \begin{enumerate} 
 \item $\lim_{\theta \to \frac{\pi}{2}} (f_1,f_2) =(c_0,0)$ \label{opt1}
 \item $\lim_{\theta \to \frac{\pi}{2}} (f_1,f_2) =(0,d_0)$.  \label{opt2}
 \end{enumerate}
A priori both quantizations of the fermions are possible, however in constructing our background  we picked the first by setting $f_2(\pi/2)=0$.  As our background should be constructed from purely normalizable modes we label mode (1) as the normalizable mode. Mode (2) should then be considered as the source which we  turn on to perturb the system. 

As the retarded Green's function is  proportional to the ratio of response to source, we wish to calculate the ratio of the asymptotic values of the linearized perturbations $\delta f_1$  and $\delta f_2$. For the purpose of identifying the location of the boundary Fermi surface it is sufficient to consider only the linearized fermionic equations.

In order to identify the singularity in the Green's function it is convenient to use shooting techniques. Once the singularity in the Green's function was identified, the behaviour in the vicinity of the pole in the $k,w$ plane can be seen in figure \eqref{fig:Green_func}. We note immediately that our Green's function is purely real and therefore lacks any information about the quasiparticle decay rate. Mechanically this  is the result of the fact that our Dirac equation, $\Gamma$ matrix algebra and eigenfunctions are purely real. This, together with the fact that our boundary conditions in IR are simple regularity conditions, ensures that our Green's function is real\footnote{In other works such as \cite{Faulkner:2009wj,Faulkner:2013bna} the ingoing boundary conditions associated with a black hole horizon render the solution complex.}. Less mechanically, the fact that we lack a black hole horizon means we do not have a mechanism for dissipation at leading order in the $\frac{1}{\mathcal{N}}$ expansion. In order to calculate the decay rate it would be necessary to calculate  1-loop diagram (for references with such 1-loop calculations, see \cite{Faulkner:2013bna, Denef:2009kn,Denef:2009yy}). The fact that the decay rate will be parametrically small indicates that this phase  is  a Fermi-like liquid, as long as perturbation theory is valid.

 Indeed, one of the motivations for the present work is identifying potential sources for such a breakdown, for example the geometry becoming non-compact in the IR. This intuition ties in with  \cite{Faulkner:2009wj} where the infinite proper distance of the $AdS_2 \times \mathcal{R}_2$ throat geometry is dual to  the low energy bosonic modes necessary to  provide fermionic quasi-particle dissipation. We hope to return to this  direction in the future.
\bigskip
\section{Concluding Remarks}
In this study we have constructed a  phase of holographic matter
with Fermi liquid like behaviour, by solving for finite density
fermion configurations on Minkowski embedding of a probe D-brane.

Possible generalizations of this setup may be useful in the study
of non-Fermi liquids. We hope that our study will help identify
the regimes of parameter space where such behaviour may be expected.
For example, a sufficiently large deformation of the embedding
may introduce large renormalization effects,  invalidating the
purely classical approximation used here. In such regime a more
careful study, along the lines of  \cite{Allais:2012ye,Allais:2013lha}, would
be needed to determine the nature of the fermionic state on the
D-brane.

More generality, the ability to construct a finite density Minkowski
brane configurations could be useful in a range of holographic models, as those are expected to be qualitatively different than the finite density black hole embeddings.
For example, the compact nature of the world volume
geometry means the charged degrees of freedom lack an efficient
mechanism of dissipation, therefore such phases can be useful in exploring the physics of holographic insulators.

\bigskip
\section*{Acknowledgements}

M.R thanks Centro de Ciencias de Benasque Pedro Pascual, IPMU at the University of  Tokyo and INI at Cambridge University for hospitality during the course of this work. This work is supported by the Natural Sciences and Engineering Research Council of Canada.
\bigskip

\section*{Appendix A: Equations of motion}\label{Charges}
The sources in the equations of motion correspond to expectation values of fermion bilinears, with factors of $k$ and $\omega$ absorbed into the definition for later numerical convenience. In combining these Grassmann quantities into bosonic fields it was important to specify a basis, which we choose to be satisfy  $gg^{\ast} \wedge gg=1$. Here $gg$ is taken to be an arbitrary spinor. The definitions of the bilinears  are then
\begin{align}\label{charges}
& Q11  = \int_{0}^{k_f} \frac{k dk}{\pi} \langle f_1^{\ast}  f_1 \rangle, \quad Q22 = \int_{0}^{k_f} \frac{k dk}{\pi} \langle  f_2^{\ast}  f_2  \rangle, \quad Q12 =\int_{0}^{k_f} \frac{k dk}{\pi} \langle  f_2^{\ast} f_1  \rangle, \nonumber \\
& Q21=Q12, \quad P11 =\int_{0}^{k_f} \frac{k dk}{\pi} k \langle f_1^{\ast}  f_1   \rangle, \quad  P22 =\int_{0}^{k_f} \frac{k dk}{\pi} k  \langle f_2^{\ast}  f_2  \rangle, \nonumber \\
& L11 =\int_{0}^{k_f} \frac{k dk}{\pi} \omega(k) \langle f_1^{\ast}   f_1 \rangle, \quad L22 =\int_{0}^{k_f} \frac{k dk}{\pi} \omega(k) \langle f_2^{\ast}   f_2 \rangle \nonumber 
\end{align}  
We identify the combinations of $Q=(\text{Q11}+\text{Q22}) $, $T_r=(\text{Q12}+\text{Q21})$, and $T_M=(\text{L11}+\text{L22}+\text{P11}-\text{P22})$ as the charge, and radial and Minkowski ``stress-energy" components, respectively. The current is the source term for the gauge field, while in analogy to the gravitational case we refer to the sources for the embedding function as ``stress-energy". Using the form\footnote{We note that without the symmeterization thus is just the usual fermion stress energy tensor, as calculated via Noether's theorem. The additional symmeterization makes the stress energy manifestly symmetric and is most simply obtained via appropriate variation of the Dirac action in terms of the metric as in \cite{Weldon:2000fr}. In our case we  must pullback the bulk form of this tensor onto the brane worldvolume.} of the fermion contribution to the stress energy 
\begin{equation}\label{ferm_stress_en}
T_{fermi}^{M N}  = \beta \frac{i}{2} (\bar{\psi} \Gamma^{(M} \mathcal{D}^{N)} \psi-\mathcal{D}^{( M} \bar{\psi} \Gamma^{N ) } \psi )
\end{equation}
and the rescaling of the fermions given in \eqref{ferm_resc}, the ``radial" and ``Minkowski" stresses are seen to result from this expression by taking the  fermion derivatives  with respect to the radial and boundary directions, respectively.

   \bigskip
\section*{Appendix B: Numerics}\label{num_meth}
Finding the solutions presented in this paper required the use of Mathematica and Matlab for symbolic and numerical work, respectively. This appendix outlines some of the key tools used, pitfalls encountered and presents some convergence tests.
\bigskip
\subsubsection*{Numerical techniques}
The derivation of the equations was done using Mathematica, utilizing the package MathTensor \cite{christensen1990mathtensor}. The Grassmann package \cite{grassmann} was  useful for consistent manipulation of anti-commuting fields. An efficient way to export the equations to Matlab is using the freely available ToMatlab package \cite{ToMatlab}.
 In Matlab the equations were solved using the procedure described in section \eqref{sols_and_anal}. Here there are several points worthy of note:
\begin{itemize}
\item The general setup of the spectral code   follows the useful pedagogical reference \cite{trefethen2000spectral}.  Once the equations were discretized, the fermionic eigenvalue problem was solved using Matlab's ``eigs" function. The source terms  were then calculated via appropriate numerical integration. In calculating the source terms it was important to normalize the eigenvalues via $\int_0^{\pi/2} \mathrm{d} \bar{\psi} \psi =1$. This is necessary to ensure a consistent definition of the fermion energy $\omega(k)$ and therefore consistent charges, $J$, $ T_r$, $T_M$. It can be easily checked that with this normalization $\int T_{00} \ dr=w(k)=E$ in the limit of a flat induced metric and zero gauge field.

\item The discretized bosonic system was solved via a generalization of the Newton's method (see \cite{boyd2001chebyshev} for an a good reference on this and other spectral methods). It was also necessary to add a line search algorithm to this solver, as outlined in \cite{press1990numerical}, to improve convergence and stability properties. 

\item It is well known that spectral methods generally respond poorly to non-analytic behaviour.  Thankfully in our case the  simple asymptotic behaviour of all fields  ensures a good  convergence of spectral methods.
\item It was found that the bosonic and fermionic equations were regular at the embedding cap-off and at the conformal boundary. Therefore it was not necessary to impose cutoffs to regulate the problem.
\end{itemize}

\subsubsection*{Convergence checks}

\begin{figure}
\center
    \vspace*{-0.5cm}
    \includegraphics[width=4cm]{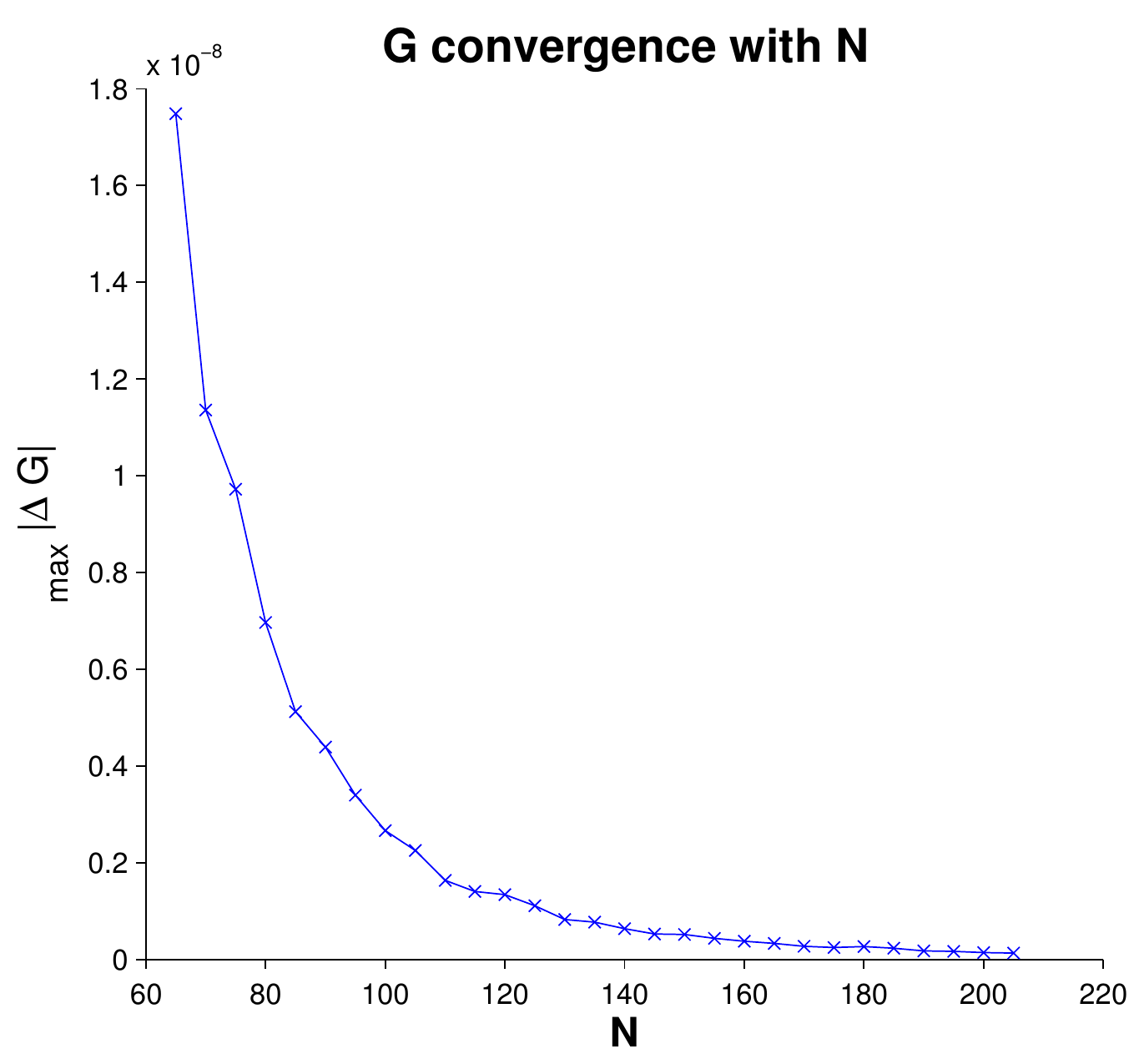}
    \includegraphics[width=4cm]{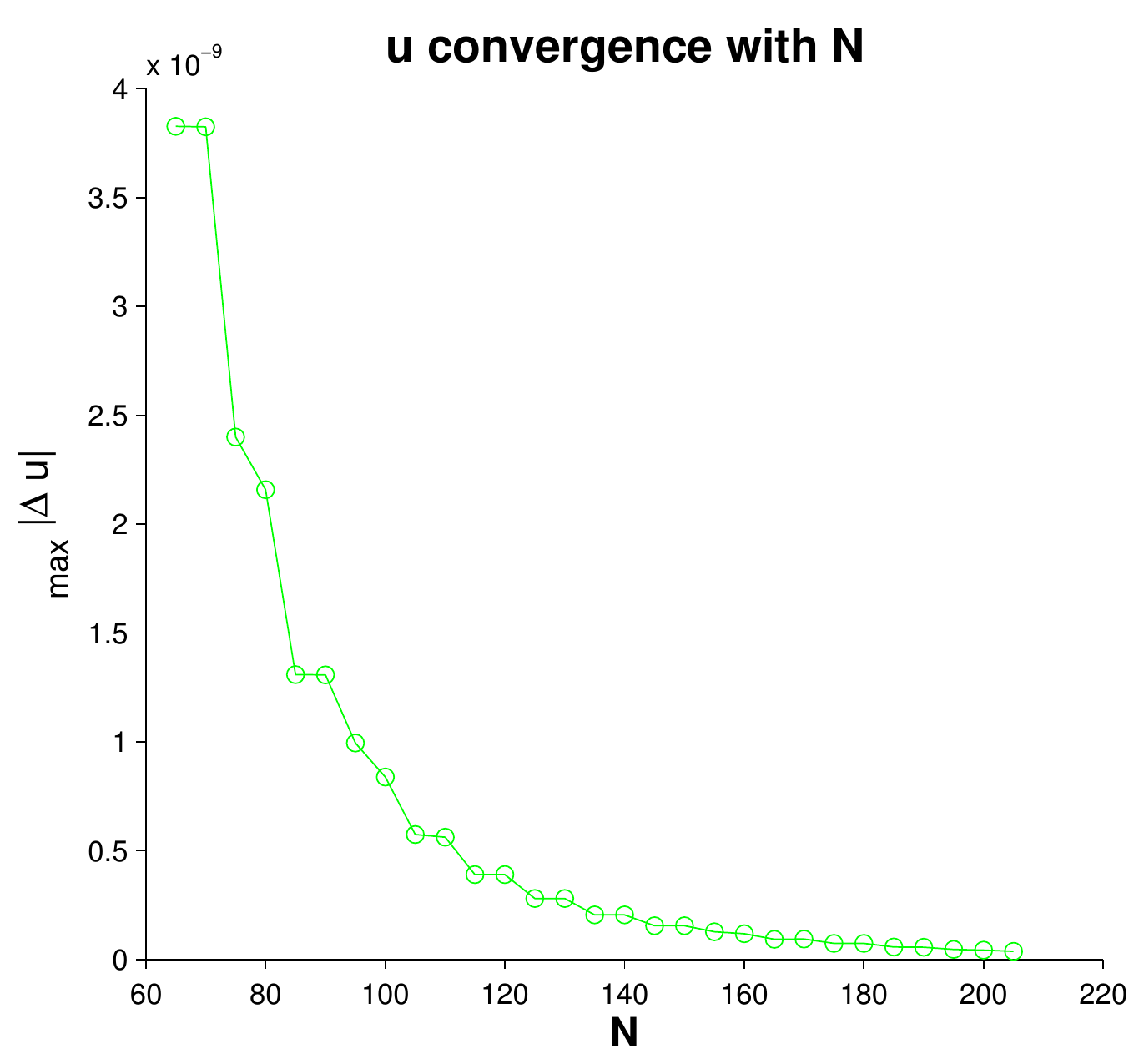}
    \includegraphics[width=4cm]{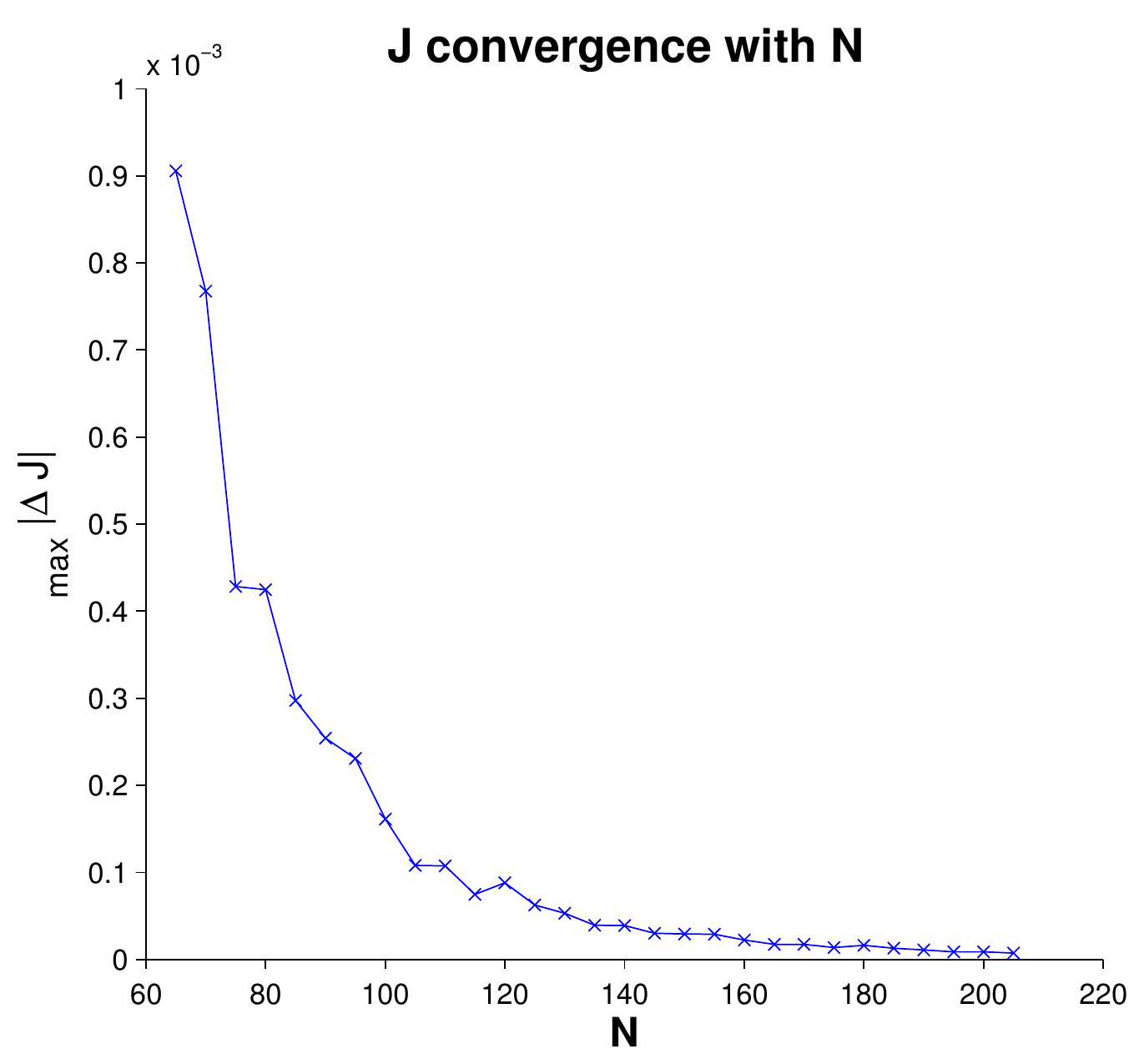}
    \includegraphics[width=4cm]{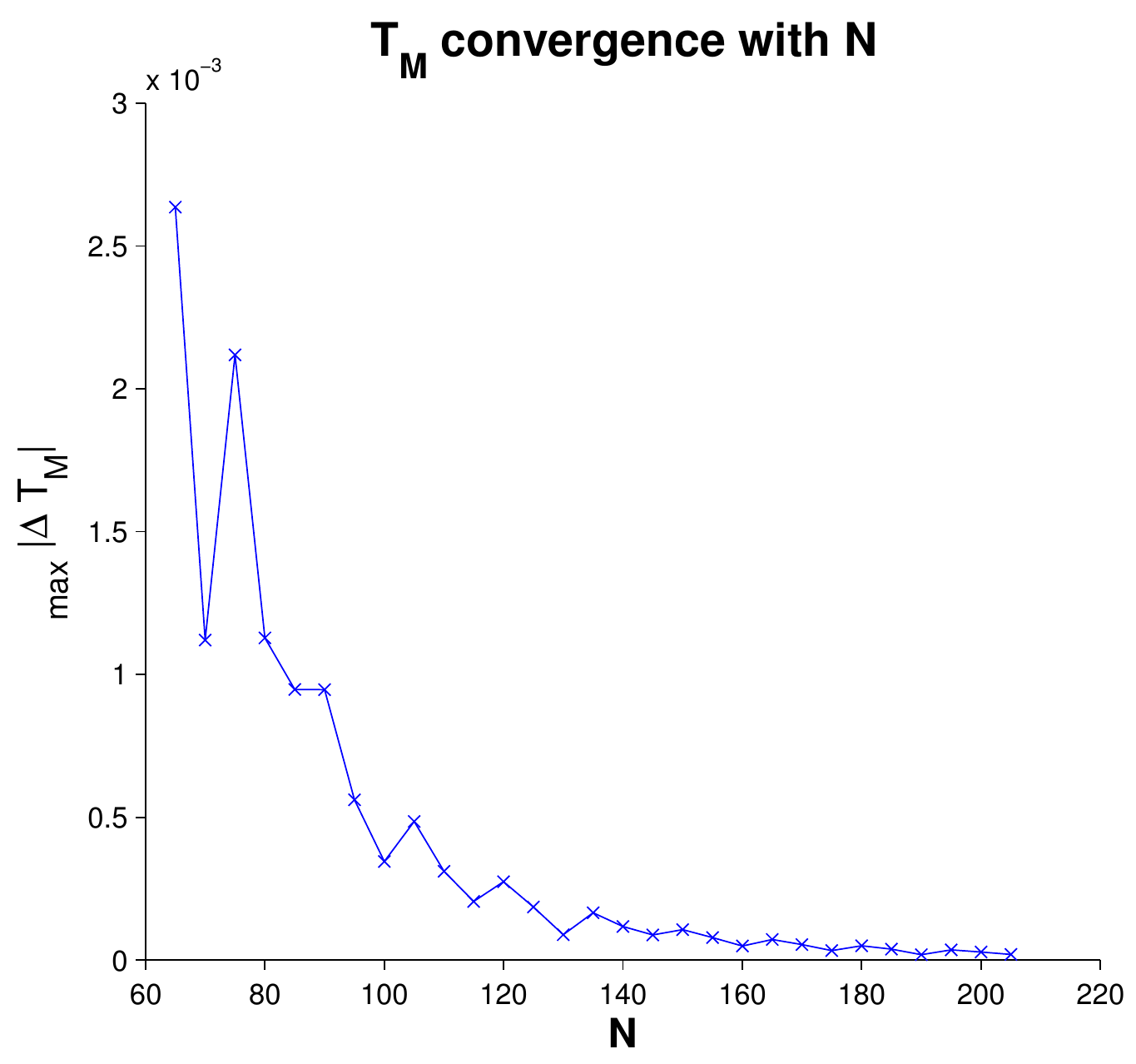}
    \includegraphics[width=4cm]{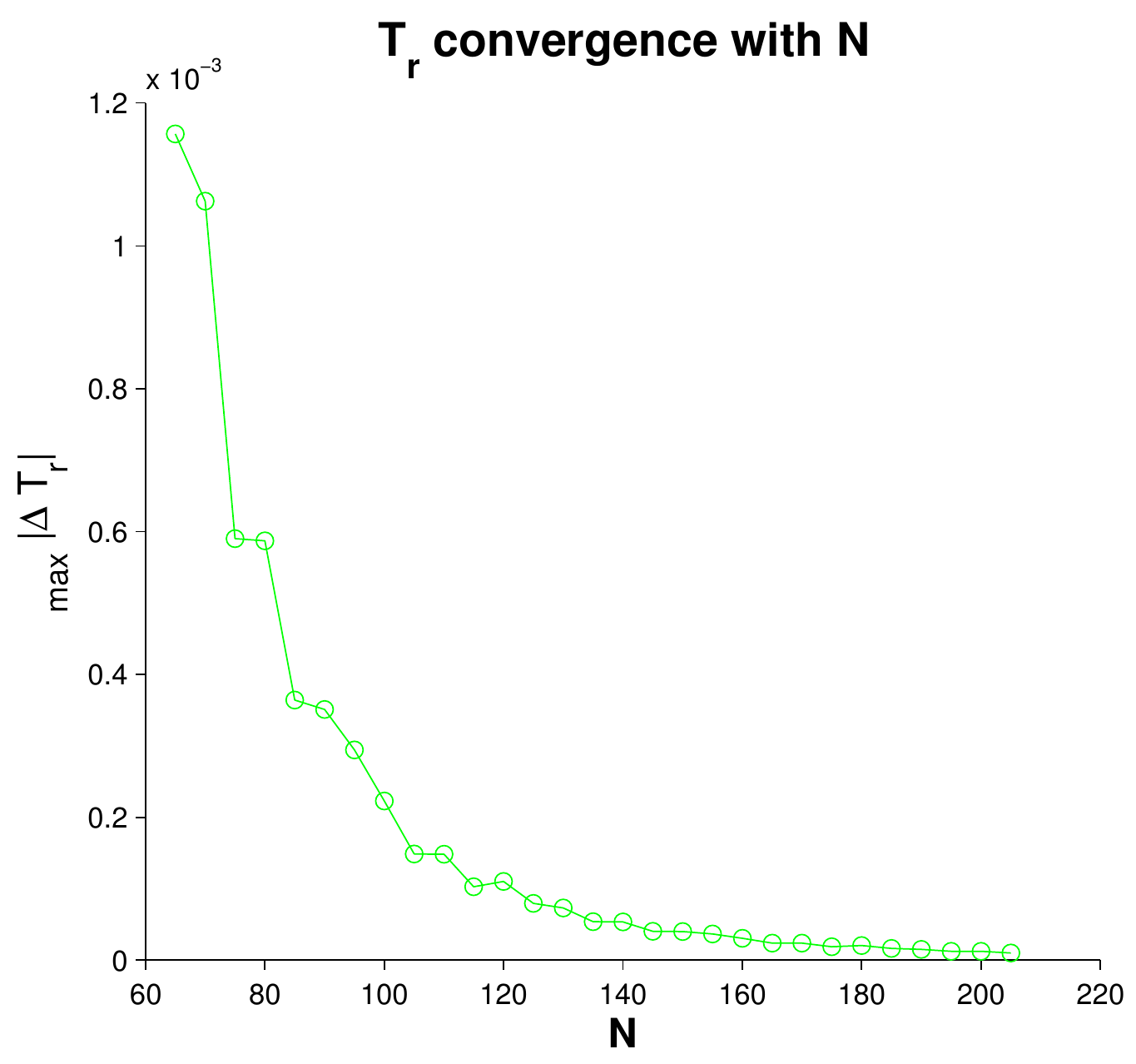}
    \caption[]{ The convergence with $N$ for the embedding function,
gauge field, and charges. The maximum difference between solutions
between successive values of $N$ is computed, and can be seen
to converge to zero as $N$ is taken large. The convergence tests
plotted were run at parameter values of $\mu=-15.7154$, $m_0=1$,
$m_{\psi}=10$, $\epsilon=0.01$, $\beta=-0.01$.}
\label{fig:N_converg}
\end{figure}

We now consider the accuracy of our numerical solutions. W present a series of plots which illustrate that our solutions converge to solutions of the integro-differential equations in the large $N$ limit (here $N$ is the order of the Chebychev approximation not the size of the adjoint gauge group). It was particularly important to verify that our solutions for the embedding function are numerically robust. The key convergence criteria were:
 
\begin{itemize}
\item The stability and accuracy of the eigenvalue routines -- this was the easiest part to check as these were based on built-in Matlab functions. The default accuracy in using the ``eigs" routine is machine precision. Therefore this was not considered a significant source of error.
\item It was necessary to ensure that the both the $\mathcal{L}^2$ norm of the residues of the bosonic system is and the maximum change in the bosonic fields after a fermion/boson iteration loop were both driven to numbers of $\mathcal{O}(10^{-12})$.
\item The solution of the discretized equations converged as N was increased therefore indicating that we may be tending to a solution of the original ODE equations- see figure \eqref{fig:N_converg}.
\item It was necessary to check that integration procedures used in computing the charges were sufficiently accurate, and the change in the solutions with increasing tolerance tended to zero.\end{itemize}

\bigskip
\bigskip

\bibliographystyle{abbrv}
\bibliography{Chapter4_library}

\end{document}